\documentclass[a4paper,11pt]{article}

\usepackage{jheppub}
\usepackage[T1]{fontenc}
\usepackage{graphicx}
\usepackage{amssymb}
\usepackage{subfigure}
\usepackage{hyperref}
\usepackage{mathrsfs}
\usepackage{latexsym}
\usepackage{amsfonts}
\usepackage{amsmath}
\usepackage{esint}
\usepackage{amsxtra}
\usepackage{pifont}
\usepackage{lscape}
\usepackage{wrapfig}
\usepackage{tocvsec2}

\DeclareMathAlphabet{\mathpzc}{OT1}{pzc}{m}{it}

\allowdisplaybreaks[0]

\def \ep1{\epsilon_1}
\def \ep2{\epsilon_2}

\def \be{\begin{equation}}
\def \ee{\end{equation}}
\def \beq{\begin{eqnarray}}
\def \eeq{\end{eqnarray}}
\def \ba{\begin{array}}
\def \ea{\end{array}}

\def \f{\frac}

\def \p{\partial}

\def \sn{\,\mbox{sn}}
\def \cn{\,\mbox{cn}}
\def \dn{\,\mbox{dn}}
\def \ep{\epsilon}

\title{\boldmath Spectra of elliptic potentials and supersymmetric gauge theories}

\author[a,b]{Wei He}

\affiliation[a]{Physics and Space Science College, China West Normal University,\\Nanchong 637002, China}
\affiliation[b]{School of Electronic Engineering, Chengdu Technological University,\\Chengdu 611730, China}

\emailAdd{weihephys@foxmail.com}

\abstract{We study a relation between asymptotic spectra of the quantum mechanics problem with a four components elliptic function potential, the Darboux-Treibich-Verdier (DTV) potential, and the Omega background deformed N=2 supersymmetric SU(2) QCD models with four massive flavors in the Nekrasov-Shatashvili limit. The weak coupling spectral solution of the DTV potential is related to the instanton partition function of supersymmetric QCD with surface operator. There are two strong coupling spectral solutions of the DTV potential, they are related to the strong coupling expansions of gauge theory prepotential at the magnetic and dyonic points in the moduli space. A set of duality transformations relate the two strong coupling expansions for spectral solution, and for gauge theory prepotential.}

\begin{document}

\maxtocdepth{subsubsection}
\maketitle
\flushbottom

\section{Introduction}

In this paper we report results about a relation between N=2 supersymmetric SU(2) QCD with $N_f$=4 hypermultiplets and asymptotic spectral solutions of stationary Schr\"{o}dinger equation with the Darboux-Treibich-Verdier (DTV) potential \cite{darboux1882, TreibichVerdier2, TreibichVerdier3, TreibichVerdier4}. This is an example of the connection between the Omega background deformed N=2 supersymmetric gauge theories and quantum integrable models discovered by Nekrasov and Shatashvili \cite{NS0908}. The theoretical models discussed here are simple enough to dig up many details, meanwhile comprise of various nontrivial ingredients to demonstrate some aspects of this connection.

The moduli space of N=2 supersymmetric gauge theory in the Coulomb branch possesses K\"{a}hler structure, with the K\"{a}hler metric given by the holomorphic prepotential.
Seiberg and Witten invented a method to compute the prepotential using Riemann surface, therefore determined the effective action of gauge theory \cite{SW9407,SW9408}.
On the other hand, in a geometric approach to integrable system, a K\"{a}hler manifold is associated with complex integrable Hamiltonian system.
Indeed, some Seiberg-Witten gauge theory models are related to well known {\em classical} integrable systems, in particular the Seiberg-Witten curve is identified with the spectral curve \cite{HokerPhong}.
Nekrasov and Shatashvili proposed the quantum version of this relation, established a correspondence between N=2 supersymmetric gauge theory deformed by the Omega background in the limit $\epsilon_1\to\epsilon,\epsilon_2\to 0$ and {\em quantum} integrable systems with the Plank constant $\epsilon$. The equations determining supersymmetric vacuum of gauge theory are identified with the Bethe Ansatz equations of integrable model, and the v.e.v of scalar operators $\langle\text{tr}\phi^n\rangle$ are identified with the eigenvalues of quantum Hamiltonians $\mathcal{E}_n$ \cite{NS0908}.

The SU(2) $N_f$=4 super QCD is one of the examples used by Seiberg and Witten to demonstrate the low energy physics of N=2 supersymmetric gauge theory \cite{SW9407,SW9408}.
As a prototype of N=2 super QCD with fundamental hypermultiplets, it is interesting to analyse various properties of this model in detail.
There exist powerful tools including the Seiberg-Witten curve, instanton counting and topological string theory,
the spectral method provides a novel approach to further reveal some features of the $N_f$=4 super QCD.
We show that the spectral data for the DTV potential precisely produce the gauge theory prepotential.
More precisely, in the weak coupling region we check explicitly the relation of $\langle\text{tr}\phi^2\rangle$ in $N_f$=4 theory and $\mathcal{E}_2$ of the Schr\"{o}dinger equation;
while in the strong coupling region we compute the dual expansion of prepotential using quantum periods calculated from the Schr\"{o}dinger equation.

The Schr\"{o}dinger equation with DTV potential is the elliptic form of Heun differential equation \cite{heun1889, NISTDLMFCHAP31}.
The Heun equation arises in a variety of physics and mathematics problems, it is important to have a complete understanding of its spectral solutions,
various methods are used to find different kinds of solutions.
The DTV potential is a four components generalization of the Lam\'{e} potential, which in turn is the elliptic generalization of the Mathieu potential. Asymptotic solutions of the Mathieu equation were known for many years, asymptotic solutions of the {\em Lam\'{e} equation} and the {\em ellipsoidal wave equation} are also known now \cite{NISTDLMF, Muller1966, MullerKirstenQUANT, wh1412, wh1608}.
But asymptotic solutions of the same nature for the Heun equation become difficult to compute.
For what is known about the Heun equation until recently, see the books \cite{Ronveaux-Heun} and \cite{NISTDLMFCHAP31}.
Among recent attempts, a study using WKB analysis can be found in \cite{Takemura2006},
the Bethe Ansatz type algebraic equations that determine the solutions of equation are discussed in \cite{Takemura2003}.
In this paper we study the spectral problem from a different point of view, by considering its relation with the $N_f$=4 super QCD.
With information extracted from results of gauge theory, asymptotic spectral solutions are computed explicitly.

The results presented here are based on partial results obtained in \cite{wh1306}, here we give a complete and consistent analysis for this problem.
The relation between the spectrum of DTV potential and the effective theory of SU(2) $N_f$=4 super QCD is more general than the relation for other SU(2) models we studied before.
Still, we can recognize that the models on both sides basically share the same feature as other SU(2) models,
and with some skills we are able to explore their properties in all interesting regions of the parameter space.
The results obtained here confirm some facts and methods observed in other examples, provide more persuasive argument for some puzzling points encountered earlier, including the monodromy relations for the strong coupling solutions \cite{wh1108,wh1412,wh1608}.

The paper is organized as follows. In section \ref{CriticalPointsAndAsymptoticSolutions}, the relation between the Heun differential equation and the $N_f$=4 super QCD is explained,
in particular, the differential equation can be considered as a way to quantize Seiberg-Witten gauge theory.
We show there are six critical points for the DTV potential, based on results obtained earlier about the relation between periodic potential and N=2 SU(2) super QCD, the nature of these critical points can be determined. Four of them are associated with the weak coupling spectral solution of Schr\"{o}dinger equation, as well as gauge theory in the weak coupling region of moduli space (hence are electric);
the other two are associated with the strong coupling spectral solutions, as well as gauge theory in the strong coupling regions (hence are magnetic and dyonic).
This observation is crucial for calculation of the asymptotic spectrum.

In section \ref{WeakCouplingSpectrumAndN=2superQCD} we perform explicit computation in the weak coupling region of parameter space to confirm the relation between the spectrum of Schr\"{o}dinger equation and instanton partition function of super QCD in the Nekrasov-Shatashvili limit. The weak coupling spectral solution is the large eigenvalue perturbative solution of the DTV potential, the eigenvalue is expressed in elliptic (quasi)modular forms and the wave function is expressed in the Weierstrass elliptic function. The partition function of gauge theory with surface operator is computed using localization formula developed for instanton computation, we show that the instanton partition function precisely matches with the asymptotic spectral solution when parameters appearing on both sides are correctly identified, their relation is given by (\ref{lambdaPrepotentialElectric}) and (\ref{psiPrepotentialElectric}).

In section \ref{StrongCouplingSpectraAndGaugeTheoryDuality} we extend the spectral solution of Schr\"{o}dinger equation to the strong coupling region.
The existence and property of strong coupling spectral solutions are inferred from the Seiberg-Witten solution of super QCD at strong coupling.
It is explained that there is a local background expansion for the DTV potential at each strong coupling critical point, the resulting local ``averaged'' potentials are asymptotic series containing infinitely many terms and maintaining elliptic periodicity, hence are {\em elliptic Hill potentials}.
Written in the Jacobian elliptic function, the potential at the magnetic critical point is given by eq. (\ref{EllipticHillOperatorSN}),
the potential at the dyonic critical point is given by eq. (\ref{EllipticHillOperatorCN}).
Then we compute the asymptotic spectral solution of the effective potential (\ref{EllipticHillOperatorSN}),
using a dynamically generated large coupling constant as the expansion parameter.
The two elliptic potentials (\ref{EllipticHillOperatorSN}) and (\ref{EllipticHillOperatorCN}) are dual to each other in the same sense of monopole-dyon duality in gauge theory,
their spectral solutions are mapped to each other by a set of duality transformations given by eqs. (\ref{EigenvalueDyonicFromMagnetic})(\ref{EigenfunctionDyonicFromMagnetic}) and (\ref{EigenvalueMagneticFromDyonic})(\ref{EigenfunctionMagneticFromDyonic}).
The Floquet theorem for the strong coupling solution of differential equation with elliptic potential, which remains unexplained in the literature of spectral theory, is clarified using the asymptotic wave functions  (\ref{EigenfunctionStrongCouplingMagnetic}) and (\ref{EigenfunctionStrongCouplingDyonic}).

In section \ref{StrongCouplingExpansionPrepotentialNf4} we compute the strong coupling expansions of prepotential for the massive $N_f$=4 super QCD around the magnetic and dyonic singularities in the moduli space.
Strong coupling dynamics are difficult for direct quantum field theory study, usually one needs some sort of analytic continuation method.
For the undeformed N=2 gauge theory the low energy physics can be extended from the weak coupling region to the strong coupling region using analytical tools provided by the Seiberg-Witten curve.
In situation with the $\epsilon$-deformation, the extension follows the same spirit, the Schr\"{o}dinger equation provides the necessary instrument for the strong coupling computation.
In this section explicit computation is restricted to the leading order of the $\epsilon$-expansion, i.e. the undeformed limit, an extension of computation to include the $\epsilon$-deformation follows the WKB perturbation of quantum mechanics. Results obtained in this section provide evidence that the Gauge/Bethe correspondence \cite{NS0908, NS0901a, NS0901b} is valid over the entire parameter space, and one would hope to have more evidence from other gauge theories and solvable models.

A few appendices provide details of some computations in the main body of the paper.

\vspace{3mm}
\noindent {\bf Notice on notations}. This paper develops some results presented in a few earlier papers \cite{wh1306,wh1108,wh1412,wh1608}.
Various forms of elliptic functions are frequently encountered, in previous papers we used different notations for the elliptic modulus/nome according to the content discussed there.
To avoid confusion, we clarify the convention of notations used in each paper.
\begin{center}
\renewcommand{\arraystretch}{1.25}
\begin{tabular}{ | c | c | c | c | c| }
\hline
Refs$\diagdown$modulus/nome & gauge theory  & Jacobi $\sn/\cn/\dn$ & Weierstrass $\wp$ & $\mathbf{E}$ and $\mathbf{K}$  \\ \hline
this paper & $q$ & $q$ & $p$ & $\mathbf{x}^2$\\ \hline
\cite{wh1306} & $q$ & $q$ & $p$ & $k^2$\\ \hline
\cite{wh1108, wh1412, wh1608} & $q$ & $k^2$ & $q$ & $k^2$\\ \hline
\end{tabular}
\end{center}
The complementary elliptic modulus is denoted by $1-q$ in \cite{wh1306} and in this paper,
and denoted by $k^{\,\prime\,2}=1-k^2$ in \cite{wh1108,wh1412,wh1608}.
In particular, notice that the instanton parameter $q$ enters elliptic potentials differently for the N=$2^*$ theory and the $N_f$=4 theory.
For the N=$2^*$ theory, $q$ appears as the nome of the Weierstrass elliptic function; for the $N_f$=4 theory, $q$ appears as the modulus of the Jacobian elliptic functions.

\section{Critical points and asymptotic solutions}\label{CriticalPointsAndAsymptoticSolutions}

\subsection{The differential equation}\label{TheDifferentialEquation}

The connection between super QCD and Schr\"{o}dinger equation is based on the fact that the instanton partition function of the Omega background deformed N=2 gauge theory with surface operator satisfies a partial differential equation, in the Nekrasov-Shatashvili (NS) limit $\epsilon_1\to\epsilon, \epsilon_2\to0$ the equation becomes an ordinary differential equation \cite{Nekrasov1711a, Nekrasov1711b}.
For SU(2) $N_f$=4 gauge theory, the ordinary differential equation is the normal form of Heun equation \cite{JeongaNekrasov1806}.
Another context of deriving the same equation is their relations to conformal field theory (CFT).
As a special case of the correspondence discovered by Alday, Gaiotto and Tachikawa \cite{AGT}, the $N_f$=4 super QCD is related to the 4-point sphere conformal block of Liouville CFT.
An extension of this relation connects super QCD model with surface operator to conformal block with a null vertex operator inserted. A conformal block including a null vertex operator satisfies a partial differential equation, the Belavin-Polyakov-Zamolodchikov equation \cite{BPZ1984}.
The quantum mechanical model is obtained by taking a further limit, which is the large central charge limit for CFT.  For the 4-point sphere conformal block with a null vertex operator, it also leads to the normal form of Heun equation \cite{MT1006}.

The Heun equation derived for the partition function as mentioned is a Fuchsian equation with four regular singularities,
adapting the notation used in this paper, it takes the form
\be
W^{''}(z)-\Big[\f{\f{1}{4}b_2\!-\!\f{3}{16}}{z^2}+\f{\f{1}{4}b_3\!-\!\f{3}{16}}{(z-q)^2}+\f{\f{1}{4}b_1\!-\!\f{3}{16}}{(z-1)^2}+\f{\f{1}{4}(b_0\!-\!b_1\!-\!b_2\!-\!b_3)\!+\!\f{3}{8}}{z(z-1)}-\f{q(1-q)C}{z(z-q)(z-1)}\Big]W(z)=0,\label{eqNormal}
\ee
where the singularity parameter $q$ is the instanton parameter of gauge theory, the dimensionless constants $b_s$ with $s=0,1,2,3$ are related to hypermultiplet masses as given by (\ref{b2murelation}).
More importantly, the accessory parameter $C$ and the eigenfunction $W(z)$ are related to the instanton partition function of gauge theory with surface operator.
One of the purpose of this paper is to examine this fact by explicitly computing the function $C$ and $W(z)$,
using both perturbative method of solving differential equation and instanton computation of $N_f$=4 gauge theory.
But for the study of asymptotic solutions, eq. (\ref{eqNormal}) is not suitable for a systematic perturbative treatment \cite{wh1306}.

The equation in the normal form is a stationary Schr\"{o}dinger equation with zero eigenvalue subject to a potential defined on the four-punctured sphere $\mathbb{C}P^1\setminus\{0, q, 1, \infty\}$ with regular poles at the marked points.
An equivalent description of the potential uses the coordinate of the double covering space of the punctured sphere, a complex torus with modulus $p$ satisfying the relation $q=\vartheta_2^4(p)/\vartheta_3^4(p)$.
Elliptic function provides a natural coordinate for the torus. Therefore, the equation can be transformed to a standard eigenvalue problem with elliptic potential and perturbation method can be applied more directly. Using the Weierstrass elliptic function, the transformation is achieved by
\be
z=\f{\wp(x)-e_2}{e_1-e_2},\label{NWCoordinatesRelation}
\ee
and
\be
W(z)=(\wp(x)-e_1)^{\f{1}{4}}(\wp(x)-e_2)^{\f{1}{4}}(\wp(x)-e_3)^{\f{1}{4}}\psi(x).
\ee
The fundamental periods are $2\omega_1,2\omega_2$, the period $2\omega_3=2\omega_1+2\omega_2$ is also useful, and the nome is $p=\exp2\pi i\f{\omega_2}{\omega_1}$.
We use the notation $\wp(x)\equiv\wp(x,p)$ and $e_i\equiv e_i(p)=\wp(\omega_i,p), i=1,2,3$.
Applying the transformation to eq. (\ref{eqNormal}), we get a stationary Schr\"{o}dinger equation
\be
\psi^{''}(x)-u(x)\psi(x)=\lambda\psi(x),\label{eqWeierstrass}
\ee
with the Treibich-Verdier potential
\begin{align}
u(x)&=b_0\widetilde{\wp}(x)+b_1\widetilde{\wp}(x+\omega_1)+b_2\widetilde{\wp}(x+\omega_2)+b_3\widetilde{\wp}(x+\omega_3) \notag\\
&=\sum_{s=0}^{3}b_s\widetilde{\wp}(x+\omega_s).\label{DTVWeierstrass}
\end{align}
The potential has second-order poles at vertices $\omega_0=0, \omega_1,\omega_2$ and $\omega_3=\omega_1+\omega_2$ of the fundamental parallelogram for the period lattice.
The eigenvalue is related to the accessory parameter by
\be
\lambda=-4q(e_1-e_3)C-\zeta_1\sum_{s=0}^3b_s+(b_0-b_2-2b_3)e_1+(b_0-b_1-2b_3)e_2.
\ee
The weak coupling asymptotic solution of this equation is directly related to the instanton partition function of $N_f$=4 super QCD, therefore the shifted Weierstrass elliptic function $\widetilde{\wp}(x;2\omega_1,2\omega_2)=\wp(x;2\omega_1,2\omega_2)+\zeta_1$ is used here because it is more convenient to compare the spectrum computed by the shifted elliptic function with the instanton computation of gauge theory, see section \ref{WeakCouplingSpectrumAndN=2superQCD}.

As we noticed for the Lam\'{e} potential, while the equation in the Weierstrass elliptic function is suitable to compute the weak coupling solution, the equation in the Jacobian elliptic function is more appropriate to compute the strong coupling solution \cite{wh1108, wh1412, wh1608}. The transformation is
\be
\f{x}{\omega_1}=\f{\eta+i\mathbf{K}^{\prime}}{\mathbf{K}},\qquad \wp(x)=(e_3-e_2)\sn^2\eta+e_2,\label{WJCoordinatesRelation}
\ee
where the modulus of the Jacobian elliptic function is $q$, so $\sn\eta\equiv\sn(\eta|q)$. The equations (\ref{eqWeierstrass})(\ref{DTVWeierstrass}) become
\be
\psi^{''}(\eta)-u(\eta)\psi(\eta)=\Lambda\psi(\eta),\label{eqJacobi}
\ee
with the Darboux potential
\be
u(\eta)=b_0q\sn^2\eta+b_1q\f{\cn^2\eta}{\dn^2\eta}+b_2\f{1}{\sn^2\eta}+b_3\f{\dn^2\eta}{\cn^2\eta}.\label{DTVJacobi}
\ee
The periods of potential $u(\eta)$ are $2\mathbf{K}, 2i\mathbf{K}^{\prime}$ and $2\mathbf{K}+2i\mathbf{K}^{\prime}$.
The eigenvalues are related by
\be
\lambda=(e_1-e_2)\Lambda-(e_2+\zeta_1)\sum_{s=0}^{3}b_s.
\ee
The computation for the strong coupling spectral solution is carried out in section \ref{StrongCouplingSpectraAndGaugeTheoryDuality}.

For later use, the variables associated with eq. (\ref{eqWeierstrass}) are called {\em Weierstrass variables}, the variables associated with eq. (\ref{eqJacobi}) are called {\em Jacobian variables}.
Among others, the Floquet exponents for eqs. (\ref{eqWeierstrass}) and (\ref{eqJacobi}), which are denoted by $\nu$ and $\mu$ respectively in later sections, are related by $\nu\omega_1=\mu\mathbf{K}$ in accordance with the relation of coordinates (\ref{WJCoordinatesRelation}).

The Heun equation can be interpreted as a quantization of the Seiberg-Witten curve of SU(2) $N_f$=4 super QCD, the reasoning goes as follows.
For the quantum mechanical problem described by the normal form of Heun equation (\ref{eqNormal}), the quantization parameter $\epsilon$ is contained in the parameters $b_s\sim\mathcal{O}(\epsilon^{-2}), C\sim\mathcal{O}(\epsilon^{-2})$ and the momentum $v(z)\sim\mathcal{O}(\epsilon^{-1})$ of eigenfunction $W(z)=\exp(\int v(z)dz)$, the classical dispersion relation is given by the leading order WKB relation.
In the classical limit, setting $\tilde{b}_s=\lim_{\epsilon\to0}b_s\epsilon^2, \widetilde{C}=\lim_{\epsilon\to0}C\epsilon^2$ and $\tilde{v}=\lim_{\epsilon\to0}v\epsilon$,
we get the following algebraic equation from the differential equation (\ref{eqNormal}),
\be
\tilde{v}^2=\f{\f{1}{4}\tilde{b}_2}{z^2}+\f{\f{1}{4}\tilde{b}_3}{(z-q)^2}+\f{\f{1}{4}\tilde{b}_1}{(z-1)^2}+\f{\f{1}{4}(\tilde{b}_0-\tilde{b}_1-\tilde{b}_2-\tilde{b}_3)}{z(z-1)}-\f{q(1-q)\widetilde{C}}{z(z-q)(z-1)}.
\label{SpectralCurveHeun1}
\ee
Using the variable $w=2z(z-q)(z-1)\tilde{v}$, we transform it to
\be w^2=az^4+bz^3+cz^2+dz+e,\label{SpectralCurveHeun2}\ee
where $a,b,c,d,e$ are polynomial of $\tilde{b}_s, q$ and $\widetilde{C}$.
The quartic polynomial (\ref{SpectralCurveHeun2}) over $\mathbb{C}$ is an elliptic curve,
in fact it is isomorphic to the Seiberg-Witten curve of $N_f$=4 super QCD which takes the form of a cubic polynomial $y^2=x^3+Ex^2+Fx+G$ \cite{SW9408},
because the two curves have the same $J$-invariant under suitable parameter identification.
Therefore, the leading order WKB computation of the Schr\"{o}dinger equation produces the same prepotential of the undeformed gauge theory calculated by the Seiberg-Witten curve;
the higher order WKB perturbation gives the quantum correction introduced by Omega background in the NS limit.
A related content to coherently interpret the classical-quantum relation is the $SL(2)$ Gaudin model on $\mathbb{C}P^1\setminus\{0, q, 1, \infty\}$, noting that the curve (\ref{SpectralCurveHeun1}), hence (\ref{SpectralCurveHeun2}), is the spectral curve of the classical model and the differential equation (\ref{eqNormal}) is equivalent to the Bethe Ansatz equations for the quantum spectrum \cite{Gaudin1976, Gaudin2014Book}.

\subsection{The critical points of elliptic potential}\label{CriticalPointsOfEllipticPotential}

We first discuss some results about the Lam\'{e} equation and the SU(2) N=$2^*$ super QCD.
The elliptic potential of the Lam\'{e} equation has three critical points,
one critical point is associated with a weak coupling spectral solution, the other two critical points are each associated with a strong coupling spectral solution.
The moduli space of SU(2) N=$2^*$ super QCD has three singular points where the electric and the dual magnetic or dyonic descriptions of effective field theory are valid.
The electric and dual expansions of super QCD turn out to be one-to-one related to the critical points of the elliptic potential and the asymptotic spectrum there \cite{wh1108}.
As we show in this paper, for the Heun equation with DTV potential and the $N_f$=4 super QCD, there is a similar relation as discussed above.
Although the DTV potential has six critical points \cite{wh1306}, the fields of $N_f$=4 super QCD in the Coulomb branch are still electric, magnetic or dyonic \cite{SW9407, SW9408};
therefore the Heun equation has only three asymptotic solutions.
In fact, four critical points are associated with a weak coupling spectral solution, for the other two critical points each of them is associated with a strong coupling solution.
Again, these spectral solutions are one-to-one related to the electric, magnetic and dyonic expansions of the $N_f$=4 super QCD prepotential.
This fact gives useful indication to compute the perturbative spectrum for the Heun equation, it is difficult to see this property by conventional approach in spectral theory.

The critical points of the DTV potential in the Darboux form (\ref{DTVJacobi}) are given by complex solutions of the equation $\p_\eta u(\eta)=0$.
It is a 12th order polynomial equation for the function $\sn\eta$, using $z$-coordinate which is $z=q\sn^2\eta$ by the relations (\ref{NWCoordinatesRelation}) and (\ref{WJCoordinatesRelation}),
we rewrite the equation as $Q_6(z)=0$ where
\be
Q_6(z)=\mathcal{C}_6z^6+\mathcal{C}_5z^5+\mathcal{C}_4z^4+\mathcal{C}_3z^3+\mathcal{C}_2z^2+\mathcal{C}_1z+\mathcal{C}_0,
\ee
with
\begin{align}
&\mathcal{C}_6=b_0,\notag\\
&\mathcal{C}_5=-2(1+q)b_0,\notag\\
&\mathcal{C}_4=(1+4q+q^2)b_0-(1-q)b_1-qb_2+q(1-q)b_3,\notag\\
&\mathcal{C}_3=-2q[(1+q)b_0-(1-q)b_1-(1+q)b_2+(1-q)b_3],\notag\\
&\mathcal{C}_2=q[qb_0-q(1-q)b_1-(1+4q+q^2)b_2+(1-q)b_3],\notag\\
&\mathcal{C}_1=2q^2(1+q)b_2,\notag\\
&\mathcal{C}_0=-q^3b_2.
\end{align}
On the gauge theory side the UV theory is assumed to be weakly coupled, so $q\ll1$, we have this condition for all perturbative computations carried out later.
Then the equation $Q_6(z)=0$ can be solved perturbatively by treating $q$ as the expansion parameter. There are six distinct solutions, the first few coefficients of the solutions are given bellow,
other coefficients can be obtained by an iterative process as shown in \cite{wh1306}.

The first pair of critical points are given by
\be
\sn^2\eta_{*1,2}=\f{b_0^{\f{1}{2}}\pm b_1^{\f{1}{2}}}{b_0^{\f{1}{2}}}\f{1}{q}\mp\f{b_1^{\f{1}{2}}[(b_0^{\f{1}{2}}\pm b_1^{\f{1}{2}})^2-b_2+b_3]}{2b_0^{\f{1}{2}}(b_0^{\f{1}{2}}\pm b_1^{\f{1}{2}})^2}+\sum_{n\geqslant1}^{\infty}s_{1,2;n}q^n,
\label{criticsn12}
\ee
where
\be
u(\eta_{*1,2})=\left(b_0^{\f{1}{2}}\pm b_1^{\f{1}{2}}\right)^2-\f{b_0^{\f{1}{2}}(b_1-b_2)\pm b_1^{\f{1}{2}}(b_0-b_3)}{b_0^{\f{1}{2}}\pm b_1^{\f{1}{2}}}q+\mathcal{O}(q^2).
\label{criticu12}
\ee
The subscript indicates $\sn^2\eta_{*1}$ and $\sn^2\eta_{*2}$ take the upper and the lower signs respectively in eq. (\ref{criticsn12}), the convention applies to other formulae (\ref{criticu12})-(\ref{criticu56}).
The coordinate $\eta_*$ of a critical point is uniquely determined by $\sn^2\eta_*$, up to the periods $2\mathbf{K}$ and $2i\mathbf{K}^{\prime}$.
The second pair of critical points are given by
\be
\sn^2\eta_{*3,4}=\f{b_2^{\f{1}{2}}}{b_2^{\f{1}{2}}\pm b_3^{\f{1}{2}}}\pm\f{b_2^{\f{1}{2}}b_3^{\f{1}{2}}[(b_2^{\f{1}{2}}\pm b_3^{\f{1}{2}})^2-b_0+b_1]}{2(b_2^{\f{1}{2}}\pm b_3^{\f{1}{2}})^4}q+\sum_{n\geqslant2}^{\infty}s_{3,4;n}q^n,
\label{criticsn34}
\ee
where
\be u(\eta_{*3,4})=\left(b_2^{\f{1}{2}}\pm b_3^{\f{1}{2}}\right)^2+\f{b_2^{\f{1}{2}}(b_0-b_3)\pm b_3^{\f{1}{2}}(b_1-b_2)}{b_2^{\f{1}{2}}\pm b_3^{\f{1}{2}}}q+\mathcal{O}(q^2).\label{criticu34}\ee
The expressions (\ref{criticu12}) and (\ref{criticu34}) indicate that the first four critical points are on the same footing, they are converted to each other by permutations of $b_s$,
or by linear transformations of physical masses $\mu_i$ according to the relations given in eq. (\ref{b2murelation}).
As shown in \cite{wh1306}, there is a single expression that represents perturbative spectral solutions around $\eta_{*1,2,3,4}$;
it is the weak coupling solution of the DTV potential, related to the low energy physics of super QCD in the weak coupling region of Coulomb branch moduli space.

There remains the last pair of critical  points,
\be
\sn^2\eta_{*5,6}=\pm\left(\f{b_2-b_3}{b_0-b_1}\right)^{\f{1}{2}}\f{1}{q^{\f{1}{2}}}+\f{b_1(b_2-b_3)^2-b_3(b_0-b_1)^2}{(b_0-b_1)^2(b_2-b_3)}
+\sum_{n\geqslant1}^{\infty}s_{5,6;n}(\pm q^{\f{1}{2}})^n,
\label{criticsn56}
\ee
where
\be
u(\eta_{*5,6})=\pm 2(b_0-b_1)^{\f{1}{2}}(b_2-b_3)^{\f{1}{2}}q^{\f{1}{2}}+\f{(b_0-b_1-b_2+b_3)(b_1b_2-b_0b_3)}{(b_0-b_1)(b_2-b_3)}q+\mathcal{O}(q^{\f{3}{2}}).
\label{criticu56}
\ee
In fact, coefficients $s_{5;n}$ and $s_{6;n}$ are the same, so $\sn^2\eta_{*5}$ and $\sn^2\eta_{*6}$ are related to each other by a simple sign inversion $q^{\f{1}{2}} \leftrightarrow -q^{\f{1}{2}}$.
They have a different nature from the previous four critical points, the spectral solutions associated with them are the strong coupling solutions, related to super QCD in the strong coupling regions.

The coefficients $s_{1;n}, s_{2;n}, s_{3;n}, \cdots s_{6;n}$ are rational functions of $b_s^{\f{1}{2}}$,
in later computations of gauge theory quantities they can be written as rational functions of mass parameters $\mu_i$.
In the main body of the paper we study models with generic coupling constants $b_s$ (equivalently generic masses $\mu_i$), so that the special values $b_0=b_1$ and $b_2=b_3$ are not applied to formulae such as (\ref{criticsn12})-(\ref{criticu56}). The computation of this paper relies on series expansion of various quantities which are not valid for these values of $b_s$.
When $b_s$ take these degenerate values, the DTV potential reduces to the Lam\'{e} potential associated with the N=$2^*$ super QCD model.

A fact that would be useful is the distance between the two strong coupling critical points. When $b_s$ take generic values, which means no coefficients $s_{5,6;n}$ in the expansion (\ref{criticsn56}) takes extremely small or large value, $\sn^2\eta_{*5,6}$ are dominated by the leading order terms of order $\pm(b_2-b_3)^{\f{1}{2}}(b_0-b_1)^{-\f{1}{2}}q^{-\f{1}{2}}\gg 1$.
It indicates the distance $\eta_{*6}-\eta_{*5}$ is very close to the half period $\mathbf{K}$. In fact, it can be shown
\be
\eta_{*6}-\eta_{*5}-\mathbf{K}=\f{(b_2+b_3)(b_0-b_1)^2-(b_0+b_1)(b_2-b_3)^2}{2i(b_0-b_1)^{\f{3}{2}}(b_2-b_3)^{\f{3}{2}}}q^{\f{1}{2}}+\mathcal{O}(q)\ll 1.
\label{DistanceOfMDPoints}
\ee

We would use ``weak coupling'' and ``strong coupling'' for both the Schr\"{o}dinger equation and the effective theory of super QCD. While their couplings are characterised by very different variables, it happens that the weak coupling solution of spectral problem is related to the weak coupling expansion of super QCD effective theory, and so is the case with the strong coupling solution. So terms like ``weak (or strong) coupling solution (or expansion)'' are understood according to the content in which results about elliptic potential or super QCD are stated.

\section{Weak coupling spectrum and $N_f$=4 super QCD}\label{WeakCouplingSpectrumAndN=2superQCD}

The solutions relevant to the $N_f$=4 super QCD are of a particular kind, they are asymptotic solutions in which the coupling constants $b_s$ take generic values.
It is the ratio of the energy $\lambda$ (or $\Lambda$) to the coupling constants $b_s$ (or their geometric average) that determines the weak or strong coupling nature of perturbation solution.
By ``weak coupling'' solution of the spectral problem, it means the kinetic energy characterised by the quasimomentum $\nu$ is much larger than the potential energy characterised by the coupling constants $b_s$,
$\nu\gg b_s^{\f{1}{2}}$.

\subsection{The spectrum at weak coupling}\label{SpectrumAtWeakCoupling}

We use the equation in the Weierstrass elliptic form (\ref{eqWeierstrass}) and (\ref{DTVWeierstrass}) to compute the weak coupling solution.
The perturbation method used to compute the weak coupling solution is closely related to the procedure of generating infinitely many conserved charges of KdV hierarchy \cite{MGK1968, classintegrable}.
Plugging the wave function $\psi(x)=\exp(\int v(x) dx)$ into eq. (\ref{eqWeierstrass}) leads to an equation for the integrand $v(x)$, $v^2(x)+v_x(x)=u(x)+\lambda$.
The weak coupling solution of $v(x)$ takes the form of large energy expansion,
\be
v(x)=\lambda^{\f{1}{2}}+\sum_{\ell=1}^{\infty}\f{v_\ell(x)}{\lambda^{\f{\ell}{2}}},
\label{Vlargelambdaexpansion}
\ee
where the coefficients are given by the KdV Hamiltonian densities $v_\ell(u,\p_xu,\p_x^2u,\cdots)$.
According to the Floquet theorem for periodic differential equation, the two fundamental eigenfunctions satisfy the relation $\psi_\pm(x+2\omega_1)=\exp(\pm i2\omega_1\nu)\psi_\pm(x)$.
The characteristic multipliers $\exp(\pm i2\omega_1\nu)$ are eigenvalues of the monodromy matrix under the shift $x\to x+2\omega_1$, and the characteristic exponent $\nu$  depends on parameters $\lambda, b_s$ and $p$. As we confirmed for other examples of equations with elliptic potential \cite{wh1108, wh1412, wh1608}, the period associated with the weak coupling solution is $2\omega_1$, this fact is also true for the DTV potential.
So the Floquet exponent is given by the integral of $v(x)$ over an interval of $2\omega_1$,
\be
\pm i\nu=\f{1}{2\omega_1}\int_{x_0}^{x_0+2\omega_1} v(x)dx.\label{Floquetexponentdefweak}
\ee
In the connection with N=2 gauge theory, the integral of $v(x)$ over $2\omega_1$ is related to the integral of the Seiberg-Witten differential over the $\alpha$-cycle on the curve of $N_f$=4 super QCD which gives the v.e.v of scalar field in the vector multiplet \cite{SW9407, SW9408}. The spectrum of the differential equation always includes two sectors labeled by ``$\pm$'' signs, they correspond to evaluating the contour integral clockwise and counter-clockwise in gauge theory respectively.

The weak coupling eigenvalue and eigenfunction are obtained from the definite and indefinite integrals of $v(x)$.
For the DTV potential, the integrands $v_\ell(x)$ contain products of elliptic function and their derivatives,
to evaluate integration some simplifications are needed, the details are explained in appendix \ref{SimplifyingWeierstrassEllipticFunction}.

The eigenvalue is computed from the monodromy relation (\ref{Floquetexponentdefweak}).
As explained in appendix \ref{Usefulrelationsofellipticfunctions}, the integrands $v_\ell(x)$ can be eventually simplified to the form given in (\ref{KdVHamiltonianDTVFinalForm}).
As $v_{2\ell}(x)$ are derivative of periodic functions, their integrals over an interval of $2\omega_1$ vanish;
and for $v_{2\ell-1}(x)$, the derivative part does not contribute to the integral, the remaining ``constant'' part contributes non-vanishing integral. So we get
\be
\varepsilon_\ell=\f{1}{2\omega_1}\int_{x_0}^{x_0+2\omega_1} v_{2\ell-1}(x)dx.
\label{IntegralDensityOfHamiltonians}
\ee
The integrals $\varepsilon_\ell$ are functions of $b_s$ and (quasi)modular functions $e_1, e_2, e_3, \zeta_1$.
The eigenvalue expansion is obtained by reversing the relation (\ref{Floquetexponentdefweak}) to obtain
\be
\lambda=-\nu^2+\sum_{l=1}^{\infty}\f{\lambda_l}{\nu^{2l}},
\label{Eigenvalueweakcoupling}
\ee
the coefficients $\lambda_l$ are polynomials of $\varepsilon_{\ell}$. The first two $\lambda_1,\lambda_2$ are
\begin{align}
\lambda_1=&\f{1}{48}\Big\lbrace(12\zeta_1^2-g_2)\sum_{s=0}^3b_s^2+24\sum_{i|j<k}[(e_i+\zeta_1)^2-(e_i-e_j)(e_i-e_k)](b_0b_i+b_jb_k)\Big\rbrace,\notag\\
\lambda_2=&\f{1}{80}\Big\lbrace(20\zeta_1^3-g_2\zeta_1-g_3)\sum_{s=0}^3b_s^3-(2g_2\zeta_1-3g_3)\sum_{s=0}^3b_s^2\notag\\
&+120\sum_{i<j<k}(e_i+\zeta_1)(e_j+\zeta_1)(e_k+\zeta_1)b_ib_jb_k\notag\\
&+120\sum_{i|j<k}(e_i+\zeta_1)(e_j+\zeta_1)(e_k+\zeta_1)b_0b_jb_k\notag\\
&+20\sum_{i|j<k}(e_i+\zeta_1)[3\zeta_1(e_i+\zeta_1)-(e_i-e_j)(e_i-e_k)](b_0b_i^2+b_0^2b_i+b_jb_k^2+b_j^2b_k)\notag\\
&+40\sum_{i|j<k}(e_i-2\zeta_1)(e_i-e_j)(e_i-e_k)(b_0b_i+b_jb_k)\Big\rbrace.
\end{align}
The expressions of $\lambda_l$ become substantially more complicated for higher order ones.
In the summation the indices $i,j,k$ take values in $\lbrace1,2,3\rbrace$.
The summation restricted by $i<j<k$ in fact requires $i=1,j=2,k=3$.
The summation restricted by $i|j<k$ means when the index $i$ takes a value in $\lbrace 1,2,3\rbrace$ the indices $j,k$ take values in the remaining set and satisfy $j<k$.

The corresponding unnormalized wave functions are obtained by evaluating the indefinite integral $\int v(x) dx$ for the integrand (\ref{Vlargelambdaexpansion}) and substituting $\lambda^{\f{1}{2}}=\pm \nu+\cdots$ derived from the dispersion relation (\ref{Eigenvalueweakcoupling}) for both $``\pm"$ signs. The two linearly independent basic wave functions are written in exponential form as
\begin{align}
\psi_\pm(x)=&\exp\Bigg(\pm i\nu x\pm\f{i\sum\limits_{s=0}^3b_s\widetilde{\zeta}(x+\omega_s)}{2\nu}-\f{\sum\limits_{s=0}^3b_s\p_x\widetilde{\zeta}(x+\omega_s)}{4\nu^2}\notag\\
&\pm\f{i\sum\limits_{s=0}^3\left[c_{3,s,0}\widetilde{\zeta}(x+\omega_s)+c_{3,s,2}\p^2_x\widetilde{\zeta}(x+\omega_s)\right]}{48\nu^3}\notag\\
&-\f{\sum\limits_{s=0}^3\left[c_{4,s,1}\p_x\widetilde{\zeta}(x+\omega_s)+c_{4,s,3}\p^3_x\widetilde{\zeta}(x+\omega_s)\right]}{48\nu^4}+\mathcal{O}(\nu^{-5})\Bigg),
\label{EigenfunctionWeakCoupling}
\end{align}
where the shifted Weierstrass zeta function $\widetilde{\zeta}(x)$ is defined by $\widetilde{\zeta}(x)=\zeta(x)-\zeta_1x$, it is a periodic function of period $2\omega_1$.
The coefficients $c_{l, s, d}$ have three indices, among them $l$ refers to the exponent of $\nu$, $s$ is in accordance with $\omega_s$, and $d$ refers to the order of derivative of $\widetilde{\zeta}(x)$.
The coefficients in the wave function are
\begin{align}
c_{3,0,0}&=c_{4,0,1}=12b_0(e_1b_1+e_2b_2+e_3b_3+\zeta_1\sum\limits_{s=0}^3b_s),\notag\\
c_{3,1,0}&=c_{4,1,1}=12b_1(e_1b_0+e_2b_3+e_3b_2+\zeta_1\sum\limits_{s=0}^3b_s),\notag\\
c_{3,2,0}&=c_{4,2,1}=12b_2(e_1b_3+e_2b_0+e_3b_1+\zeta_1\sum\limits_{s=0}^3b_s),\notag\\
c_{3,3,0}&=c_{4,3,1}=12b_3(e_1b_2+e_2b_1+e_3b_0+\zeta_1\sum\limits_{s=0}^3b_s),\notag\\
c_{3,s,2}&=b_s(b_s-6),\notag\\
c_{4,s,3}&=b_s(b_s-3).
\end{align}
Notice that the wave functions (\ref{EigenfunctionWeakCoupling}) are already in the Floquet form,
\be
\psi_\pm(x)=\exp(\pm i\nu x)\mathbf{p}_1^{\pm}(\nu,x),
\ee
where $\mathbf{p}_1^{\pm}(\nu,x)$ are periodic functions, $\mathbf{p}_1^{\pm}(\nu,x+2\omega_1)=\mathbf{p}_1^{\pm}(\nu,x)$, they are related by $\mathbf{p}_1^{+}(\nu,x)=\mathbf{p}_1^{-}(-\nu,x)$.

\subsection{Instanton partition function of $N_f$=4 super QCD}\label{InstantonPartitionFunctionNf=4SuperQCD}

The criteria for weak coupling in spectral solution is $\nu\gg b_s^{\f{1}{2}}$; under appropriate parameter identification it corresponds to the semiclassical region for the gauge theory in Coulomb branch.
In this region, the crucial object of gauge theory is the instanton partition function. The Gauge/Bethe correspondence proposes a direct relation between the spectral solution and the low energy dynamics of gauge theory. Here we explain how it works for parameters in the weak coupling region; in the next two sections we extend the relation for parameters in the strong coupling region.

\subsubsection{Instanton partition function of super QCD with surface operator}

Nonperturbative effects induced by solitons have important consequence for quantum field theories. Various technics are developed to compute nonperturbative effects,
in particular for N=2 supersymmetric gauge theory the multi-instanton contributions are exactly computed in the Coulomb vacuum \cite{SW9407, SW9408}.
There is a refinement for N=2 gauge theory, the Omega background field is introduced to deform gauge theory,
the instanton moduli space is tamed so that the localization technic can be used to compute the instanton partition function in the weak coupling region \cite{LNS9711, MNS9712, LNS9801, Nekrasov0206}.
For the SU(2) $N_f$=4 super QCD, the partition function depends on the following parameters, the scalar v.e.v of the vector multiplet $\langle\phi\rangle=a$,
the masses of the hypermultiplets $\mu_i$ with $i=1,2,3,4$, the Omega background field parameterized by $\epsilon_1,\epsilon_2$, and the ultraviolet gauge coupling characterised by the instanton parameter $q$.

Besides local operators, there are finer probes for gauge theories such as line operators and surface operators which detect more structure \cite{AGGTV2010, Gaiotto0911, Gukov1412}.
In the presence of a surface operator preserving supersymmetry, such as a plane surface in the $x_1, x_2$ direction of spacetime, the instanton partition function can be computed as well \cite{Braverman0401, BravermanEtingof0409, AT1005, AFKMY1008, wyllard1012, KannoTachikawa1105, Nekrasov1711a, ABFJ1901}.
Correspondingly, the instanton counting parameter $q$ splits up into multiple counting parameters, for the SU(2) gauge group there are two, $x_1, x_2$ (not the spacetime coordinates) which are subject to the relation $x_1x_2=q$.
The relevant properties of the instanton partition function with surface operator for the SU(2) $N_f$=4 super QCD are discussed in \cite{AT1005, AFKMY1008},
using the equivariant character developed in \cite{FFNR0812}.
In the context of Gauge/Bethe correspondence, instanton partition function with surface operator in the NS limit contains information about the weak coupling spectrum of the associated quantum mechanical problem \cite{NS0908, AT1005, MT1006, AFKMY1008, JeongaNekrasov1806}.

The expressions of instanton partition function for the $N_f$=4 super QCD are pretty lengthy, but a few explicit expansions are already enough to explain the relation.
The more useful quantity here is the logarithm of instanton partition function, the prepotential.
So when the instanton partition function is written in the exponential form to manifest the pole structure of the deformation parameters $\epsilon_1,\epsilon_2$, it takes the general form
\be
Z^{inst}=\exp\left(\f{\mathcal{F}^{inst}}{\epsilon_1\epsilon_2}+\f{\mathcal{G}^{inst}}{\epsilon_1}\right).
\ee
The function $\mathcal{F}^{inst}$ is a polynomial of $\epsilon_2$, therefore the pole structure alone does not fully fix $\mathcal{F}^{inst}$ and $\mathcal{G}^{inst}$.
The functions $\mathcal{F}^{inst}$ and $\mathcal{G}^{inst}$ are both expanded as double series in the parameters $x_1$ and $x_2$, they can be uniquely determined by imposing a condition about $x_1,x_2$.
The first piece is required to contain only terms with the same exponent for $x_1,x_2$, i.e. it is a $q$-series.
To compare with the spectral solution, we expand the prepotential for large $a$.
The first piece is given by
\be
\mathcal{F}^{inst}(a,\mu_i,\epsilon_{1,2},q)=\sum_{l=-1}^{\infty}\frac{1}{a^{2l}}\left(\sum_{k=1}^{\infty}\mathcal{F}_{l,k}q^k\right),
\label{instFLargeA}
\ee
with $\mathcal{F}_{l,k}$ rational functions of $\mu_i$ and $\epsilon_{1}, \epsilon_{2}$. When $\epsilon_{1}, \epsilon_{2}$ are assigned with mass dimension one, $\mathcal{F}_{l,k}$ has mass dimension $2l+2$.
The second piece is required to contain terms with different exponents for $x_1,x_2$, it is expanded as
\be
\mathcal{G}^{inst}(a,\mu_i,\epsilon_{1,2},x_1,x_2)=\sum_{l=-1}^{\infty}\frac{1}{a^{l}}\left(\sum_{\substack{k_1,k_2=1\\k_1\ne k_2}}^{\infty}\mathcal{G}_{l,k_1,k_2}x_1^{k_1}x_2^{k_2}\right),
\label{instGLargeA}
\ee
with $\mathcal{G}_{l,k_1,k_2}$ also rational functions of $\mu_i$ and $\epsilon_{1}, \epsilon_{2}$, has mass dimension $l+1$.
In the following discussion, the useful objects are the coefficient function in the expansion of $\mathcal{F}^{inst}$ and $\mathcal{G}^{inst}$ for large $a$, namely the series in the brackets of (\ref{instFLargeA}) and (\ref{instGLargeA}), they are denoted by $\mathcal{F}_l^{inst}$ and $\mathcal{G}_l^{inst}$.

In fact, besides the instanton contribution, the full gauge theory partition function also contains the classical and the perturbative contributions, $Z=Z^{cl}Z^{pert}Z^{inst}$.
Correspondingly, the full prepotential is given by $\mathcal{F}=\mathcal{F}^{cl}+\mathcal{F}^{pert}+\mathcal{F}^{inst}$ and a similar formula for $\mathcal{G}$.
In the rest of this subsection we show that the function $\mathcal{F}$ in the NS limit is related to the eigenvalue (\ref{Eigenvalueweakcoupling}) and the function $\mathcal{G}$ is related to the unnormalized wave function (\ref{EigenfunctionWeakCoupling}).
As it would be clear, the relation between the spectrum of DTV potential and the prepotential of $N_f$=4 super QCD is not quite straight to identify because the weak coupling spectral solution is expressed in the Weierstrass variables while the instanton partition function is primarily expressed in the Jacobian variables, this causes some extra twists compared to the simpler example discussed in \cite{wh1608}.

\subsubsection{Eigenvalue from instanton partition function}

For quantum mechanical problem with the DTV potential, by the Gauge/Bethe correspondence the quasimomentum $\nu$ is related to v.e.v of the scalar $\langle\phi\rangle=a$,  the energy eigenvalue $\lambda$ is related to v.e.v of the chiral operator $\f{1}{2}\langle\text{tr}\phi^2\rangle$. Moreover, $\f{1}{2}\langle\text{tr}\phi^2\rangle$ is related to the prepotential through $\f{1}{2}\langle\text{tr}\phi^2\rangle=q\p_q\mathcal{F}$ \cite{NS0908, JeongaNekrasov1806}. In this section the precise relation of $\lambda$ and $q\p_q\mathcal{F}$ would be fixed.

The relation of coupling constants $b_s$ and flavor masses $\mu_i$ are fixed through deriving the Heun differential equation from gauge theory or CFT as
\begin{align}
&b_0=\left(\f{\mu_1-\mu_2}{\epsilon}-\f{1}{2}\right)\left(\f{\mu_1-\mu_2}{\epsilon}+\f{1}{2}\right),\quad b_1=\left(\f{\mu_1+\mu_2}{\epsilon}-\f{3}{2}\right)\left(\f{\mu_1+\mu_2}{\epsilon}-\f{1}{2}\right),\notag\\
&b_2=\left(\f{\mu_3-\mu_4}{\epsilon}-\f{1}{2}\right)\left(\f{\mu_3-\mu_4}{\epsilon}+\f{1}{2}\right),\quad b_3=\left(\f{\mu_3+\mu_4}{\epsilon}-\f{3}{2}\right)\left(\f{\mu_3+\mu_4}{\epsilon}-\f{1}{2}\right).
\label{b2murelation}
\end{align}
The coupling constants can be written as $b_s=n_s(n_s-1)$, with $n_s$ related to $\mu_i$ by
\begin{align}
&n_0=\f{\mu_1-\mu_2}{\epsilon}+\f{1}{2},\qquad n_1=\f{\mu_1+\mu_2}{\epsilon}-\f{1}{2},\notag\\
&n_2=\f{\mu_3-\mu_4}{\epsilon}+\f{1}{2},\qquad n_3=\f{\mu_3+\mu_4}{\epsilon}-\f{1}{2}.
\end{align}
For gauge theory prepotential, $n_s$ are more primary parameters because a few coefficients in the formulae (\ref{aIndependentTermEigenvalue}) and (\ref{aIndependentTermWavefunction}) given below
do not appear in combinations $n_s(n_s-1)$. By these relations, the weak coupling condition for spectral solution $\nu\gg b_s^{\f{1}{2}}$ is the same as the weak coupling condition for super QCD in the electric region  $a\gg\mu_i$.

The classical contribution to the prepotential is $\mathcal{F}^{cl}=-a^2\ln q$.
The one-loop contribution can be computed as the regularized universal denominator extracted from Nekrasov instanton partition \cite{NekrasovOkounkov0306},
it is a $q$-independent function and therefore irrelevant to the discussion here. Apart from the $q$-independent piece, the prepotential $\mathcal{F}$ is
\be
\mathcal{F}=-a^2\ln q+\sum_{l=-1}^{\infty}\frac{\mathcal{F}^{inst}_{l}}{a^{2l}}=\sum_{l=-1}^{\infty}\frac{\mathcal{F}_{l}}{a^{2l}}.
\ee
From the point of view 2-dimensional field theory on the surface defect, $\mathcal{F}$ in the NS limit is the effective twisted superpotential \cite{NS0908, NS0901a, NS0901b}.
Notice that there are two elliptic variables involved, gauge theory quantities use the modulus $q$, spectral data in the Weierstrass variables use the nome $p$.
To see their connection, we rewrite the prepotential as a series in $p$ using the relation $q=\vartheta_2^4(p)/\vartheta_3^4(p)$.
For $N_f$=4 gauge theory, the instanton contribution $a^2\mathcal{F}_{-1}^{inst}$ is an infinite series $-a^2(\f{1}{2}q+\f{13}{64}q^2+\f{23}{192}q^3+\cdots)$,
combined with an extra piece from the classical contribution $-a^2\ln q$,
it leads to $a^2\mathcal{F}_{-1}=-a^2(\ln q+\f{1}{2}q+\f{13}{64}q^2+\f{23}{192}q^3+\cdots)=-\f{1}{2}a^2(\ln p+8\ln 2)$.
When $a$ and $\nu$ are identified by $\pi a=\epsilon\omega_1\nu$, the leading order terms of the power series $\mathcal{F}$ for large $a$ and the power series $\lambda$ for large $\nu$ are related by $2\pi^2p\f{\p}{\p p}(a^2\mathcal{F}_{-1})=-\epsilon^2\omega_1^2\nu^2$.

The nome $p$ has a role in gauge theory. The massive $N_f$=4 super QCD can be constructed from the massless and superconformal theory by adding mass deformation,
the nome $p$ is (exponential of) the complex coupling of Coulomb branch effective theory for the massless theory \cite{SW9408}.
The low energy effective coupling is renormalized by instanton effects, encoded in the relation of $p$ and $q$ \cite{gkmw07, AGT}.

In the next order, there is an $a$-independent piece in the prepotential $\mathcal{F}$, there is no counterpart in the eigenvalue $\lambda$. This piece is summarised in the function $\Phi$ given below.
For terms of order $a^{-2l}$, we substitute the series relation $q=16p^{\f{1}{2}}-128p+704p^{\f{3}{2}}+\cdots$ for the parameter $q$,
then up to a scaling factor $p\f{\p}{\p p}\mathcal{F}_{l}$ precisely matches the corresponding order coefficient $\lambda_l$ of the eigenvalue expansion.

In summary, the relation between the eigenvalue $\lambda$ and the prepotential $\mathcal{F}$ is
\be
\f{\lambda+\Phi}{4(e_1-e_3)}=\f{1}{\epsilon_1^2}q\f{\p}{\p q}\mathcal{F}\Bigr|_{\substack{\epsilon_1\to\epsilon\\ \epsilon_2\to0}},
\label{lambdaPrepotentialElectric}
\ee
with
\be
\Phi=(e_3+\zeta_1)\sum_{s=0}^3b_s+(e_2-e_3)(n_1+n_3)^2.
\label{aIndependentTermEigenvalue}
\ee
In eq. (\ref{lambdaPrepotentialElectric}) the left side is expressed in (quasi)modular functions $e_i,\zeta_1$ with nome $p$, the right side is expressed as a $q$-series.
Therefore, on the right side instead of using the derivative $p\f{\p}{\p p}$ we use
\be
q\f{\p}{\p q}=\f{\pi^2}{2\omega_1^2}\f{1}{e_1-e_3}p\f{\p}{\p p}.
\ee

\subsubsection{Wave function from instanton partition function}

First we need to identify relation between the instanton parameters $x_1,x_2$ and the coordinate of wave function $x$.
The condition $x_1x_2=q$ does not give enough information. The next hint comes from the leading order coefficient of the series expansion $\mathcal{G}^{inst}$ for large $a$ in the NS limit.
The function $\mathcal{G}_{-1}^{inst}$ is an infinite series in $x_1,x_2$ expanded as $\f{1}{2}x_1-\f{1}{2}x_2-\f{3}{16}x_1^2+\f{3}{16}x_2^2+\cdots$,
but the leading order term in the (logarithm of) wave function contains only one term $-i\nu x$.
The situation is similar to the case of matching $\mathcal{F}$ and the eigenvalue $\lambda$. Inspecting the series expansion of $\mathcal{G}_{-1}^{inst}$,
we conclude that an extra piece is needed to convert it into a single term.
The extra piece can be interpreted as the classical contribution of the surface defect, the only reasonable form is $\f{1}{2}a\ln(x_2/x_1)$, up to terms independent of $x_1,x_2$.
Therefore, the prepotential $\mathcal{G}$ is
\be
\mathcal{G}=\f{1}{2}a\ln\f{x_2}{x_1}+\sum_{l=-1}^{\infty}\frac{\mathcal{G}^{inst}_{l}}{a^l}=\sum_{l=-1}^{\infty}\frac{\mathcal{G}_{l}}{a^l}.
\ee
The leading order coefficient is
\begin{align}
\mathcal{G}_{-1}=&-\f{1}{2}\ln x_1+\f{1}{2}\ln x_2+\f{1}{2}x_1-\f{1}{2}x_2-\f{3}{16}x_1^2+\f{3}{16}x_2^2+\f{5}{48}x_1^3+\f{1}{16}x_1^2x_2-\f{1}{16}x_1x_2^2\notag\\
&-\f{5}{48}x_2^3-\f{35}{512}x_1^4-\f{1}{32}x_1^3x_2+\f{1}{32}x_1x_2^3+\f{35}{512}x_2^4+\f{63}{1280}x_1^5+\f{5}{256}x_1^4x_2+\f{1}{32}x_1^3x_2^2\notag\\
&-\f{1}{32}x_1^2x_2^3-\f{5}{256}x_1x_2^4-\f{63}{1280}x_2^5-\f{77}{2018}x_1^6-\f{7}{512}x_1^5x_2-\f{35}{2018}x_1^4x_2^2+\f{35}{2048}x_1^2x_2^4\notag\\
&+\f{7}{512}x_1x_2^5+\f{77}{2048}x_2^6+\cdots
\label{WaveFunctionG-1}
\end{align}
Define the normalized shifted coordinate $x_{+}=\f{\pi}{2\omega_1}(x+\f{\omega_2}{2})$, then $x_1$ and $x_2$ are related to $x$ through elliptic theta function,
\be
x_1=-\f{\vartheta_2^2(0,p)}{\vartheta_3^2(0,p)}\f{\vartheta_1^2(x_+,p)}{\vartheta_4^2(x_+,p)},\quad x_2=-\f{\vartheta_2^2(0,p)}{\vartheta_3^2(0,p)}\f{\vartheta_4^2(x_+,p)}{\vartheta_1^2(x_+,p)}.
\label{WavefunctionParameterIndentify}
\ee
They satisfy the relation $x_1x_2=\vartheta_2^4(0,p)/\vartheta_3^4(0,p)=q$. The series expansion of $x_1, x_2$ are
\begin{align}
x_1=&4e^{-\f{i\pi x}{\omega_1}}p^{\f{1}{4}}-8(1-e^{-\f{2i\pi x}{\omega_1}})p^{\f{1}{2}}+4(3e^{-\f{3i\pi x}{\omega_1}}-8e^{-\f{i\pi x}{\omega_1}}+e^{\f{i\pi x}{\omega_1}})p^{\f{3}{4}}+\cdots,\notag\\
x_2=&4e^{\f{i\pi x}{\omega_1}}p^{\f{1}{4}}-8(1-e^{\f{2i\pi x}{\omega_1}})p^{\f{1}{2}}+4(3e^{\f{3i\pi x}{\omega_1}}-8e^{\f{i\pi x}{\omega_1}}+e^{-\f{i\pi x}{\omega_1}})p^{\f{3}{4}}+\cdots.
\label{x1x2SeriesExpansion}
\end{align}
Substitution of the expansions of $x_1, x_2$ into $\mathcal{G}_{-1}$ leads to $\mathcal{G}_{-1}=i\pi x/\omega_1$. Therefore, the leading order piece of prepotential $\mathcal{G}$ is
\be
\f{\mathcal{G}_{-1}}{\epsilon a^{-1}}=i\nu x,
\ee
which is precisely the first piece in the wave function.

The next order piece $\mathcal{G}_{0}/a^0$ is $a$-independent, there is no counterpart in the wave function. It is given by
\be
\epsilon^{-1}\mathcal{G}_{0}=-\f{1}{2}n_3\ln(1+x_1)-\f{1}{2}n_1\ln(1+x_2).
\ee
When (\ref{WavefunctionParameterIndentify}) is used, $\mathcal{G}_{0}$ is expressed in theta function as
\be
\f{\mathcal{G}_{0}}{\epsilon}=-n_3\ln\left[\f{\vartheta_3(x_+,p)}{\vartheta_4(x_+,p)}\right]
-n_1\ln\bigg[\f{\vartheta_2(x_+,p)}{\vartheta_1(x_+,p)}\bigg],
\label{aIndependentTermWavefunction}
\ee
where some constant terms are dropped because they do not affect comparison with the unnormalized wave function.

The third order coefficient, after every term is distributed to the right place, is a series expanded as
\begin{align}
\epsilon^{-2}\mathcal{G}_{1}&=b_0\left(\f{1}{8}x_2\!-\!\f{1}{32}x_2^2\!+\!\f{1}{64}x_2^3\!+\!\f{1}{32}x_1x_2^2\!-\!\f{1}{64}x_1^2x_2\!-\!\f{5}{512}x_2^4\!-\!\f{1}{128}x_1x_2^3\!+\!\f{1}{128}x_1^3x_2+\cdots\right)\notag\\
&+b_1\left(-\f{1}{8}x_2\!+\!\f{3}{32}x_2^2\!-\!\f{5}{64}x_2^3\!+\!\f{1}{32}x_1x_2^2\!+\!\f{1}{64}x_1^2x_2\!+\!\f{35}{512}x_2^4\!-\!\f{3}{128}x_1x_2^3\!-\!\f{1}{128}x_1^3x_2+\cdots\right)\notag\\
&+b_2\left(-\f{1}{8}x_1\!+\!\f{1}{32}x_1^2\!-\!\f{1}{64}x_1^3\!-\!\f{1}{32}x_1^2x_2\!+\!\f{1}{64}x_1x_2^2\!+\!\f{5}{512}x_1^4\!+\!\f{1}{128}x_1^3x_2\!-\!\f{1}{128}x_1x_2^3+\cdots\right)\notag\\
&+b_3\left(\f{1}{8}x_1\!-\!\f{3}{32}x_1^2\!+\!\f{5}{64}x_1^3\!-\!\f{1}{32}x_1^2x_2\!-\!\f{1}{64}x_1x_2^2\!-\!\f{35}{512}x_1^4\!+\!\f{3}{128}x_1^3x_2\!+\!\f{1}{128}x_1x_2^3+\cdots\right).
\end{align}
Nicely enough, each piece in brackets is summed up in a theta function, in sequence they are $\f{i\omega_1}{2\pi}\p_x\ln\vartheta_r(x_+,p)$ for $r=1,2,4,3$.
In fact, when the series (\ref{x1x2SeriesExpansion}) is substituted into $\mathcal{G}_{1}$ some constant terms arise, again they are neglected for comparison with the unnormalized wave function.
So apart from constant terms we have
\be
\epsilon^{-2}\mathcal{G}_{1}=\f{i\omega_1}{2\pi}\p_x\left[b_0\ln\vartheta_1(x_+,p)+b_1\ln\vartheta_2(x_+,p)+b_2\ln\vartheta_4(x_+,p)+b_3\ln\vartheta_3(x_+,p)\right].
\ee
The shifted Weierstrass zeta-function is related to theta function by $\widetilde{\zeta}(x)=\p_x\ln\vartheta_1(\pi x/2\omega_1,p)$, so the third order piece in $\mathcal{G}$ precisely gives the corresponding order terms in the wave function,
\be
\f{\mathcal{G}_{1}}{\epsilon a}=\f{i\sum\limits_{s=0}^3b_s\widetilde{\zeta}(x+\f{1}{2}\omega_2+\omega_s)}{2\nu}.
\ee

The fourth order coefficient shares the same structure as the third order one,
\begin{align}
\epsilon^{-3}\mathcal{G}_{2}&=b_0\left(x_2+\f{1}{2}x_1x_2^2+\f{11}{32}x_1^2x_2^3+\cdots\right)\notag\\
&+b_1\left(-x_2+x_2^2-x_2^3+\f{1}{2}x_1x_2^2+x_2^4-\f{1}{2}x_1x_2^3-x_2^5+\f{1}{2}x_1x_2^4+\f{5}{32}x_1^2x_2^3+\cdots\right)\notag\\
&+b_2\left(-x_1+x_1^2-x_1^3+\f{1}{2}x_1^2x_2+x_1^4-\f{1}{2}x_1^3x_2-x_1^5+\f{1}{2}x_1^4x_2+\f{5}{32}x_1^3x_2^2+\cdots\right)\notag\\
&+b_3\left(x_1+\f{1}{2}x_1^2x_2+\f{11}{32}x_1^3x_2^2+\cdots\right).
\end{align}
It is recognized that $\mathcal{G}_{2}\sim\p_x\mathcal{G}_{1}$, then apart from constant terms
\be
\epsilon^{-3}\mathcal{G}_{2}=\left(\f{i\omega_1}{2\pi}\right)^2\p_x^2\left[b_0\ln\vartheta_1(x_+,p)+b_1\ln\vartheta_2(x_+,p)+b_2\ln\vartheta_4(x_+,p)+b_3\ln\vartheta_3(x_+,p)\right].
\ee
The fourth order piece of $\mathcal{G}$ gives the corresponding terms in the wave function,
\be
\f{\mathcal{G}_{2}}{\epsilon a^2}=-\f{\sum\limits_{s=0}^3b_s\p_x\widetilde{\zeta}(x+\f{1}{2}\omega_2+\omega_s)}{4\nu^2}.
\ee

The explicit verification can be continued to higher order expansion terms. In summary, under the identification of parameters (\ref{WavefunctionParameterIndentify}),
the wave functions $\psi_\pm(x)$ are related to the prepotential $\mathcal{G}$ through the relation
\be
\psi_{\pm}\left(\pm x\pm\f{1}{2}\omega_2,\nu,b_s,p\right)\left[\f{\vartheta_4(x_+,p)}{\vartheta_3(x_+,p)}\right]^{n_3}\left[\f{\vartheta_1(x_+,p)}{\vartheta_2(x_+,p)}\right]^{n_1}
\Leftarrow\exp\left[\f{\mathcal{G}(a,\mu_i,\epsilon_{1,2},x_{1,2})}{\epsilon_1}\right]\biggr|_{\substack{\epsilon_1\to\epsilon\\ \epsilon_2\to0}}.
\label{psiPrepotentialElectric}
\ee
Their relation is represented by ``$\Leftarrow$'' because when the relation (\ref{WavefunctionParameterIndentify}) is substituted for the variables $x_1, x_2$ on the right side some constant terms are dropped to match the left side.

There are limits for the parameters $b_s$ under which the weak coupling results of the DTV potential and the $N_f$=4 super QCD reduce to the weak coupling results of the Lam\'{e} potential and the N=$2^*$ super QCD,
the necessary Landen transformation of elliptic functions are given in appendix \ref{LandenTransformations}.

\section{Strong coupling spectra and gauge theory duality}\label{StrongCouplingSpectraAndGaugeTheoryDuality}

In this section we compute the strong coupling expansion of spectrum for the DTV potential.
For the Schr\"{o}dinger equation, strong coupling means the four coupling constants $b_s$ of the potential are large quantities, presumably one should seek perturbative spectral solution for large $b_s$.
But the method of perturbative computation for the strong coupling spectrum of the Lam\'{e} equation and the ellipsoidal wave equation \cite{wh1412,wh1608},
both of which have only one large coupling constant, cannot be straightly applied in this case.
Plugging the wave function $\psi(\eta)=\exp(\int v(\eta) d\eta)$ into eq. (\ref{eqJacobi}), we get the relation $v^2(\eta)+v_\eta(\eta)=u(\eta)+\Lambda$.
For the DTV potential in its original form (\ref{DTVJacobi}), it is difficult to find the series expansion of $v(\eta)$ for large $b_s$ that solves the nonlinear relation.
In fact, as discussed in \cite{wh1306, wh1412} with help from the results of gauge theory, in this instance the strong coupling expansion parameter is $(b_0-b_1)^{\f{1}{2}}(b_2-b_3)^{\f{1}{2}}q^{\f{1}{2}}$ which does not naturally appear in any simple manipulation of $u(\eta)$. This expansion parameter is interpreted as {\em dynamically generated}, hence cannot be determined unless we have some prior knowledge about the strong coupling solution, the connection with the $N_f$=4 super QCD provides the key clue.

A few properties of the strong coupling solutions can be expected based on results obtained for other elliptic potentials.
We should use equation in the Jacobian elliptic form (\ref{eqJacobi}) and (\ref{DTVJacobi}) to study the strong coupling solution, so that expressions of the eigenvalue and eigenfunction appear in compact form,
and the duality relation manifests more clearly as we discuss later.
The Floquet exponent is denoted by $\mu$, it is different from the exponent $\nu$ of the weak coupling solution computed using the equation in the Weierstrass elliptic form. In the connection with gauge theory, $\mu$ is proportional to the magnetic dual of the scalar v.e.v $a_D$, so it is dual to $\nu$ in the sense of electric-magnetic duality of N=2 super QCD \cite{SW9407,SW9408}. Computation of the strong coupling spectral solution also provides necessary materials to derive the dual expansion of prepotential for the $N_f$=4 super QCD, presented in section \ref{StrongCouplingExpansionPrepotentialNf4}.

\subsection{Effective potential at the magnetic critical points $\eta_{*5}$} \label{EffectivePotentialEta5}

The leading order (in the $\epsilon$-expansion) of eigenvalue is computed in \cite{wh1306} using the normal form of Heun equation. The result obtained there implies that the coupling constants $b_s$, which are assumed to be large now, are not the correct expansion parameter. The correct perturbation parameter is a single quantity of order $(b_0-b_1)^{\f{1}{2}}(b_2-b_3)^{\f{1}{2}}q^{\f{1}{2}}$.
Meanwhile, the periodicity of elliptic potential is not affected by perturbative treatment. These facts suggest that in the strong coupling region the four components potential $u(\eta)$ can be ``averaged'' so that it behaves like a single term elliptic potential with periods $2\mathbf{K}$ and $2i\mathbf{K}^{\prime}$.

It is helpful to learn from the simpler examples in previous study \cite{wh1412}, to have a clue about how to ``average'' the potential $u(\eta)$. For elliptic potentials with multiple critical points, the strong coupling spectral solutions are eigenstates whose (complex valued) energies are small compared to the strength of potential.  Their eigenfunctions are spatially localised around the corresponding critical points of the  potential. That means the critical points looks like {\em local vacuum}, and the potential around the critical points behaves like a binding potential. To see more clearly the shape of the potential around a critical point one should change to a new coordinate system that the critical point is at its origin, then expand the potential at the critical point to rewrite it as an ``effective potential'' in the new coordinate.

\begin{figure}[ht]
\begin{center}
\includegraphics[width=9cm]{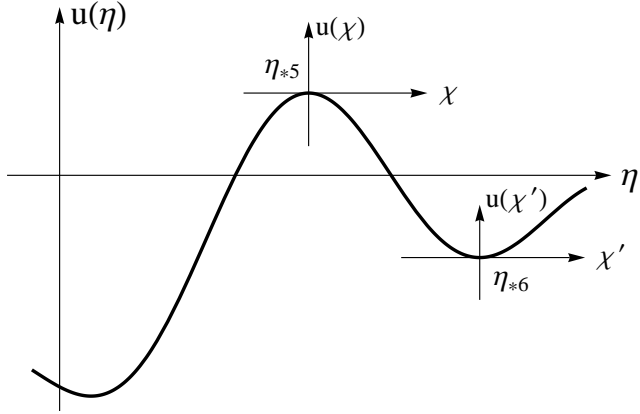}
\end{center}
\caption{Local coordinates at strong coupling critical points.} \label{Localcoordinates1}
\end{figure}

Figure \ref{Localcoordinates1} schematically illustrates the local coordinate systems around the strong coupling critical points $\eta_{*5}$ and $\eta_{*6}$, using a real potential with multiple critical points.
In the region around $\eta_{*5}$, the local $\chi$-coordinate is put at the critical point, as shown below.
When the DTV potential is rewritten in the $\chi$-coordinate and expanded at $\chi=0$, the effective potential contains a single dominant term with the largest coupling constant.
Then one can compute the perturbative spectral solution for this large coupling constant.
The situation is similar for the region around $\eta_{*6}$ when the potentia is rewritten in the local $\chi^{\,\prime}$-coordinate and expanded at $\chi^{\,\prime}=0$.
A fact should be stressed, the effective potential used to compute spectrum at the critical point $\eta_{*5}$ cannot be used to compute spectrum at the critical point $\eta_{*6}$, and vice versa. For the Lam\'{e} potential and the ellipsoidal potential discussed in \cite{wh1412, wh1608}, they are not generic enough to reveal this feature because their strong coupling critical points are at special values of the  coordinate so that their spectral solutions can be computed in the same coordinate system.

Regarding the relation with N=2 super QCD, {\em change coordinate systems} for the spectral problem is analogical to {\em change degrees of freedom} for the corresponding gauge theory models. In quantum field theory, from the viewpoint of effective theory, the proper degrees of freedom used to formulate the theory vary within the parameter space. In particular for some field theory models, particles and solitons are used to formulate the effective theory at weak and strong coupling regimes, respectively. The duality phenomenon of this kind on which the solution of SU(2) $N_f$=4 super QCD relies \cite{SW9407, SW9408}, constitutes an important development of quantum field theory.

In the rest of this section we compute the effective potentials at the critical points $\eta_{*5}, \eta_{*6}$ and their spectral solutions. To expand the DTV potential around the critical points $\eta_{*5}$,
introduce the coordinate $\chi$ and the corresponding energy relative to the local vacuum $\delta$ by
\be
\eta=\eta_{*5}+\chi, \qquad \Lambda=\Lambda_{*5}+\delta,
\ee
where $\Lambda_{*5}=-u(\eta_{*5})$, $\delta$ satisfies the condition $\delta\ll\Lambda_{*5}$ so that it is a small perturbation around $\eta_{*5}$.
Then eq. (\ref{eqJacobi}) is rewritten as
\be
\psi^{''}(\chi)-u(\chi)\psi(\chi)=\delta\psi(\chi),\label{eqJacobichisn}
\ee
with the effective potential $u(\chi)$ given by
\be
u(\chi)=u(\eta_{*5}+\chi)-u(\eta_{*5}).\label{effectivepotentialdef}
\ee
To perform perturbative computation, we need to simplify the right side of (\ref{effectivepotentialdef}) using the series solution of $\sn^2\eta_{*5}$ obtained in subsection \ref{CriticalPointsOfEllipticPotential}.
To keep the periodicity of the potential, the effective potential $u(\chi)$ must be an elliptic function made of $\sn\chi, \cn\chi$ and $\dn\chi$.

As the perturbation results at $\eta_{*5}$ and $\eta_{*6}$ are related in a simple way by the sign inversion $q^{\f{1}{2}}\to-q^{\f{1}{2}}$,
in this section we just need to discuss the spectral solution at $\eta_{*5}$.
The spectral solution at $\eta_{*6}$ is obtained by the same sign inversion. So in the discussion below the numerical subscriptions are dropped.

The periodicity of the potential plays a crucial role in deriving the correct expression for $u(\chi)$.
The periods of the potentials $u(\eta)$ and $u(\chi)$ are $2\mathbf{K}, 2i\mathbf{K}^{\prime}, 2\mathbf{K}+2i\mathbf{K}^{\prime}$,
we need to keep every elliptic function expression with the right periodic behavior when we change their forms.
For example, the difference $\sn^2(\eta_*+\chi)-\sn^2\eta_*$ is on the right side of eq. (\ref{effectivepotentialdef}),
to simplify it we need the addition formula for elliptic function,
\be
\sn(\eta_*+\chi)=\f{\sn\eta_*\cn\chi\dn\chi+\sn\chi\cn\eta_*\dn\eta_*}{1-q\sn^2\eta_*\sn^2\chi}.
\ee
In the expression of  $\sn^2(\eta_*+\chi)$, the factor $(1-q\sn^2\eta_*\sn^2\chi)^{-2}$ can be expanded as a $q$-series, the coefficient of each term is a non-negative integer power of $\sn^2\eta_*\sn^2\chi$.
The factor $(\sn\eta_*\cn\chi\dn\chi+\sn\chi\cn\eta_*\dn\eta_*)^2$ contains three terms
\begin{align}
&\sn^2\eta_*\cn^2\chi\dn^2\chi,\notag\\
&\sn^2\chi\cn^2\eta_*\dn^2\eta_*, \notag\\
&\sn\eta_*\cn\eta_*\dn\eta_*\sn\chi\cn\chi\dn\chi.
\end{align}
The first two can be transformed to contain only non-negative integer powers of $\sn^2\eta_*$ and $\sn^2\chi$,
using relations of the Jacobian elliptic functions. The third term needs careful treatment.
The series solutions of $\sn\eta_*, \cn\eta_*$ and $\dn\eta_*$ can be obtained from the solution of $\sn^2\eta_*$ given in (\ref{criticsn56}), by taking either $``+"$ or $``-"$ sign for square roots.
But we shall {\em not} rewrite $\sn\chi\cn\chi\dn\chi$ in a similar way as $\sn\chi\cn\chi\dn\chi \to \pm\sn\chi(1-\sn^2\chi)^{\f{1}{2}}(1-q\sn^2\chi)^{\f{1}{2}}$,
because this substitution would spoil the periodicity of the potential $u(\chi)$.
The factor $\sn\chi\cn\chi\dn\chi$ on the left side of ``$\to$'' is invariant under the translation of $\chi$ by periods $2\mathbf{K}, 2i\mathbf{K}^{\prime}$ and $2\mathbf{K}+2i\mathbf{K}^{\prime}$.
But the substituted expression on the right side of ``$\to$'' does not maintain periodicity for the translation of $\chi$ by $2\mathbf{K}$ because $\sn(\chi+2\mathbf{K})=-\sn\chi$ while $\cn\chi$ and $\dn\chi$ are invariant, so the overall sign flips.
The situation is similar for the other three difference terms in (\ref{effectivepotentialdef}), such as $\cn^2(\eta_*+\chi)/\dn^2(\eta_*+\chi)-\cn^2\eta_*/\dn^2\eta_*$.
Whenever a term with powers of $\cn\chi$ and $\dn\chi$ arises, the correct substitution rules that preserve periodicity of the potential are the following,
\begin{align}
\cn^{2m-1}\chi&=(1-\sn^2\chi)^{m-1}\cn\chi,\notag\\
\cn^{2m}\chi&=(1-\sn^2\chi)^m, \notag\\
\dn^{2m-1}\chi&=(1-q\sn^2\chi)^{m-1}\dn\chi, \notag\\
\dn^{2m}\chi&=(1-q\sn^2\chi)^m,
\label{substitutionright1}
\end{align}
for integer $m\geqslant 1$.

The effective potential obtained in this way contains infinitely many terms,
\be
u(\chi)=\sum\limits_{m=1}^{\infty}\left(\Omega_{2m}\sn^{2m}\chi+\Omega_{2m+1}\sn^{2m+1}\chi\cn\chi\dn\chi \right),
\label{EllipticHillOperatorSN}
\ee
where the effective coupling constants $\Omega_{2m}$ and $\Omega_{2m+1}$ are functions of $b_s$ and $q$, determined by the solution of $\sn\eta_*$.
The coupling constants are $q$-series of mass dimension zero, by counting the powers of $q$ we have $\Omega_2\gg\Omega_3\sim\Omega_4\gg\Omega_5\sim\Omega_6 \cdots$.
So among the infinitely many coupling constants there is a single largest one $\Omega_2\sim-4(b_0-b_1)^{\f{1}{2}}(b_2-b_3)^{\f{1}{2}}q^{\f{1}{2}}$ which can be used as the perturbation parameter.
Moreover, in the region $-\mathbf{K}\lesssim\chi\lesssim\mathbf{K}$, $\vert\sn\chi\vert\lesssim 1$, so that later terms in the effective potential with increasing $m$-index become less important.
Therefor, the effective potential indeed behaves like the Lam\'{e} potential, $u(\chi)\sim\Omega_2\sn^2\chi$, with higher order correction terms.
This fact explains why the strong coupling solution of the DTV potential (\ref{EigenvalueStrongCouplingMagnetic}) and (\ref{EigenfunctionStrongCouplingMagnetic}) has a similar form as the strong coupling solution of the Lam\'{e} potential \cite{wh1412, wh1608}, with higher order perturbations $\Omega_{m\geqslant3}$ included accordingly. Notice that $\Omega_2$ and $\Lambda_{*5}$ have the same magnitude, $\Omega_2\sim2\Lambda_{*5}$.

To compute the dual expansion of prepotential for the $N_f$=4 super QCD,
in appendix \ref{EffectiveCouplingOmegaMassRelation} we give explicit expressions for the first few coupling constants $\Omega_m$ in the leading order of the $\epsilon$-expansion,
in terms of physical masses using the relation of $b_s$ and $\mu_i$ given in eq. (\ref{b2murelation}).

\subsection{The asymptotic eigenvalue}\label{TheAsymptoticEigenvalueEta5}

Now we can proceed to find the strong coupling solution for the potential (\ref{EllipticHillOperatorSN}) associated with the magnetic critical point at $\chi_{*5}=0$ where $u(\chi_{*5})=0$.
The existence of a single dominant coupling constant of $u(\chi)$ implies that the equation (\ref{eqJacobichisn}) is in the same class of elliptic spectral problem including the Lam\'{e} equation and the ellipsoidal wave equation studied in \cite{wh1412,wh1608}, the method used there can be applied straightforwardly.

Plugging the wave function $\psi(\chi)=\exp(\int v(\chi)d\chi)$ into eq. (\ref{eqJacobichisn}) leads to the relation $v^2(\chi)+v_\chi(\chi)=u(\chi)+\delta$. We are seeking a strong coupling expansion of $v(\chi)$, with $\Omega_2^{\f{1}{2}}$ as the large parameter. The solution is expanded as
\be
v(\chi)=\sum_{\ell=-1}^{\infty}\f{v_\ell(\chi)}{(\pm\Omega_2^{\f{1}{2}})^\ell}, \label{Vlargeomega2expansion}
\ee
it includes two sectors depending on the choice of the signs. In the process of recursively solving $v_\ell(\chi)$,
the substitution rules (\ref{substitutionright1}) are applied whenever a term with $\cn^n\chi$ or $\dn^n\chi$ with $n\geqslant2$ arises. The first few $v_\ell(\chi)$ are
\begin{align}
v_{-1}(\chi)=&\sn\chi,\notag\\
v_0(\chi)=&-\f{1}{2}\p_\chi(\ln\sn\chi),\notag\\
v_1(\chi)=&-\f{3}{2^3\sn^3\chi}+\f{1+q+4\delta}{2^3\sn\chi}+\f{q}{2^3}\sn\chi+\dots+\p_\chi\left(\f{\Omega_3}{6}\sn^3\chi+\f{\Omega_5}{10}\sn^5\chi+\cdots\right),\notag\\
v_2(\chi)=&\p_\chi\left(\f{3}{2^4\sn^4\chi}-\f{1+q+4\delta}{2^4\sn^2\chi}-\f{\Omega_4}{2^2}\sn^2\chi+\cdots\right)+\p_\chi^2\left(-\f{\Omega_3}{2^3}\sn^2\chi-\f{\Omega_5}{2^4}\sn^4\chi+\cdots\right), \notag\\
v_3(\chi)=&-\f{297}{2^7\sn^7\chi}+\f{139(1+q)+76\delta}{2^6\sn^5\chi}-\f{25+236q+25q^2+104(1+q)\delta+16\delta^2}{2^7\sn^3\chi}+\cdots\notag\\
\quad&+\p_\chi\left(-\f{\Omega_3}{2^4\sn\chi}-\f{[5(1+q)+4\delta]\Omega_3-9\Omega_5}{2^4}\sn\chi+\cdots\right).
\label{Vlargeomega2expansioncoeff}
\end{align}
There are two noteworthy properties for the expansion of $v_\ell(\chi)$.
The coordinate dependence is shown only in the function $\sn\chi$, this feature is crucial for the discussion in subsection \ref{Dualityforstrongcouplingspectra} about the duality transformation of magnetic and dyonic wave functions.
This expansion is valid in the neighbourhood of the point $\chi\sim0$, and $v_\ell(\chi)$ are Laurent series in $\sn\chi$ which would be used to evaluate the contour integrals in the next subsection.
The general form of $v_{2\ell}(\chi)$ can be written as a derivative,
\be
v_{2\ell}(\chi)=\p_\chi\left(c_{2\ell}^{\;\prime}\ln\sn\chi+\sum_{n=-\infty}^{2\ell}\f{c_{2\ell,2n}^{\;\prime}}{\sn^{2n}\chi}\right)+\p^2_\chi\left(c_{2\ell}^{\;\prime\prime}\ln\sn\chi+\sum_{n=-\infty}^{2\ell-3}\f{c_{2\ell,2n}^{\;\prime\prime}}{\sn^{2n}\chi}\right).
\label{vevenmagnetic}
\ee
In the summations, the coefficient of constant terms $c_{2\ell,0}^{\;\prime}$ and $c_{2\ell,0}^{\;\prime\prime}$ are all zero, and $c_{2\ell}^{\;\prime}=0$ for $\ell \geqslant 1$.
The general form of $v_{2\ell+1}(\chi)$ is
\be
v_{2\ell+1}(\chi)=\sum_{n=-\infty}^{2\ell+1}\f{c_{2\ell+1,2n+1}}{\sn^{2n+1}\chi}+\p_\chi\left(\sum_{n=-\infty}^{2\ell-2}\f{c_{2\ell+1,2n+1}^{\;\prime}}{\sn^{2n+1}\chi}\right).
\label{voddmagnetic}
\ee
Nonzero coefficients $c_{2\ell}^{\;\prime}, c_{2\ell,2n}^{\;\prime}, c_{2\ell}^{\;\prime\prime}, c_{2\ell,2n}^{\;\prime\prime}$ and $c_{2\ell+1,2n+1}, c_{2\ell+1,2n+1}^{\;\prime}$ are polynomials of $\delta, \Omega_{m\geqslant3}$ and $q$.
In the following subsections \ref{TheAsymptoticEigenvalueEta5} and \ref{TheAsymptoticEigenfunctionEta5},
eigenvalue and eigenfunction are obtained by evaluating the definite and indefinite integrals of $v(\chi)$ respectively.

The eigenvalue $\delta(\mu)$ at the strong coupling point $\chi_{*5}$ is computed from the monodromy of wave function over an interval of the period $2i\mathbf{K}^{\prime}$.
As we showed for other examples in \cite{wh1108, wh1412}, the definition of the Floquet exponent $\mu$ does not follow a direct application of the classical Floquet theory.
The correct relation of $\mu$ and the monodromy of wave function is
\be
\mu=\f{1}{\pi}\int_{\chi_0}^{\chi_0+2i\mathbf{K}^{\prime}} v(\chi)d\chi.\label{FloquetExponentDefMagnetic}
\ee
An explanation of the relation is given in subsection \ref{TheAsymptoticEigenfunctionEta5} by rewriting the corresponding wave function in the canonical coordinate.
To actually evaluate the integration, the integral over an interval of $2i\mathbf{K}^{\prime}$ in the $\chi$-plane is mapped to the integral along the $\beta$-contour in the $\xi$-plane defined by $\xi=\sn^2\chi$,
as explained in appendix B of \cite{wh1412}.
In the connection with gauge theory, it is related to the integral of the Seiberg-Witten differential over the $\beta$-cycle on the curve of $N_f$=4 super QCD \cite{SW9407,SW9408}.

As shown in eq. (\ref{vevenmagnetic}), $v_{2\ell}$ are derivative of series in $\sn\chi$, therefore are invariant under the translation $\chi\to\chi+2i\mathbf{K}^{\prime}$,
so we have
\be
\int_{\chi_0}^{\chi_0+2i\mathbf{K}^{\prime}} v_{2\ell}(\chi)d\chi=0.
\ee
As shown in eq. (\ref{voddmagnetic}), each $v_{2\ell+1}(\chi)$ contains two parts.
The series in the first part are made of $\sn^m\chi$ with $m=\mp(2n+1)$, so the corresponding integrals are given by
$\mathrm{I}_{m}=\int_{\chi_0}^{\chi_0+2i\mathbf{K}^{\prime}} \sn^m\chi d\chi$.
According to the prescription of contour integration compatible with the Floquet theorem, described in \cite{wh1108,wh1412}, for positive exponent $\mathrm{I}_{2n+1}=0$,
and for negative exponent $\mathrm{I}_{-(2n+1)}\ne0$.
In the second part, the series in the bracket made of $\sn^{2n+1}\chi$ are invariant under the translation $\chi\to\chi+2i\mathbf{K}^{\prime}$,
hence their integrals over an interval of $2i\mathbf{K}^{\prime}$ vanish.
Therefore, in the series of $v_{2\ell+1}(\chi)$  given by (\ref{voddmagnetic}), only terms in the first part with $2n+1>0$ (i.e. the principle part of the Laurent series) contribute nonzero integrals.
When all terms of vanishing integral are dropped, the Floquet exponent (\ref{FloquetExponentDefMagnetic}) is computed by
\be
\mu=\f{1}{\pi}\sum_{\ell=0}^{\infty}\sum_{n=0}^{2\ell+1}\f{c_{2\ell+1,2n+1}}{(\pm\Omega_2^{\f{1}{2}})^{2\ell+1}}\int_{\chi_0}^{\chi_0+2i\mathbf{K}^{\prime}} \f{d\chi}{\sn^{2n+1}\chi}.
\label{FloquetExponentDefMagneticsimplified}
\ee
The integrals $\mathrm{I}_{-(2n+1)}$ can be computed by recurrence relations with a choice for the integral contour in the $\xi$-plane as explained in appendix B of \cite{wh1412}.
The summation involves only odd integer powers of $\pm\Omega_2^{-\f{1}{2}}$, so the sign inversion for $\Omega_2^{\f{1}{2}}$ leads to the sign inversion for $\mu$.

The integral (\ref{FloquetExponentDefMagneticsimplified}) gives the functional relation $\mu(\delta)$ as a series expansion for large $\Omega_2$,
reversing the series leads to the strong coupling dispersion relation.
We give the series expansion of the $``+"$ sign sector eigenvalue up to order $\mathcal{O}(\Omega_2^{-2})$,
\begin{align}
\delta=&-2i\Omega_2^{\f{1}{2}}\mu-\f{1}{2^3}(1+q)(4\mu^2-1)\notag\\
&-\f{i}{2^5\Omega_2^{\f{1}{2}}}\Big\{(1+q)^2(4\mu^3-3\mu)-4q(4\mu^3-5\mu)\Big\}\notag\\
&+\f{1}{2^{10}\Omega_2}\Big\{(1+q)(1-q)^2(80\mu^4-136\mu^2+9)+384\Omega_4(4\mu^2-1)\Big\}\notag\\
&+\f{i}{2^{13}\Omega_2^{\f{3}{2}}}\Big\{(1+q)^4(528\mu^5-1640\mu^3+405\mu)-24q(1+q)^2(112\mu^5-360\mu^3+95\mu)\notag\\
&\quad+16q^2(144\mu^5-520\mu^3+173\mu)-512\Omega_4(1+q)(4\mu^3-11\mu)+5120\Omega_6(4\mu^3-5\mu)\Big\}\notag\\
&-\f{1}{2^{15}\Omega_2^2}\Big\{(1+q)(1-q)^2\big[9(1+q^2)(224\mu^6-1120\mu^4+654\mu^2-27)\notag\\
&\quad-2q(224\mu^6-2080\mu^4+2286\mu^2-135)\big]+1024\Omega_3^2(60\mu^2-7)\notag\\
&\quad+32\Omega_4\big[(1-q)^2(80\mu^4-136\mu^2+9)-288q(4\mu^2-1)\big]\notag\\
&\quad-1920\Omega_6(1+q)(16\mu^4-104\mu^2+21)+8960\Omega_8(16\mu^4-56\mu^2+9)\Big\}\notag\\
&+\mathcal{O}(\Omega_2^{-\f{5}{2}})
\label{EigenvalueStrongCouplingMagnetic}
\end{align}
The $``-"$ sign sector eigenvalue is obtained by inversion $\Omega_2^{\f{1}{2}}\to-\Omega_2^{\f{1}{2}}$ or by inversion $\mu\to-\mu$.
The condition $\delta\ll\Lambda_{*5}$ is satisfied for $\mu\ll(b_0-b_1)^{\f{1}{2}}(b_2-b_3)^{\f{1}{4}}q^{\f{1}{4}}\sim\Omega_2^{\f{1}{2}}$. This spectral solution is considered as a strong coupling solution in the sense that the strength of potential is much larger than the eigenvalue, the condition is $\Omega_2\gg\delta$, or equivalently characterised by the condition $\Omega_2^{\f{1}{2}}\gg\mu$.
The eigenvalue (\ref{EigenvalueStrongCouplingMagnetic}) generalises the strong coupling solution for the Lam\'{e} equation and the ellipsoidal wave equation.

The dominant coupling constant is $\Omega_2\sim(b_0-b_1)^{\f{1}{2}}(b_2-b_3)^{\f{1}{2}}q^{\f{1}{2}}+\mathcal{O}(q)$, therefore either $\Omega_2^{\f{1}{2}}$ or the leading term   $(b_0-b_1)^{\f{1}{2}}(b_2-b_3)^{\f{1}{4}}q^{\f{1}{4}}$ can be used as the strong coupling expansion parameter. In the spectral problem $\Omega_2$ is a more convenient expansion parameter for the eigenvalue (\ref{EigenvalueStrongCouplingMagnetic}) and the eigenfunction (\ref{EigenfunctionStrongCouplingMagnetic}), but in the related problem about the dual expansion of prepotential for the $N_f$=4 super QCD, discussed in section \ref{StrongCouplingExpansionPrepotentialNf4}, the proper expansion parameter is $(b_0-b_1)^{\f{1}{2}}(b_2-b_3)^{\f{1}{4}}q^{\f{1}{4}}\sim(\mu_1\mu_2\mu_3\mu_4)q^{\f{1}{4}}$.

\subsection{The asymptotic eigenfunction}\label{TheAsymptoticEigenfunctionEta5}

The wave function $\psi(\chi)=\exp(\int v(\chi) d\chi)$ associated with the eigenvalue (\ref{EigenvalueStrongCouplingMagnetic}) is obtained from the indefinite integral of $v(\chi)$ given by eqs. (\ref{Vlargeomega2expansion}) and (\ref{Vlargeomega2expansioncoeff}). We substitute the solution (\ref{EigenvalueStrongCouplingMagnetic}) for the eigenvalue $\delta$,
then the wave function is written as exponential of a series for large $\Omega_2^{\f{1}{2}}$. Taking into account both $``\pm"$ sign sectors $\delta=\pm i\Omega_2^{\f{1}{2}}\mu+\cdots$,
we obtain the unnormalized linearly independent basic wave functions up to order $\mathcal{O}(\Omega_2^{-1})$,
\begin{align}
\psi_\pm(\chi)=&\exp\Bigg(\pm\Omega_2^{\f{1}{2}}q^{-\f{1}{2}}\ln(\dn\chi-q^{\f{1}{2}}\cn\chi)-\f{1}{2}\Big\{\ln\sn\chi\mp 2i\mu\ln\f{\dn\chi+\cn\chi}{\sn\chi}\Big\}\notag\\
&\pm\f{1}{2^4\Omega_2^{\f{1}{2}}}\Big\{\f{\pm 8i\mu+(3-4\mu^2)\cn\chi\dn\chi}{\sn^2\chi}+2q^{\f{1}{2}}\ln(\dn\chi-q^{\f{1}{2}}\cn\chi)+\f{8}{3}\Omega_3\sn^3\chi\notag\\
&\quad+4\Omega_4\big[q^{-1}\cn\chi\dn\chi+q^{-\f{3}{2}}(1+q)\ln(\dn\chi-q^{\f{1}{2}}\cn\chi)\big]+\mathcal{O}(\Omega_5)\Big\}\notag\\
&+\f{1}{2^6\Omega_2}\Big\{\f{12\!-\!32\mu^2\pm i(38\mu\!-\!8\mu^3)\cn\chi\dn\chi}{\sn^4\chi}+\f{(1\!+\!q)(3\!-\!4\mu^2)(\pm i\mu\cn\chi\dn\chi\!-\!2)}{\sn^2\chi}\notag\\
&\quad\!+\!16\Omega_3(\pm2i\mu\!-\!\cn\chi\dn\chi)\sn\chi\!-\!16\Omega_4\big[\sn^2\chi\mp 2i\mu q^{-\f{1}{2}}\ln(\dn\chi\!-\!q^{\f{1}{2}}\cn\chi)\big]\!+\!\mathcal{O}(\Omega_5)\Big\}\notag\\
&+\mathcal{O}(\Omega_2^{-\f{3}{2}})\Bigg).
\label{EigenfunctionStrongCouplingMagnetic}
\end{align}

The wave functions provide a more transparent explanation for the monodromy relation (\ref{FloquetExponentDefMagnetic}).
According to the prescription of integral over an interval of $2\mathbf{K}$ explained in appendix B of \cite{wh1412},
the corresponding path of integration in the $\chi$-plane crosses the branch cut of $\ln[(\dn\chi+\cn\chi)/\sn\chi]$ and produces a monodromy $i\pi$,
but it avoids the branch cut of $\ln(\dn\chi-q^{\f{1}{2}}\cn\chi)$.
The rest of the wave function containing $\ln\sn\chi, \sn^m\chi,\cn^{2m}\chi$ and $\cn\chi\dn\chi$, with $m\in\mathbb{Z}$, are invariant under the translation $\chi\to\chi+2i\mathbf{K}^{\prime}$.
Therefore, the monodromy of wave function comes from a single term $\pm i\mu\ln[(\dn\chi+\cn\chi)/\sn\chi]$, the rest of the wave function is periodic.
This property indicates $\psi_\pm(\chi)$ given by (\ref{EigenfunctionStrongCouplingMagnetic}) are Floquet functions, but they are not yet in a standard form.
When $\psi_\pm(\chi)$ are regarded as the wave functions of a quantum particle, $\mu$ is interpreted as the quasimomentum, so we define the canonically conjugate coordinate $\rho$ by
\be
\ln\frac{\dn\chi+\cn\chi}{(1-q)^{\f{1}{2}}\sn\chi}=\rho.\label{CanonCoordMagneticDef}
\ee
The factor $(1-q)^{\f{1}{2}}$ is included so that all terms of elliptic function in the wave functions can be expressed in hyperbolic functions $\sinh\rho$ and $\cosh\rho$,
it causes a coordinate independent constant that does not affect the unnormalized wave functions.
Then the wave functions are transformed to the standard form of Floquet function,
\be
\psi_\pm(\chi)=\exp(\pm i\mu\rho)\mathbf{p}_2^{\pm}(\mu,\rho),
\label{MagneticWavefunctionFloquetForm}
\ee
where $\mathbf{p}_2^{\pm}(\mu,\rho)$ are periodic functions, $\mathbf{p}_2^{\pm}(\mu,\rho+i\pi)=\mathbf{p}_2^{\pm}(\mu,\rho)$, they are related by $\mathbf{p}_2^{+}(\mu,\rho)=\mathbf{p}_2^{-}(-\mu,\rho)$.

The wave functions written in the canonical coordinate explain a puzzling point in previous study \cite{wh1108,wh1412,wh1608}. There the monodromy relation (\ref{FloquetExponentDefMagnetic}) for elliptic potentials is determined by requiring the solutions of elliptic potential have the right limit to the solutions of trigonometric potentials when taking the limit $q\to 0$ and fixing other coupling constants properly. This requirement implies that the naive relation between the Floquet exponent and the monodromy of wave function that follows from the translation $\chi\to\chi+2i\mathbf{K}^{\prime}$,
\be
\mp2\mu\mathbf{K}^{\prime}\stackrel{?}{=}\ln\f{\psi_{\pm}(\chi+2i\mathbf{K}^{\prime})}{\psi_{\pm}(\chi)},
\ee
is {\em not} correct. This is because $\mu$ is the conjugate momentum not to the original coordinate $\chi$ but to the canonical coordinate $\rho$, so the correct monodromy relation follows from the translation $\rho\to \rho+i\pi$ is
\be
\mp\mu\pi=\ln\f{\psi_{\pm}(\rho+i\pi)}{\psi_{\pm}(\rho)}.
\ee
It is equivalent to the relation in the $\chi$-coordinate
\be
\mp\mu\pi=\ln\f{\psi_{\pm}(\chi+2i\mathbf{K}^{\prime})}{\psi_{\pm}(\chi)},
\ee
which is the monodromy relation (\ref{FloquetExponentDefMagnetic}).
A similar argument applies to the dyonic strong coupling solution obtained in subsection \ref{secondstrongcouplingsolutionfromduality}, the explicit form of wave function is discussed in appendix \ref{MoreAboutDyonicSpectrum}.

In the strong coupling expansion of the eigenvalue (\ref{EigenvalueStrongCouplingMagnetic}) and the eigenfunction (\ref{EigenfunctionStrongCouplingMagnetic}), higher order coupling constants $\Omega_{m\geqslant5}$ would enter the expressions in later terms. Setting $\Omega_2, \Omega_4$ nonzero and all other coupling constants to zero, we get the corresponding solution for the {\em ellipsoidal wave equation}.

\subsection{The second strong coupling solution from duality}\label{secondstrongcouplingsolutionfromduality}

\subsubsection{Effective potential at the dyonic critical point $\eta_{*6}$}\label{EffectivePotentialEta6}

The second strong coupling solution is associated with the dyonic critical point $\eta_{*6}$ of the DTV potential (\ref{DTVJacobi}).
The distance of the magnetic and dyonic critical points is $\eta_{*6}-\eta_{*5}\sim\mathbf{K}$ as shown in (\ref{DistanceOfMDPoints}),
therefore for the effective potential (\ref{EllipticHillOperatorSN}) the dyonic critical point is at $\chi_{*6}=\eta_{*6}-\eta_{*5}\sim\mathbf{K}$ where $\sn\chi\sim 1$.
The effective potential (\ref{EllipticHillOperatorSN}) has infinitely many terms, although the magnitude of coupling constants $\Omega_m$ decreases with increasing $m$-index,
the magnitude of $\sn^m\chi$ grows unbounded with increasing $m$-index for $\sn\chi \gtrsim 1$.
As a result, the potential (\ref{EllipticHillOperatorSN}) is valid only in the neighbourhood of the critical point $\chi_{*5}$ where $\sn\chi\sim 0$,
and cannot be directly used to compute the spectrum around $\chi_{*6}$.

To find the effective potential locally valid in the neighbourhood of the dyonic critical point, recall that the corresponding strong coupling solution of the Lam\'{e} potential is computed by rewriting the potential in the Jacobi {\itshape \cn}-function \cite{wh1412, wh1608}.
But here it is {\em not} correct to get the potential by simply applying the substitution $\sn^{2m}\chi\to (1-\cn^2\chi)^m$ to (\ref{EllipticHillOperatorSN}) as
\be
\sum_{m=1}^\infty\Big(\Omega_{2m}(1-\cn^2\chi)^m+\Omega_{2m+1}(1-\cn^2\chi)^m\sn\chi\cn\chi\dn\chi\Big),
\label{WrongSubstitutionSn2Cn}
\ee
and expanding it as a polynomial in $\cn\chi$. This is because the summation in potential (\ref{EllipticHillOperatorSN}) is valid only for $\sn\chi\sim 0$, hence $\cn\chi\sim 1$ is already assumed,
so the series expansion in $\cn\chi$ obtained from (\ref{WrongSubstitutionSn2Cn}) does not make sense as an asymptotic series.

We rewrite the effective potential in another coordinate system, denoted by $\varkappa$, in which the dyonic critical point is at $\varkappa_{*6}=\mathbf{K}$.
Then in the region around the point $\varkappa\sim\mathbf{K}$ the condition $\cn\varkappa\sim 0$ is satisfied, so an asymptotic series in $\cn\varkappa$ does make sense as long as the coefficients do not grow too fast.
The effective potential at $\varkappa_{*6}$ is computed as described below. Define a new coordinate by
\be
\eta=\eta_{*6}+\chi^{\,\prime},\qquad \Lambda=\Lambda_{*6}+\sigma,
\label{etachiprimerelation}
\ee
where $\Lambda_{*6}=-u(\eta_{*6})$, and $\sigma$ is a small perturbation around the critical point $\eta_{*6}$ so that $\sigma\ll\Lambda_{*6}$ is satisfied.
The equation (\ref{eqJacobi}) can be rewritten as the spectral problem for an effective potential $u(\chi^{\,\prime})$,
\be
\psi^{''}(\chi^{\,\prime})-u(\chi^{\,\prime})\psi(\chi^{\,\prime})=\sigma\psi(\chi^{\,\prime}),
\label{eqJacobichiprime}
\ee
where $u(\chi^{\,\prime})=u(\eta_{*6}+\chi^{\,\prime})-u(\eta_{*6})$.
The effective potential $u(\chi^{\,\prime})$ can be simplified following the procedure in subsection \ref{EffectivePotentialEta5} using the series solution of $\sn^2\eta_{*6}$.
When the substitution rules given in (\ref{substitutionright1}) are taken into account, $u(\chi^{\,\prime})$ is a polynomial of $\sn\chi^{\,\prime}, \cn\chi^{\,\prime} $ and $\dn\chi^{\,\prime}$.
Because solutions of the critical points $\eta_{*5}$ and $\eta_{*6}$ differ only by a sign inversion $q^{\f{1}{2}}\leftrightarrow -q^{\f{1}{2}}$,
the effective potential $u(\chi^{\,\prime})$ takes the same form as the effective potential $u(\chi)$ given in (\ref{EllipticHillOperatorSN}) with the effective coupling constants replaced by $\Omega_m(-q^{\f{1}{2}},\mu_i)$. Then around the critical point $\eta_{*6}$ the effective potential is
\be
u(\chi^{\,\prime})=\sum\limits_{m=1}^{\infty}\Big(\Omega_{2m}^{\,\prime}\sn^{2m}\chi^{\,\prime}+\Omega_{2m+1}^{\,\prime}\sn^{2m+1}\chi^{\,\prime}\cn\chi^{\,\prime}\dn\chi^{\,\prime} \Big),\label{EllipticHillOperatorSN2}
\ee
where $\Omega_m^{\,\prime}\equiv\Omega_m^{\,\prime}(q^{\f{1}{2}})=\Omega_m(-q^{\f{1}{2}})$, the hierarchical structure of coupling constants persists.

One should not, however, continue as in subsection \ref{TheAsymptoticEigenvalueEta5} to compute the eigenvalue in the $\chi^{\,\prime}$-coordinate.
In the approach of this paper two elements are crucial for computation of asymptotic solutions for differential equation with elliptic potentials.
One is deriving the integrand under proper assumption of a large parameter;
the other is choosing the correct integral contour associated to a period. The integrand and the integral contour must fit with each other so that the procedure works.
As shown in \cite{wh1412}, the spectrum at the dyonic critical point is obtained by evaluating integrals over an interval of the period $2\mathbf{K}+2i\mathbf{K}^{\prime}$.
The integrands obtained for the potential $u(\chi)$ and the potential $u(\chi^{\,\prime})$ are Laurent series in $\sn\chi$ and $\sn\chi^{\,\prime}$ respectively,
they are compatible with the integral over an interval of $2i\mathbf{K}^{\prime}$, but incompatible with the integral over an interval of $2\mathbf{K}+2i\mathbf{K}^{\prime}$.
We have to work in the $\varkappa$-coordinate promised before, in doing so, the integrands are expressed in $\cn\varkappa^{\,\prime}$ and become compatible with the integral over an interval of  $2\mathbf{K}+2i\mathbf{K}^{\prime}$.

\begin{figure}[ht]
\begin{center}
\includegraphics[width=9cm]{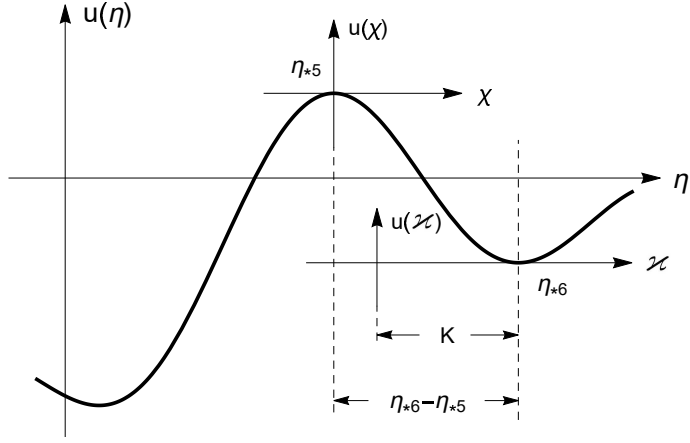}
\end{center}
\caption{The $\varkappa$-coordinate replaces the $\chi^{\,\prime}$-coordinate.} \label{Localcoordinates2}
\end{figure}

The $\eta$-coordinate and the $\varkappa$-coordinate are related by $\eta=\eta_{*6}+\varkappa-\mathbf{K}$, so from (\ref{etachiprimerelation}) the $\chi^{\,\prime}$-coordinate and the $\varkappa$-coordinate are related by $\chi^{\,\prime}=\varkappa-\mathbf{K}$. The region around $\chi^{\,\prime}\sim 0$ becomes the region around $\varkappa\sim \mathbf{K}$, as desired, see figure \ref{Localcoordinates2}.
To rewrite the potential (\ref{EllipticHillOperatorSN2}) in the $\varkappa$-coordinate, the following relations with the lower sign are used,
\begin{align}
\sn(\varkappa\pm\mathbf{K})&=\pm\f{\cn\varkappa}{\dn\varkappa},\notag\\
\cn(\varkappa\pm\mathbf{K})&=\mp(1-q)^{\f{1}{2}}\f{\sn\varkappa}{\dn\varkappa},\notag\\
\dn(\varkappa\pm\mathbf{K})&=(1-q)^{\f{1}{2}}\f{1}{\dn\varkappa}.
\label{ellipticfunctionschishiftK}
\end{align}
Substitution of these relations into the potential (\ref{EllipticHillOperatorSN2}) leads to an expression involving negative integer powers of $\dn\varkappa$,
the following substitution rules convert it to an expression in $\cn\varkappa$,
\begin{align}
\dn^{-2m}\varkappa&=(1-q+q\cn^2\varkappa)^{-m},\notag\\
\dn^{-(2m-1)}\varkappa&=(1-q+q\cn^2\varkappa)^{-m}\dn\varkappa.
\label{substitutionright2}
\end{align}
Expanding the obtained expression with respect to $\cn\varkappa$, we get the effective potential valid in the neighborhood of the point $\varkappa\sim\mathbf{K}$ which is an infinite series in $\cn\varkappa$,
\be
u(\varkappa)=\sum_{m=1}^{\infty}\Big(\mho_{2m}\cn^{2m}\varkappa+\mho_{2m+1}\cn^{2m+1}\varkappa\sn\varkappa\dn\varkappa\Big).
\label{EllipticHillOperatorCN}
\ee
The effective coupling constants $\mho_m$ are combinations of the coupling constants $\Omega_m^{\,\prime}$,
\begin{align}
\mho_{2m}&=\sum_{n=1}^{m}\f{n}{m}C^{m}_{n}\f{(-q)^{m-n}}{(1-q)^m}\Omega_{2n}(-q^{\f{1}{2}}),\notag\\
\mho_{2m+1}&=-\sum_{n=1}^{m}\f{m+1}{n+1}C^{m}_{n}\f{(-q)^{m-n}}{(1-q)^{m+1}}\Omega_{2n+1}(-q^{\f{1}{2}}),
\label{mhofromOmega}
\end{align}
for $m\geqslant n\geqslant1$, where $C^{m}_{n}$ are the binomial coefficients.
The effective coupling constants also have hierarchial values $\mho_2\gg\mho_3\sim\mho_4\gg\mho_5\sim\mho_6\cdots$, so the dominant one $\mho_2$ is the large expansion parameter.
In the region $-\mathbf{K}\lesssim\varkappa-\mathbf{K}\lesssim\mathbf{K}$, $\vert\cn\varkappa\vert\lesssim 1$, so that in the effective potential terms with increasing $m$-index become less important.
As the magnitude of $\mho_2\sim4(b_0-b_1)^{\f{1}{2}}(b_2-b_3)^{\f{1}{2}}q^{\f{1}{2}}$, the relation $\mho_2\sim2\Lambda_{*6}$ holds. The corresponding eigenvalue problem becomes
\be
\psi^{''}(\varkappa)-u(\varkappa)\psi(\varkappa)=\sigma\psi(\varkappa).\label{eqJacobichicn}
\ee

\subsubsection{Perturbative spectrum}

The computation of the strong coupling solution for the potential (\ref{EllipticHillOperatorCN}) is similar to that in subsections \ref{EffectivePotentialEta5} and \ref{TheAsymptoticEigenvalueEta5}.
The $``+"$ and $``-"$ sign sectors of solution for the relation $v^2(\varkappa)+v_\varkappa(\varkappa)=u(\varkappa)+\sigma$ are given by series for large $\mho_2$,
\be
v(\varkappa)=\sum_{\ell=-1}^{\infty}\f{v_\ell(\varkappa)}{(\pm\mho_2^{\f{1}{2}})^\ell}, \label{Vlargemho2expansion}
\ee
with $v_{-1}(\varkappa)=\cn\varkappa, v_0(\varkappa)=-\f{1}{2}\p_\varkappa(\ln\cn\varkappa)$, etc.
This strong coupling spectral solution is related to the monodromy of wave function over an interval of the period $2\mathbf{K}+2i\mathbf{K}^{\prime}$, so the Floquet exponent is given by
\be
\mu=\f{i(1-q)^{\f{1}{2}}}{\pi}\int_{\varkappa_0}^{\varkappa_0+2\mathbf{K}+2i\mathbf{K}^{\prime}} v(\varkappa)d\varkappa.\label{FloquetExponentDefDyonic}
\ee
It is the same relation that holds for a few other elliptic potential spectral problems \cite{wh1108,wh1412,wh1608}.
The reason to include a factor $(1-q)^{\f{1}{2}}$ is further explained in appendix \ref{MoreAboutDyonicSpectrum} by rewriting the corresponding wave function in the canonical coordinate.
The integral over an interval of $2\mathbf{K}+2i\mathbf{K}^{\prime}$ in the $\varkappa$-plane is mapped to the integral along the $\gamma$-contour in the $\xi$-plane defined by $\xi=\sn^2\varkappa$.
It is related to the integral of the Seiberg-Witten differential over the $\gamma$-cycle on the curve of $N_f$=4 super QCD \cite{SW9407,SW9408}.

For computation of the integrals in (\ref{FloquetExponentDefDyonic}), the integrands $v_\ell(\varkappa)$ need to be simplified, explicit expressions of $v_\ell(\varkappa)$ are given in appendix \ref{MoreAboutDyonicSpectrum}.
The integrands $v_{2\ell}(\varkappa)$ can be written as derivatives, $v_{2\ell}(\varkappa)\sim\p_\varkappa(\cdots)+\p^2_\varkappa(\cdots)$, where the series in brackets are made of $\ln\cn\varkappa, \cn^{2n}\varkappa$ with $n$ taking both positive and negative integer value. These series are invariant under the translation $\varkappa \to \varkappa+2\mathbf{K}+2i\mathbf{K}^{\prime}$, so that the integrals of $v_{2\ell}(\varkappa)$ vanish,
\be
\int_{\varkappa_0}^{\varkappa_0+2\mathbf{K}+2i\mathbf{K}^{\prime}} v_{2\ell}(\varkappa)d\varkappa=0.
\ee
Each integrand $v_{2\ell+1}(\varkappa)$ also contains two parts. \;The first part contains series made of $\cn^m\varkappa$ with $m=\mp(2n+1)$,
\;and the corresponding integrals are given by $\mathrm{J}_{m}=\int_{\varkappa_0}^{\varkappa_0+2\mathbf{K}+2i\mathbf{K}^{\prime}}\cn^m\varkappa d\varkappa$.
By the same argument for contour integration used in \cite{wh1412}, similar to the case of computing the spectrum at $\eta_{*5}$, for positive exponent $\mathrm{J}_{2n+1}=0$, and for negative exponent  $\mathrm{J}_{-(2n+1)}\ne0$. The second part can be written as a derivative $\p_\varkappa(\cdots)$,
the series in the bracket has vanishing monodromy under the translation $\varkappa \to \varkappa+2\mathbf{K}+2i\mathbf{K}^{\prime}$,
so the second part does not contribute to the integral (\ref{FloquetExponentDefDyonic}).
Taken together, the non-vanishing contributions to the integral of $v(\chi)$ come only from terms in the first part of $v_{2\ell+1}(\varkappa)$ with negative power of $\cn\varkappa$.
The monodromy relation (\ref{FloquetExponentDefDyonic}) is simplified to
\be
\mu=\f{i(1-q)^{\f{1}{2}}}{\pi}\sum_{\ell=1}^{\infty}\sum_{n=0}^{2\ell+1}\f{\tilde{c}_{2\ell+1,2n+1}}{(\pm\mho_2^{\f{1}{2}})^{2\ell+1}}
\int_{\varkappa_0}^{\varkappa_0+2\mathbf{K}+2i\mathbf{K}^{\prime}} \f{d\varkappa}{\cn^{2n+1}\varkappa},
\label{FloquetExponentDefDyonicsimplified}
\ee
where $\tilde{c}_{2\ell+1,2n+1}$ are another set of coefficients, they are polynomials of $\sigma, \mho_{m\geqslant3}$ and $q$.
The integrals $\mathrm{J}_{-(2n+1)}$ are computed by recurrence relations as explained in appendix B of \cite{wh1412}.

The definite integral (\ref{FloquetExponentDefDyonicsimplified}) gives the functional relation $\mu(\sigma)$, from which the eigenvalue expansion $\sigma(\mu)$ can be derived.
The corresponding wave function is obtained from the indefinite integral $\int v(\varkappa)d\varkappa$, see appendix \ref{MoreAboutDyonicSpectrum} for the explicit expression.

\subsubsection{Duality for strong coupling spectra}\label{Dualityforstrongcouplingspectra}

We have obtained two strong coupling spectral solutions, it turns out that the two sets of spectral data are related by simple transformations.
In the terminology of gauge theory, these transformations are the monopole-dyon duality of $N_f$=4 super QCD.

\vspace{3mm}
\noindent {\bf Duality transformation for eigenvalues}.
The eigenvalue $\sigma$ is related to the eigenvalue $\delta$ by a transformation given by (\ref{EigenvalueDyonicFromMagnetic}). To explain the transformation, now we treat parameters $\mu,q$ and $\Omega_m$ in (\ref{EigenvalueStrongCouplingMagnetic}) as independent variables. That is to say, as the duality transformations discussed here are limited to spectral data,
the connection with the $N_f$=4 super QCD is ignored, the coupling constants $\Omega_m$ are regarded as free parameters.
The explicit form of $\Omega_m$ given in appendix \ref{EffectiveCouplingOmegaMassRelation} is a particular case for $\Omega_m$, they are $q$-series determined by the connection with the $N_f$=4 super QCD.
Then expansion of the eigenvalue $\sigma(\mu,\mho_{2m}, \mho_{2m+1},q)$ is given by expansion of the eigenvalue $\delta(\mu,\Omega_{2m}, \Omega_{2m+1},q)$ with the arguments transformed as
\be
\sigma=(1-q)\delta\left(\f{i\mu}{(1-q)^{\f{1}{2}}},\f{\mho_{2m}}{1-q},\f{\mho_{2m+1}}{(1-q)^{\f{1}{2}}},\f{-q}{1-q}\right),
\label{EigenvalueDyonicFromMagnetic}
\ee
with $m\geqslant1$. Notice the disparity of the transformation rules for the coupling constants $\Omega_{2m}$ and $\Omega_{2m+1}$.
This kind of relation between spectra data at the magnetic and the dyonic expansion regions is also verified for other examples of the relation between periodic spectral problem and super gauge theory  \cite{wh1108,wh1412}.
But in those simpler cases the transformation does not involve transformation for the coupling constants $\Omega_m$.

\vspace{3mm}
\noindent {\bf Duality transformation for integrands of eigenfunctions}. There should be a duality transformation to relate the magnetic and the dyonic wave functions. The wave functions at the magnetic point are given by (\ref{EigenfunctionStrongCouplingMagnetic}), the wave functions at the dyonic point are given by (\ref{EigenfunctionStrongCouplingDyonic}) in appendix \ref{MoreAboutDyonicSpectrum}. In this subsection we use notations $\psi_{\pm}^D(\chi)$ and $\psi_{\pm}^T(\varkappa)$ to distinguish the wave functions, use notations $v^D(\chi)$ and $v^T(\varkappa)$ to distinguish the corresponding integrands,
at the magnetic and the dyonic points respectively. Inspecting the wave functions (\ref{EigenfunctionStrongCouplingMagnetic}) and (\ref{EigenfunctionStrongCouplingDyonic}),
we find no simple way to transform $\psi_{\pm}^D(\chi)$ to $\psi_{\pm}^T(\varkappa)$. Various combinations of $\sn\chi,\cn\chi$ and $\dn\chi$ that appear in the wave functions $\psi_{\pm}^D(\chi)$ defeat a simple transformation.

Things would appear clearer when we examine more primitive objects, in fact there is a simple duality transformation for the integrands $v^D(\chi)$ and $v^T(\varkappa)$.
More precisely, it is the coefficient functions in the strong coupling expansion of the integrands that possess the duality transformation.
For briefness we call these coefficient functions, $v_\ell^D(\chi)$ in (\ref{Vlargeomega2expansion}) and $v_\ell^T(\varkappa)$ in (\ref{Vlargemho2expansion}), integrands of wave function.
The integrands of magnetic wave function contain only terms made of $\sn\chi$, and the integrands of dyonic wave function contain only terms made of $\cn\varkappa$.
With the arguments explicitly displayed, $v_\ell^T(\cn\varkappa,\sigma,\mho_{m\geqslant3},q)$ can be transformed from $v_\ell^D(\sn\chi,\delta,\Omega_{m\geqslant3},q)$ by
\be
v_\ell^T=(1-q)^{\lceil\f{\ell}{2}\rceil}v_\ell^D\left(\cn\varkappa,\f{\sigma}{1-q},\f{(-1)^m\mho_m}{1-q},\f{-q}{1-q}\right),
\label{EigenfunctionDyonicFromMagnetic}
\ee
where $\lceil * \rceil$ is the ceiling function.
The integrands $v_\ell^D$ contain only $\Omega_m$ with $m\geqslant3$, the dominant one $\Omega_2$ is the expansion parameter for the total integrand $v^D$; the situation is similar for $v_\ell^T$.
The transformation (\ref{EigenfunctionDyonicFromMagnetic}) combined with $\Omega_2\to\mho_2$ transforms the total integrands as $v^D\to v^T$. In particular, the duality transformation for elliptic function is a simple substitution $\sn\chi\to\cn\varkappa$, the simplicity becomes obscured after integration.

\vspace{3mm}
\noindent {\bf Inverse duality transformations}. The duality discussed above are transformations that map the magnetic solution to the dyonic solution.
Because the magnetic and dyonic points are really symmetric, one can derive the strong coupling solutions in a different order, start from the potential and its spectrum at the dyonic point,
then use the inverse transformations to obtain the potential and spectrum at the magnetic point.

We first derive the inverse map of the relation (\ref{mhofromOmega}) between the effective coupling constants.
We start from the effective potential at the critical point $\eta_{*6}$ which is written in the $\varkappa$-coordinate as given by eq. (\ref{EllipticHillOperatorCN}).
Following the argument similar to that in subsection \ref{EffectivePotentialEta6}, we can rewrite the effective potential at the critical point $\eta_{*5}$ (or $\varkappa_{*5}$) in another local coordinate $\varkappa^{\,\prime}$ which is fixed by $\eta=\eta_{*5}+\varkappa^{\,\prime}-\mathbf{K}$. Then it takes the same form as (\ref{EllipticHillOperatorCN}),
\be
u(\varkappa^{\,\prime})=\sum\limits_{m=1}^{\infty}\Big(\mho_{2m}^{\,\prime}\cn^{2m}\varkappa^{\,\prime}+\mho_{2m+1}^{\,\prime}\cn^{2m+1}\varkappa^{\,\prime}\sn\varkappa^{\,\prime}\dn\varkappa^{\,\prime} \Big),\label{EllipticHillOperatorCN2}
\ee
with $\mho_m^{\,\prime}\equiv\mho_m^{\,\prime}(q^{\f{1}{2}})=\mho_m(-q^{\f{1}{2}})$. Around the point $\eta_{*5}$ we have $\varkappa^{\,\prime}\sim\mathbf{K}$, so $\cn\varkappa^{\,\prime}\sim 0$ is satisfied,
the potential is an asymptotic series in $\cn\varkappa^{\,\prime}$.

To rewrite the potential (\ref{EllipticHillOperatorCN2}) in the $\chi$-coordinate, notice that by definition $\eta=\eta_{*5}+\chi$, so that $\varkappa^{\,\prime}=\chi+\mathbf{K}$.
Apply the elliptic function relations given in (\ref{ellipticfunctionschishiftK}) with the upper sign,
the potential (\ref{EllipticHillOperatorCN2}) becomes an expression with negative integer powers of $\dn\chi$.
Then the following substitution rules are used to convert it to an expression in $\sn\chi$,
\begin{align}
\dn^{-2m}\chi&=(1-q\sn^2\chi)^{-m},\notag\\
\dn^{-(2m-1)}\chi&=(1-q\sn^2\chi)^{-m}\dn\chi.
\label{substitutionright3}
\end{align}
The region around $\eta_{*5}$ is the same region around $\chi\sim 0$ in the $\chi$-coordinate where $\sn\chi\sim 0$. The potential, when expanded in $\sn\chi$,
is an infinite series given by (\ref{EllipticHillOperatorSN}).
The effective coupling constants $\Omega_m$ are related to $\mho_m^{\,\prime}$ by the inverse transformations
\begin{align}
\Omega_{2m}&=\sum_{n=1}^{m}\f{n}{m}C^{m}_{n}q^{m-n}(1-q)^n\mho_{2n}(-q^{\f{1}{2}}),\notag\\
\Omega_{2m+1}&=-\sum_{n=1}^{m}\f{m+1}{n+1}C^{m}_{n}q^{m-n}(1-q)^{n+1}\mho_{2n+1}(-q^{\f{1}{2}}).
\label{Omegafrommho}
\end{align}
In fact, (\ref{Omegafrommho}) can be obtained from (\ref{mhofromOmega}) by substitutions $\Omega_\ell(\pm q^{\f{1}{2}})\leftrightarrow\mho_\ell(\pm q^{\f{1}{2}})$ and $q\to-\f{q}{1-q}$.

The inverse duality transformations that map the dyonic spectral data to the magnetic ones are similar to the duality transformations just presented.
The inverse transformation between the eigenvalues $\delta(\mu,\Omega_{2m},\Omega_{2m+1},q)$ and $\sigma(\mu,\mho_{2m},\mho_{2m+1},q)$ is
\be
\delta=(1-q)\sigma\left(\f{-i\mu}{(1-q)^{\f{1}{2}}},\f{\Omega_{2m}}{1-q},\f{\Omega_{2m+1}}{(1-q)^{\f{1}{2}}},\f{-q}{1-q}\right).
\label{EigenvalueMagneticFromDyonic}
\ee
The inverse transformation between the integrands $v_\ell^D(\sn\chi,\delta,\Omega_{m\geqslant3},q)$ and $v_\ell^T(\cn\varkappa,\sigma,\mho_{m\geqslant3},q)$ of wave functions is
\be
v_\ell^D=(1-q)^{\lceil\f{\ell}{2}\rceil}v_\ell^T\left(\sn\chi,\f{\delta}{1-q},\f{(-1)^m\Omega_m}{1-q},\f{-q}{1-q}\right).
\label{EigenfunctionMagneticFromDyonic}
\ee
The transformation (\ref{EigenfunctionMagneticFromDyonic}), combined with $\mho_2\to\Omega_2$, transforms the total integrand as $v^T\to v^D$.

\section{Strong coupling expansion of gauge theory prepotential}\label{StrongCouplingExpansionPrepotentialNf4}

For gauge theory in the strong coupling region of moduli space, the proper degrees of freedom are the massless solitons with magnetic or dyonic charges,
therefore the instanton partition function computed by localization method is inapplicable.
The work of Seiberg and Witten provides the machinery of analytical continuation to explore the strong coupling physics.
It relies on analytical property of N=2 supersymmetric field theory prepotential that follows from non-renormalization and duality.
Riemann surface provides the concrete tool to realise the physical reasoning, for SU(2) gauge theories the surfaces are elliptic curve.

In the presence of Omega background deformation, the analytical continuation to the strong coupling region is not completely clear, but evidence indicates it is physically reasonable.
It is demonstrated in \cite{NekrasovOkounkov0306} that the Seiberg-Witten curve can be recovered from instanton computation in the classical limit $\epsilon_1\to 0, \epsilon_2\to 0$,
that means the instanton partition function at least knows the existence of strong coupling region.
It is not yet clear how to extend the prepotential to strong coupling region for general gauge theories with full Omega background deformation,
but for the special case of SU(2) gauge theory with partial deformation $\epsilon_1\to\epsilon, \epsilon_2=0$ the extension is possible.
In the Gauge/Bethe correspondence $\epsilon$ is interpreted as the quantization parameter,
to extend the $\epsilon$-deformed gauge theory quantity to the strong coupling region one needs a quantum version of the Seiberg-Witten curve.
The instrument for this task is the Schr\"{o}dinger equation, the quantum counterpart of the elliptic curve for the SU(2) $N_f$=4 super QCD, as explained in subsection \ref{TheDifferentialEquation}.

For the undeformed gauge theory, the dual variables $a$ and $a_D$ (or $a$ and $a_T$) are holomorphic section over the moduli space,
they are computed by the integrals of the Seiberg-Witten differential over dual cycles of the elliptic curve.
As shown in section \ref{WeakCouplingSpectrumAndN=2superQCD} and section \ref{StrongCouplingSpectraAndGaugeTheoryDuality},
the $\epsilon$-deformed version of $a$ and $a_D$ are identified with the Floquet exponents $\nu$ and $\mu$ respectively which are computed by the integrals of spectral data over intervals of a period of the elliptic potential.
As we are dealing with the strong coupling region, the proper form of elliptic potential is the Hill potential (\ref{EllipticHillOperatorSN}) and (\ref{EllipticHillOperatorCN}).
The quantum correction introduced by the $\epsilon$-deformation are contained in the integrand $v(\chi)$ and $v(\varkappa)$ for the Schr\"{o}dinger equation.
In the following we explain how to derive the correct series solutions of $v(\chi)$ and $v(\varkappa)$, then evaluate integrals consistently to produce the strong coupling expansion of gauge theory prepotential.

For brevity, explicit computation in this section is restricted to the leading order of the $\epsilon$-expansion, i.e. for super QCD without the $\epsilon$-deformation.
The computation can be extended to the deformed super QCD by including higher orders of the $\epsilon$-expansion as explained in appendix \ref{Details4DualPrepotentials}.

\subsection{Magnetic expansion of prepotential for $N_f$=4 super QCD}\label{MagneticPrepotential4SuperQCD}

The magnetic dual prepotential $\mathcal{F}_D(a_D)$ is computed by the relation between the scalar v.e.v $a$ and its magnetic dual $a_D$,
\be
a=\f{\p}{\p a_D}\mathcal{F}_D(a_D),
\label{aaDrelation}
\ee
where $a$ and $a_D$ are given by the integrals of the Seiberg-Witten differential over the conjugate cycles $\alpha$ and $\beta$ on the curve of $N_f$=4 super QCD \cite{SW9407,SW9408}.
By the connection with spectral problem, $a$ and $a_D$ are given by the integrals of $v(\chi)$ associated with the potential (\ref{EllipticHillOperatorSN}) over intervals of periods $2\mathbf{K}$ and $2i\mathbf{K}^{\prime}$ respectively.
As the undeformed super QCD is the classical limit $\epsilon\to0$ of the deformed theory, the leading order of the $\epsilon$-expansion for $v(\chi)$ is needed to compute $a$ and $a_D$, they are given by
\be
a_{(0)}=\f{\epsilon}{2\pi i}\int_{\chi_0}^{\chi_0+2\mathbf{K}}\sqrt{\delta+u(\chi)}d\chi \label{adefmagnetic}
\ee
and
\be
a_{D(0)}=\f{\epsilon}{2\pi i}\int_{\chi_0}^{\chi_0+2i\mathbf{K}^{\prime}}\sqrt{\delta+u(\chi)}d\chi.\label{aDdefmagnetic}
\ee
In the square root the eigenvalue $\delta$ and the coupling constants $\Omega_m$ take the leading order part of their $\epsilon$-expansion, the subscriptions are dropped for simplicity.

The dual integral $a_D$ is related to the Floquet exponent $\mu$ of the first strong coupling spectral solution by the relation
\be
\mu=\f{2ia_D}{\epsilon}.
\label{muADrelation}
\ee
So the relation (\ref{aDdefmagnetic}) is the classical limit of the monodromy relation (\ref{FloquetExponentDefMagnetic}).
The leading order of the $\epsilon$-expansion for the eigenvalue $\delta$ given by (\ref{EigenvalueStrongCouplingMagnetic}) is
\be
\delta_{(0)}=-2i\Omega_2^{\f{1}{2}}\mu-\bigg(\f{1}{2}(1+q)-\f{3\Omega_4}{2\Omega_2}+\f{15\Omega_3^2}{2^3\Omega_2^2}\bigg)\mu^2+\mathcal{O}(\mu^3).
\label{EigenvalueStrongCouplingMagneticepsilon0}
\ee
Substituting the coefficients $\Omega_m$ expressed in flavor masses of super QCD given in appendix {\ref{EffectiveCouplingOmegaMassRelation}}, we get
\begin{align}
\delta_{(0)}=&-4\Bigg(-2(\mu_1\mu_2\mu_3\mu_4)^{1/4}q^{1/4}+\f{\sum_{i<j<k}\mu_i^2\mu_j^2\mu_k^2}{2(\mu_1\mu_2\mu_3\mu_4)^{5/4}}q^{3/4}\notag\\
&+\f{9\sum_{i<j<k}\mu_i^4\mu_j^4\mu_k^4\!-\!10(\mu_1\mu_2\mu_3\mu_4)^2\sum_{i<j}\mu_i^2\mu_j^2\!+\!24(\mu_1\mu_2\mu_3\mu_4)^3}{32(\mu_1\mu_2\mu_3\mu_4)^{11/4}}q^{5/4}+\mathcal{O}(q^{7/4})\Bigg)\mu\epsilon^{-1}\notag\\
&-4\Bigg(\f{1}{8}-\f{3\sum_{i<j<k}\mu_i^2\mu_j^2\mu_k^2}{16(\mu_1\mu_2\mu_3\mu_4)^{3/2}}q^{1/2}\notag\\
&-\f{39\sum_{i<j<k}\mu_i^4\mu_j^4\mu_k^4-30(\mu_1\mu_2\mu_3\mu_4)^2\sum_{i<j}\mu_i^2\mu_j^2+8(\mu_1\mu_2\mu_3\mu_4)^3}{128(\mu_1\mu_2\mu_3\mu_4)^3}q+\mathcal{O}(q^{3/2})\Bigg)\mu^2\notag\\
&+\mathcal{O}(\mu^3\epsilon).
\label{EigenvalueStrongCouplingMagneticepsilon0physicalmass}
\end{align}
This expression reproduces the result computed from the normal form of Heun equation presented in \cite{wh1306}, where the eigenvalue is characterized by another quantity $\Delta$.
The variable $a_D$ in \cite{wh1306} is computed by integral over the $\beta$-cycle on the elliptic curve (\ref{SpectralCurveHeun2}), it is twice of the variable $a_D$ used in this paper.
Taking into account the factor two discrepancy for $a_D$, we find $\delta$ is related to $\Delta$ by
\be
\delta_{(0)}=-4(1-q)\f{\Delta}{\epsilon^2}.
\label{deltaDelta}
\ee

Now we need to evaluate the integration in (\ref{adefmagnetic}) to get the functional relation $a(\delta)$, substitute the eigenvalue $\delta(\mu)$ into $a(\delta)$ to get the functional relation $a(a_D)$,
then integrate the relation (\ref{aaDrelation}) to obtain the magnetic dual expansion of prepotential.
Notice that in this section the integral in (\ref{adefmagnetic}) is computed in the strong coupling region, so one should not bring here the result of weak coupling solution obtained in section \ref{WeakCouplingSpectrumAndN=2superQCD}. Neither should one use $\Omega_2^{\f{1}{2}}$ as the large expansion parameter, because the series obtained in this way are the classical limit of $v_{\ell}(\chi)$ given in (\ref{Vlargeomega2expansioncoeff}) used for the integral over an interval of $2i\mathbf{K}^{\prime}$, they cannot be used for the integral over an interval of $2\mathbf{K}$.

In the effective elliptic potential, terms with coefficients $\Omega_{m\geqslant3}$ are treated as perturbation to the dominant term $\Omega_2\sn^2\chi$.
This consideration leads to the correct form of expansion for the integrand of (\ref{adefmagnetic}),
\be
\sqrt{\delta+u(\chi)}=\sum_{\ell=-1}^{\infty}\f{\mathsf{P}^{\alpha}_{(0);\ell}(\chi)}{(\delta+\Omega_2\sn^2\chi)^{\f{\ell}{2}}}.\label{p0alphaExpansionMagnetic}
\ee
In the numerator of the right side, power functions of $\cn\chi$ and $\dn\chi$ are simplified by the relations (\ref{substitutionright1}).
The coefficient series with even indices actually vanish, $\mathsf{P}^{\alpha}_{(0);2\ell}(\varkappa)=0$.
Here we keep them in the expansion because when we generalise the computation to include the $\epsilon$-deformation, these coefficients series are non-vanishing at higher orders of the $\epsilon$-expansion, see appendix \ref{Details4DualPrepotentials}. The coefficient series with odd indices $\mathsf{P}^{\alpha}_{(0);2\ell+1}(\chi)$ are expanded in $\sn\chi$, the first few are
\begin{align}
\mathsf{P}^\alpha_{(0);-1}(\chi)=&1,\notag\\
\mathsf{P}^\alpha_{(0);1}(\chi)=&\f{1}{2}\Big\{\Omega_3\sn^3\chi\cn\chi\dn\chi+\Omega_4\sn^4\chi+\Omega_5\sn^5\chi\cn\chi\dn\chi+\mathcal{O}(\sn^6\chi)\Big\},\notag\\
\mathsf{P}^\alpha_{(0);3}(\chi)=&-\f{1}{2^3}\Big\{\Omega_3^2\sn^6\chi+2\Omega_3\Omega_4\sn^7\chi\cn\chi\dn\chi-[(1+q)\Omega_3^2-\Omega_4^2-2\Omega_3\Omega_5]\sn^8\chi\notag\\
&+\mathcal{O}(\sn^9\chi\cn\chi\dn\chi)\Big\},\notag\\
\mathsf{P}^\alpha_{(0);5}(\chi)=&\f{1}{2^4}\Big\{\Omega_3^3\sn^9\chi\cn\chi\dn\chi\!+\!3\Omega_3^2\Omega_4\sn^{10}\chi
\!-\![(1\!+\!q)\Omega_3^3\!-\!3\Omega_3\Omega_4^2\!-\!3\Omega_3^2\Omega_5]\sn^{11}\chi\cn\chi\dn\chi\notag\\
&+\mathcal{O}(\sn^{12}\chi)\Big\},\notag\\
\mathsf{P}^\alpha_{(0);7}(\chi)=&-\f{5}{2^7}\Big\{\Omega_3^{4}\sn^{12}\chi+4\Omega_3^3\Omega_4\sn^{13}\chi\cn\chi\dn\chi+\mathcal{O}(\sn^{14}\chi)\Big\}.
\label{p0alphaExpansionMagneticCoeff}
\end{align}

In the integrands given by (\ref{p0alphaExpansionMagneticCoeff}) the elliptic functions that appear in the numerator $\mathsf{P}^{\alpha}_{(0);2\ell+1}(\chi)$ are either $\sn^{2m+1}\chi\cn\chi\dn\chi$ or $\sn^{2m}\chi$, they lead to two classes of integrals over an interval of $2\mathbf{K}$.
The first class integrals actually vanish,
\be
\mathrm{S}_{2\ell+1,2m+1}=\int_{\chi_0}^{\chi_0+2\mathbf{K}}\f{\sn^{2m+1}\chi\cn\chi\dn\chi}{(\delta+\Omega_2\sn^2\chi)^{\f{2\ell+1}{2}}}d\chi=0,
\label{IntegralS2m+1}
\ee
with $ 2m+1\geqslant 3(\ell+1)\geqslant 0$.
The second class integrals give non-vanishing results,
\be
\mathrm{S}_{2\ell+1,2m}=\int_{\chi_0}^{\chi_0+2\mathbf{K}}\f{\sn^{2m}\chi}{(\delta+\Omega_2\sn^2\chi)^{\f{2\ell+1}{2}}}d\chi,
\label{IntegralS2m}
\ee
with $2m\geqslant 3(\ell+1)\geqslant 0$.
An easier way to see the vanishing or non-vanishing of these integrals is to rewrite the integrals over an interval of a period as integrals along a closed contour in the $\xi$-coordinate defined by $\xi=\sn^2\chi$.
The details are explained in appendix \ref{ContourIntegral}.

Therefore, for computational task terms with $\sn^{2m+1}\chi\cn\chi\dn\chi$ in the series $\mathsf{P}^{\alpha}_{(0);2\ell+1}(\chi)$ can be dropped.
The resulting series, denoted by $\widetilde{\mathsf{P}}_{(0);2\ell+1}^\alpha(\chi)$, contain only terms with $\sn^{2m}\chi$.
When elliptic functions are written in the amplitude $\varphi=\mbox{am}\chi$, the non-vanishing integrals for (\ref{p0alphaExpansionMagnetic}) become
\begin{align}
\int_{\chi_0}^{\chi_0+2\mathbf{K}}\f{\widetilde{\mathsf{P}}_{(0);2\ell+1}^\alpha(\chi)}{(\delta+\Omega_2\sn^2\chi)^{\f{2\ell+1}{2}}}d\chi=&2\int_0^{\f{\pi}{2}}\f{\widetilde{\mathsf{P}}_{(0);2\ell+1}^\alpha(\varphi)}{(\delta+\Omega_2\sin^2\varphi)^{\f{2\ell+1}{2}}}\f{d\varphi}{\sqrt{1-q\sin^2\varphi}}\notag\\
=&2\int_0^{\f{\pi}{2}}\f{\widetilde{\mathsf{P}}_{(0);2\ell+1}^\alpha(\varphi)(1+\frac{1}{2}q\sin^2\varphi+\frac{3}{8}q^2\sin^4\varphi+\cdots)}{\delta^{\f{2\ell+1}{2}}(1-\mathbf{x}^2\sin^2\varphi)^{\f{2\ell+1}{2}}}d\varphi,
\label{p0alphaExpansionMagneticSimplified}
\end{align}
where $\mathbf{x}^2=-\Omega_2/\delta$ with magnitude $\vert\mathbf{x}^2\vert\gg1$.
The numerator in (\ref{p0alphaExpansionMagneticSimplified}) can be expanded as a series in $\sin^2\varphi$.
Then computation of the integral in (\ref{adefmagnetic}) boils down to computation of the integrals in (\ref{p0alphaExpansionMagneticSimplified}). The basis integrals are
\be
\mathrm{M}_{2\ell+1,2m}=\int_0^{\f{\pi}{2}}\f{\sin^{2m}\varphi}{(1-\mathbf{x}^2\sin^2\varphi)^{\f{2\ell+1}{2}}}d\varphi,
\label{IntegralM2L+12m}
\ee
where the indices satisfy $2m\geqslant 3(\ell+1)$. Their values are given by the complete elliptic integrals with elliptic modulus $\mathbf{x}^2$.
The details are explained in appendix \ref{RecurrenceRelationMellm} where the recurrence relations for $\mathrm{M}_{2\ell+1,2m}$ are given.

Following the procedure from (\ref{p0alphaExpansionMagnetic}) to (\ref{p0alphaExpansionMagneticSimplified}), we get the function $a_{(0)}(\delta)$ in the form
\be
a_{(0)}=\f{\epsilon}{\pi i}\sum_{\ell=0}^{\infty}\left[\f{\mbox{CE}_{2\ell+1}(\Omega_{m},q)}{\mathbf{x}^{2\ell+1}}\mathbf{E}(\mathbf{x}^2)+\f{\mbox{CK}_{2\ell+1}(\Omega_{m},q)}{\mathbf{x}^{2\ell+1}}\mathbf{K}(\mathbf{x}^2)\right],
\label{a0ExpansionMagnetic}
\ee
which can be expanded in the variable $\mathbf{x}$ using the expansion of $\mathbf{E}(\mathbf{x}^2)$ and $\mathbf{K}(\mathbf{x}^2)$ for large $\mathbf{x}$.
The coefficients $\mbox{CE}_{2\ell+1}(\Omega_{m\geqslant2},q)$ and $\mbox{CK}_{2\ell+1}(\Omega_{m\geqslant2},q)$ can be expanded as series for large $\Omega_2$,
the first few of them are given in appendix \ref{Details4DualPrepotentials}.
Substituting the relation $\delta(a_{D(0)})$ given by (\ref{EigenvalueStrongCouplingMagneticepsilon0}) into (\ref{a0ExpansionMagnetic}),
we get the relation $a_{(0)}(a_{D(0)})$. The magnetic dual expansion of prepotential is determined from the relation (\ref{aaDrelation}) by integration with respect to $a_{D(0)}$, in terms of the coupling constants $\Omega_m$ it is
\begin{align}
\pi i\mathcal{F}_D^{(0)}=&\Bigg(\mathbf{C}^1_{2^{-1}}\Omega_2^{\f{1}{2}}+\Big[\mathbf{C}^1_{2^14^1}\Omega_4+\mathbf{C}^1_{2^16^1}\Omega_6+\mathcal{O}(\Omega_8)\Big]\Omega_2^{-\f{1}{2}}\notag\\
&+\Big[\mathbf{C}^1_{2^33^2}\Omega_3^2+\mathbf{C}^1_{2^34^2}\Omega_4^2+\mathbf{C}^1_{2^33^15^1}\Omega_3\Omega_5+\mathcal{O}(\Omega_3\Omega_7)\Big]\Omega_2^{-\f{3}{2}}+\mathcal{O}(\Omega_2^{-\f{5}{2}})\Bigg)\hat{a}_D\epsilon\notag\\
&+\Bigg(\mathbf{C}^2_{2^0}+\f{1}{2}\ln\f{\hat{a}_D}{4i\Omega_2^{\f{1}{2}}}+\Big[\mathbf{C}^2_{2^24^1}\Omega_4+\mathbf{C}^2_{2^26^1}\Omega_6+\mathcal{O}(\Omega_8)\Big]\Omega_2^{-1}\notag\\
&+\Big[\mathbf{C}^2_{2^43^2}\Omega_3^2+\mathbf{C}^2_{2^44^2}\Omega_4^2+\mathbf{C}^2_{2^43^15^1}\Omega_3\Omega_5+\mathcal{O}(\Omega_3\Omega_7)\Big]\Omega_2^{-2}+\mathcal{O}(\Omega_2^{-3})\Bigg)\hat{a}_D^2\notag\\
&+\mathcal{O}(\hat{a}_D^3\epsilon^{-1}),
\label{MagneticPrepotentialInOmega}
\end{align}
where $\hat{a}_D=ia_D$. The coefficients $\mathbf{C}^{n}_{2^{\ell}m_1^{\ell_1}m_2^{\ell_2}\cdots}(q)$ are $q$-polynomials where $n$ refer to the exponent of $\hat{a}_D$, $\ell$ refer to the exponent of $\Omega_2^{-\f{1}{2}}$ and $\ell_i$ refer to the exponent of higher order coupling constants $\Omega_{m_i\geqslant3}$. See appendix \ref{Details4DualPrepotentials} for the first few of them needed to produce (\ref{MagneticPrepotential}) given below. Notice that there is only one logarithm term in the prepotential, this fact depends on precise cancellations between many logarithm terms produced by the expansion of the complete elliptic integrals for large $\mathbf{x}$.

Substituting the expressions of $\Omega_m$ given in appendix \ref{EffectiveCouplingOmegaMassRelation}, we get the dual expansion of prepotential,
\begin{align}
\pi i\mathcal{F}_D^{(0)}=&-\Bigg(4(\mu_1\mu_2\mu_3\mu_4)^{1/4}q^{1/4}-\f{\sum_{i<j<k}\mu_i^2\mu_j^2\mu_k^2}{3(\mu_1\mu_2\mu_3\mu_4)^{5/4}}q^{3/4}\notag\\
&-\f{9\sum_{i<j<k}\mu_i^4\mu_j^4\mu_k^4\!-\!10(\mu_1\mu_2\mu_3\mu_4)^2\sum_{i<j}\mu_i^2\mu_j^2\!-\!40(\mu_1\mu_2\mu_3\mu_4)^3}{80(\mu_1\mu_2\mu_3\mu_4)^{11/4}}q^{5/4}\!+\!\mathcal{O}(q^{7/4})\Bigg)\hat{a}_D\notag\\
&+\Bigg(\f{3}{4}-\f{1}{2}\ln\Big(-\f{\hat{a}_D}{2^4(\mu_1\mu_2\mu_3\mu_4)^{1/4}q^{1/4}}\Big)-\f{3\sum_{i<j<k}\mu_i^2\mu_j^2\mu_k^2}{8(\mu_1\mu_2\mu_3\mu_4)^{3/2}}q^{1/2}\notag\\
&-\f{39\sum_{i<j<k}\mu_i^4\mu_j^4\mu_k^4-30(\mu_1\mu_2\mu_3\mu_4)^2\sum_{i<j}\mu_i^2\mu_j^2-8(\mu_1\mu_2\mu_3\mu_4)^3}{128(\mu_1\mu_2\mu_3\mu_4)^3}q+\mathcal{O}(q^{3/2})\Bigg)\hat{a}_D^2\notag\\
&+\mathcal{O}(\hat{a}_D^3).
\label{MagneticPrepotential}
\end{align}
The expansions of the eigenvalue (\ref{EigenvalueStrongCouplingMagneticepsilon0physicalmass}) and the magnetic dual prepotential (\ref{MagneticPrepotential}) are related by the relation
\be
\delta_{(0)}=-\f{16\pi i}{\epsilon^2}(1-q)q\f{\p}{\p q}\mathcal{F}_D^{(0)},\label{deltaPrepotentialMagnetic}
\ee
taking into account the relation $\hat{a}_D=\f{1}{2}\mu\epsilon$. This relation is the magnetic dual version of the relation in N=2 gauge theories that connects moduli and prepotential \cite{matone1}.

\subsection{Dyonic expansion of prepotential for $N_f$=4 super QCD}\label{DyonicPrepotential4SuperQCD}

The dyonic dual prepotential is defined by the relation
\be
a=\f{\p}{\p a_T}\mathcal{F}_T(a_T),
\label{aaTrelation}
\ee
where $a_T$ is the dyonic dual of scalar v.e.v, it is given by the integral of the Seiberg-Witten differential over the $\gamma$-cycle on the curve of $N_f$=4 super QCD.
By the connection with spectral problem, $a$ and $a_T$ are given by the integrals of $v(\varkappa)$ associated with the potential (\ref{EllipticHillOperatorCN}) over intervals of periods $2\mathbf{K}$ and $2\mathbf{K}+2i\mathbf{K}^{\prime}$ respectively.
At the leading order of the $\epsilon$-expansion they are given by
\be
a_{(0)}=\f{\epsilon}{2\pi i}\int_{\varkappa_0}^{\varkappa_0+2\mathbf{K}}\sqrt{\sigma+u(\varkappa)}d\varkappa \label{adefdyonic}
\ee
and
\be
a_{T(0)}=\f{\epsilon}{2\pi i}\int_{\varkappa_0}^{\varkappa_0+2\mathbf{K}+2i\mathbf{K}^{\prime}}\sqrt{\sigma+u(\varkappa)}d\varkappa. \label{aTdefdyonic}
\ee
In the square root the eigenvalue $\sigma$ and the coupling constants $\mho_m$ take the leading order part of their $\epsilon$-expansion.

The dual integral $a_{T}$ is related to the Floquet exponent of the second strong coupling spectral solution by
\be
\mu=-\f{2(1-q)^{\f{1}{2}}a_T}{\epsilon}.
\label{muATrelation}
\ee
So the integral in (\ref{aTdefdyonic}) is the classical limit of the monodromy relation (\ref{FloquetExponentDefDyonic}).

The classical limit of the eigenvalue are obtained by applying the duality transformation (\ref{EigenvalueDyonicFromMagnetic}) to the classical eigenvalue $\delta_{(0)}$ given in eq. (\ref{EigenvalueStrongCouplingMagneticepsilon0}), up to order $\mathcal{O}(\mu^2)$ it is
\be
\sigma_{(0)}=2\mho_2^{\f{1}{2}}\mu+\bigg(\f{1-2q}{2(1-q)}-\f{3\mho_4}{2\mho_2}+\f{15(1-q)\mho_3^2}{2^3\mho_2^2}\bigg)\mu^2+\mathcal{O}(\mu^3).
\label{EigenvalueStrongCouplingDyonicepsilon0}
\ee
The effective coupling constants $\mho_m$ are computed from $\Omega_m$ using the relation (\ref{mhofromOmega}), they can be expressed in flavor masses of super QCD.
Then (\ref{EigenvalueStrongCouplingDyonicepsilon0}) becomes
\begin{align}
\sigma_{(0)}=&-4\Bigg(-2(\mu_1\mu_2\mu_3\mu_4)^{1/4}q^{1/4}-\f{\sum_{i<j<k}\mu_i^2\mu_j^2\mu_k^2}{2(\mu_1\mu_2\mu_3\mu_4)^{5/4}}q^{3/4}\notag\\
&+\f{9\sum_{i<j<k}\mu_i^4\mu_j^4\mu_k^4\!-\!10(\mu_1\mu_2\mu_3\mu_4)^2\sum_{i<j}\mu_i^2\mu_j^2\!-\!8(\mu_1\mu_2\mu_3\mu_4)^3}{32(\mu_1\mu_2\mu_3\mu_4)^{11/4}}q^{5/4}+\mathcal{O}(q^{7/4})\Bigg)\mu\epsilon^{-1}\notag\\
&-4\Bigg(-\f{1}{8}-\f{3\sum_{i<j<k}\mu_i^2\mu_j^2\mu_k^2}{16(\mu_1\mu_2\mu_3\mu_4)^{3/2}}q^{1/2}\notag\\
&+\f{39\sum_{i<j<k}\mu_i^4\mu_j^4\mu_k^4-30(\mu_1\mu_2\mu_3\mu_4)^2\sum_{i<j}\mu_i^2\mu_j^2-8(\mu_1\mu_2\mu_3\mu_4)^3}{128(\mu_1\mu_2\mu_3\mu_4)^3}q+\mathcal{O}(q^{3/2})\Bigg)\mu^2\notag\\
&+\mathcal{O}(\mu^3\epsilon).
\label{EigenvalueStrongCouplingDyonicepsilon0physicalmass}
\end{align}

The integral in (\ref{adefdyonic}) is computed in a similar way as in the magnetic dual case.
In the dyonic expansion region $\mho_2^{\f{1}{2}}\gg\sigma$, but simply expanding the integrand $\sqrt{\sigma+u(\varkappa)}$ for large $\mho_2$ leads to an expression incompatible with the integral over an interval of $2\mathbf{K}$. The correct form of expansion is
\be
\sqrt{\sigma+u(\varkappa)}=\sum_{\ell=-1}^{\infty}\f{\mathsf{Q}^{\alpha}_{(0);\ell}(\varkappa)}{(\sigma+\mho_2\cn^2\varkappa)^{\f{\ell}{2}}}.
\label{p0alphaExpansionDyonic}
\ee
The coefficient series with even indices vanish, $\mathsf{Q}^{\alpha}_{(0);2\ell}(\varkappa)=0$.
The coefficient series with odd indices $\mathsf{Q}^{\alpha}_{(0);2\ell+1}(\varkappa)$ are expanded in $\cn\varkappa$, the first few are
\begin{align}
\mathsf{Q}^\alpha_{(0);-1}(\varkappa)=&1,\notag\\
\mathsf{Q}^\alpha_{(0);1}(\varkappa)=&\f{1}{2}\Big\{\mho_3\cn^3\varkappa\sn\varkappa\dn\varkappa+\mho_4\cn^4\varkappa+\mho_5\cn^5\varkappa\sn\varkappa\dn\varkappa+\mathcal{O}(\cn^6\varkappa)\Big\},\notag\\
\mathsf{Q}^\alpha_{(0);3}(\varkappa)=&-\f{1}{2^3}\Big\{(1-q)\mho_3^2\cn^6\varkappa+2\mho_3\mho_4\cn^7\varkappa\sn\varkappa\dn\varkappa\notag\\
&\quad-[(1-2q)\mho_3^2-\mho_4^2-2(1-q)\mho_3\mho_5]\cn^8\varkappa+\mathcal{O}(\cn^9\varkappa\sn\varkappa\dn\varkappa)\Big\},\notag\\
\mathsf{Q}^\alpha_{(0);5}(\varkappa)=&\f{1}{2^4}\Big\{(1-q)\mho_3^3\cn^9\varkappa\sn\varkappa\dn\varkappa+3(1-q)\mho_3^2\mho_4\cn^{10}\varkappa\notag\\
&\quad-[(1-2q)\mho_3^3-3\mho_3\mho_4^2-3(1-q)\mho_3^2\mho_5]\cn^{11}\sn\varkappa\dn\varkappa+\mathcal{O}(\cn^{12}\varkappa)\Big\},\notag\\
\mathsf{Q}^\alpha_{(0);7}(\varkappa)=&-\f{5}{2^7}\Big\{(1-q)^2\mho_3^4\cn^{12}\varkappa+4(1-q)\mho_3^3\mho_4\cn^{13}\varkappa\sn\varkappa\dn\varkappa+\mathcal{O}(\cn^{14}\varkappa)\Big\}.
\label{q0alphaExpansionDyonicCoeff}
\end{align}

Because the elliptic functions in $\mathsf{Q}^{\alpha}_{(0);2\ell+1}(\varkappa)$ are $\cn^{2m+1}\varkappa\sn\varkappa\dn\varkappa$ and $\cn^{2m}\varkappa$,
there are also two classes of integrals over an interval of $2\mathbf{K}$ that follow from (\ref{p0alphaExpansionDyonic}) and (\ref{q0alphaExpansionDyonicCoeff}). The first class integrals vanish,
\be
\mathrm{T}_{2\ell+1,2m+1}=\int_{\varkappa_0}^{\varkappa_0+2\mathbf{K}}\f{\cn^{2m+1}\varkappa\sn\varkappa\dn\varkappa}{(\sigma+\mho_2\cn^2\varkappa)^{\f{2\ell+1}{2}}}d\varkappa=0,
\label{IntegralT2m+1}
\ee
with $2m+1\geqslant 3(\ell+1)\geqslant 0$.
The second class  integrals are nonzero,
\be
\mathrm{T}_{2\ell+1,2m}=\int_{\varkappa_0}^{\varkappa_0+2\mathbf{K}}\f{\cn^{2m}\varkappa}{(\sigma+\mho_2\cn^2\varkappa)^{\f{2\ell+1}{2}}}d\varkappa,
\label{IntegralT2m}
\ee
with $2m\geqslant 3(\ell+1)\geqslant 0$.
This fact is also explained in appendix \ref{ContourIntegral} using contour integrals in the $\xi$-coordinate defined by $\xi=\sn^2\varkappa$.
So terms with $\cn^{2m+1}\varkappa\sn\varkappa\dn\varkappa$ in the expansion of $\mathsf{Q}^{\alpha}_{(0);2\ell+1}(\varkappa)$ can be dropped,
the resulting series are denoted by $\widetilde{\mathsf{Q}}^{\alpha}_{(0);2\ell+1}(\varkappa)$ which contain only terms with $\cn^{2m}\varkappa$.
To evaluate the remaining integrals, rewrite elliptic functions in the amplitude $\varphi=\mbox{am}\varkappa$, the non-vanishing integrals for (\ref{p0alphaExpansionDyonic}) become
\begin{align}
\int_{\varkappa_0}^{\varkappa_0+2\mathbf{K}}\f{\widetilde{\mathsf{Q}}_{(0);2\ell+1}^\alpha(\varkappa)}{(\sigma+\mho_2\cn^2\varkappa)^{\f{2\ell+1}{2}}}d\varkappa=&2\int_0^{\f{\pi}{2}}\f{\widetilde{\mathsf{Q}}_{(0);2\ell+1}^\alpha(\varphi)}{(\sigma+\mho_2-\mho_2\sin^2\varphi)^{\f{2\ell+1}{2}}}\f{d\varphi}{\sqrt{1-q\sin^2\varphi}}\notag\\
=&2\int_0^{\f{\pi}{2}}\f{\widetilde{\mathsf{Q}}_{(0);2\ell+1}^\alpha(\varphi)(1+\frac{1}{2}q\sin^2\varphi+\frac{3}{8}q^2\sin^4\varphi+\cdots)}{(\sigma+\mho_2)^{\f{2\ell+1}{2}}(1-\mathbf{x}^2\sin^2\varphi)^{\f{2\ell+1}{2}}}d\varphi,
\label{q0alphaExpansionDyonicSimplified}
\end{align}
where in this instance $\mathbf{x}^2=\mho_2/(\mho_2+\sigma)$. As $\widetilde{\mathsf{Q}}_{(0);2\ell+1}^\alpha$ is a series in $\sin^2\varphi$ by the substitution $\cn^{2m}\varkappa=(1-\sin\varphi^2)^{m}$,
the numerator in (\ref{q0alphaExpansionDyonicSimplified}) is an infinite series in $\sin^2\varphi$.
Therefore, to compute $a_{(0)}$ we need the same basis integrals as in (\ref{IntegralM2L+12m}),
\be
\mathrm{M}_{2\ell+1,2m}=\int_0^{\f{\pi}{2}}\f{\sin^{2m}\varphi}{(1-\mathbf{x}^2\sin^2\varphi)^{\f{2\ell+1}{2}}}d\varphi,
\ee
now the indices satisfy $m\geqslant 0$, the elliptic modulus satisfies $\vert\mathbf{x}^2\vert\sim1$.

The procedure from (\ref{p0alphaExpansionDyonic}) to (\ref{q0alphaExpansionDyonicSimplified}) leads to the function $a_{(0)}(\sigma)$ expanded in variable $\mathbf{x}$ as
\be
a_{(0)}=\f{\epsilon}{\pi i}\sum_{\ell=0}^{\infty}\left[\widetilde{\mbox{CE}}_{\ell}(\mho_{m},q)(\mathbf{x}-1)^{\ell}\mathbf{E}(\mathbf{x}^2)+\widetilde{\mbox{CK}}_{\ell+1}(\mho_{m},q)(\mathbf{x}-1)^{\ell+1}\mathbf{K}(\mathbf{x}^2)\right].
\label{a0ExpansionDyonic}
\ee
The coefficients $\widetilde{\mbox{CE}}_{\ell}(\mho_{m\geqslant2},q)$ and $\widetilde{\mbox{CK}}_{\ell+1}(\mho_{m\geqslant2},q)$ can be expanded as series for large $\mho_2$,
see appendix \ref{Details4DualPrepotentials}.
Substituting the expansion of $\mathbf{E}(\mathbf{x}^2)$ and $\mathbf{K}(\mathbf{x}^2)$ for $\mathbf{x}\sim1$ and the series of $\sigma(\mu)$ given in (\ref{EigenvalueStrongCouplingDyonicepsilon0}),
we get the relation $a_{(0)}(a_{T(0)})$.
The dyonic dual expansion of prepotential is determined from the relation (\ref{aaTrelation}) by integration with respect to $a_{T(0)}$, in terms of the coupling constants $\mho_m$ it is
\begin{align}
\pi i\mathcal{F}_T^{(0)}=&\Bigg(\mathbf{D}^1_{2^{-1}}\mho_2^{\f{1}{2}}+\Big[\mathbf{D}^1_{2^14^1}\mho_4+\mathbf{D}^1_{2^16^1}\mho_6+\mathcal{O}(\mho_8)\Big]\mho_2^{-\f{1}{2}}\notag\\
&+\Big[\mathbf{D}^1_{2^33^2}\mho_3^2+\mathbf{D}^1_{2^34^2}\mho_4^2+\mathbf{D}^1_{2^33^15^1}\mho_3\mho_5+\mathcal{O}(\mho_3\mho_7)\Big]\mho_2^{-\f{3}{2}}+\mathcal{O}(\mho_2^{-\f{5}{2}})\Bigg)a_T\epsilon\notag\\
&+\Bigg(\mathbf{D}^2_{2^0}+\f{1}{2}\ln\big(-\f{a_T}{4\mho_2^{\f{1}{2}}}\big)+\Big[\mathbf{D}^2_{2^24^1}\mho_4+\mathbf{D}^2_{2^26^1}\mho_6+\mathcal{O}(\mho_8)\Big]\mho_2^{-1}\notag\\
&+\Big[\mathbf{D}^2_{2^43^2}\mho_3^2+\mathbf{D}^2_{2^44^2}\mho_4^2+\mathbf{D}^2_{2^43^15^1}\mho_3\mho_5+\mathcal{O}(\mho_3\mho_7)\Big]\mho_2^{-2}+\mathcal{O}(\mho_2^{-3})\Bigg)a_T^2\notag\\
&+\mathcal{O}(a_T^3\epsilon^{-1}).
\label{DyonicPrepotentialInMho}
\end{align}
The coefficients $\mathbf{D}^{n}_{2^{\ell}m_1^{\ell_1}m_2^{\ell_2}\cdots}(q)$ are $q$-polynomials, their indices are arranged in the same way as for $\mathbf{C}^{n}_{2^{\ell}m_1^{\ell_1}m_2^{\ell_2}\cdots}(q)$.
The first few of them are given in appendix \ref{Details4DualPrepotentials}. The complete elliptic integrals expanded for $\mathbf{x}\sim1$ produce many logarithm terms with $\ln(\mathbf{x}-1)$, they cancel out  almost completely when we compute the prepotential, leave a single logarithm term in the end.

Substituting the coupling constants $\mho_m$ expressed in super QCD masses into (\ref{DyonicPrepotentialInMho}), we get
\begin{align}
\pi i\mathcal{F}_T^{(0)}=&\Bigg(4(\mu_1\mu_2\mu_3\mu_4)^{1/4}q^{1/4}+\f{\sum_{i<j<k}\mu_i^2\mu_j^2\mu_k^2}{3(\mu_1\mu_2\mu_3\mu_4)^{5/4}}q^{3/4}\notag\\
&-\f{9\sum_{i<j<k}\mu_i^4\mu_j^4\mu_k^4\!-\!10(\mu_1\mu_2\mu_3\mu_4)^2\sum_{i<j}\mu_i^2\mu_j^2\!-\!40(\mu_1\mu_2\mu_3\mu_4)^3}{80(\mu_1\mu_2\mu_3\mu_4)^{11/4}}q^{5/4}\!+\!\mathcal{O}(q^{7/4})\Bigg)a_T\notag\\
&+\Bigg(-\f{3}{4}+\f{1}{2}\ln\Big(-\f{a_T}{2^4(\mu_1\mu_2\mu_3\mu_4)^{1/4}q^{1/4}}\Big)-\f{3\sum_{i<j<k}\mu_i^2\mu_j^2\mu_k^2}{8(\mu_1\mu_2\mu_3\mu_4)^{3/2}}q^{1/2}\notag\\
&+\f{39\sum_{i<j<k}\mu_i^4\mu_j^4\mu_k^4-30(\mu_1\mu_2\mu_3\mu_4)^2\sum_{i<j}\mu_i^2\mu_j^2-8(\mu_1\mu_2\mu_3\mu_4)^3}{128(\mu_1\mu_2\mu_3\mu_4)^3}q+\mathcal{O}(q^{3/2})\Bigg)a_T^2\notag\\
&+\mathcal{O}(a_T^3).
\label{DyonicPrepotential}
\end{align}

There is also a relation between the expansions of the eigenvalue (\ref{EigenvalueStrongCouplingDyonicepsilon0physicalmass}) and the dyonic dual prepotential (\ref{DyonicPrepotential}),
\be
\sigma_{(0)}=-\f{16\pi i}{\epsilon^2}(1-q)q\f{\p}{\p q}\mathcal{F}_T^{(0)}.
\ee
On the left side the relation (\ref{muATrelation}) is substituted for the argument of $\sigma_{(0)}(\mu)$.

Various quantities at the magnetic and dyonic points are related by monopole-dyon duality transformations.
The expansion of eigenvalue $\delta$ given in (\ref{EigenvalueStrongCouplingMagneticepsilon0physicalmass}) and the expansion of eigenvalue $\sigma$ given in (\ref{EigenvalueStrongCouplingDyonicepsilon0physicalmass}) are mapped to each other by
\be
\mu\to \f{-i\mu}{(1-q)^{\f{1}{2}}},\qquad q^{\f{1}{2}}\to -q^{\f{1}{2}}.
\label{Transformation4StrongCouplingEigenvalues}
\ee
The transformation of $q$ looks different from the transformation in (\ref{EigenvalueDyonicFromMagnetic}), but they are equivalent.
The eigenvalue $\delta$ given in eq. (\ref{EigenvalueStrongCouplingMagnetic}) has an explicit $q$-dependence, and it also has an implicit $q$-dependence through the coupling constants $\Omega_m$.
By contrast, the expression of $\delta$ given in eq. (\ref{EigenvalueStrongCouplingMagneticepsilon0physicalmass}) has only an explicit $q$-dependence.
Taking into account the difference, the combination of transformations for $q$ and $\Omega_m$ in eq. (\ref{EigenvalueDyonicFromMagnetic}) is equivalent to the transformation for $q$ in  (\ref{Transformation4StrongCouplingEigenvalues}).
The magnetic dual expansion of prepotential in (\ref{MagneticPrepotential}) is transformed to the dyonic dual expansion of prepotential in (\ref{DyonicPrepotential}) by
\be
\hat{a}_D\to ia_T,\qquad q^{\f{1}{2}}\to -q^{\f{1}{2}}.
\ee
And the inverse transformation is
\be
a_T\to i\hat{a}_D,\qquad q^{\f{1}{2}}\to -q^{\f{1}{2}}.
\ee

Various mass decoupling limits for the magnetic and dyonic expansions of prepotential can be taken, they lead to corresponding dual expansions of prepotential for super QCD with fewer hypermultiplets
and pure gauge theory. There is another class of limits, associated with the Landen transformation, which do not decouple any hypermultiplet, provide a connection between the $N_f$=4 super QCD and the N=$2^*$ super QCD.

\section{Summary and discussion}

Among the relations we studied between supersymmetric SU(2) gauge theory models (pure YM, N=2$^*$ and $N_f$=4 super QCD models) and Schr\"{o}dinger equation with periodic potentials (Mathieu, Lam\'{e} and Heun equations), the case studied in this paper is the richest one of this class. Various features of gauge theory have a corresponding interpretation in the content of spectral solutions for periodic potential. There is one aspect of this connection still missing, at the moment there are no gauge theory quantities that correspond to the strong coupling wave functions of the quantum mechanical models.
For example, are there gauge theory quantities of $N_f$=4 theory that can be identified with the wave functions given by (\ref{EigenfunctionStrongCouplingMagnetic}) and (\ref{EigenfunctionStrongCouplingDyonic})?
In the electric region, the instanton partition function with surface defect is related to the weak coupling wave function, as in (\ref{psiPrepotentialElectric}) for $N_f$=4 theory (and in formula (51) of ref. \cite{wh1608} for N=2$^*$ theory). Probably partition functions of effective gauge theory with surface defect evaluated at the monopole and dyon singularities in the moduli space are related to the strong coupling wave functions.

All the spectral solutions can be verified by plugging eigenvalue and wave function into the corresponding equation. Another way to verify is by checking a relation for second-order
differential equations. For example, for the weak coupling solution obtained in subsection \ref{SpectrumAtWeakCoupling}, the quotient function $f(x)=\psi_{\pm}(x)/\psi_{\mp}(x)$ with either the upper or the lower sign wave function satisfies the relation
\be
u(x)+\lambda=-\f{1}{2}\mathcal{S}(f(x)),
\ee
where $\mathcal{S}(f(x))$ is the Schwarzian derivative of $f(x)$.
The strong coupling spectra obtained in section \ref{StrongCouplingSpectraAndGaugeTheoryDuality} also satisfy similar relations.

The effective potentials (\ref{EllipticHillOperatorSN}) and (\ref{EllipticHillOperatorCN}) are derived by expanding the DTV potential around its magnetic and dyonic critical points.
They are very generic elliptic Hill potentials compatible with the doubly-periodic Floquet theory, generalise other widely used trigonometric and elliptic potentials.
The classical Floquet theory of periodic potential remains valid for elliptic potentials.
The weak coupling solution is a Floquet solution, but for the two strong coupling solutions this fact becomes obvious only when the wave functions are written in the canonical coordinates.
A question it raises is, whether the DTV potential, or in the guise of effective potentials (\ref{EllipticHillOperatorSN}) and (\ref{EllipticHillOperatorCN}), are the most generic elliptic potentials of this kind.

The example studied here is a non-relativistic two particles quantum mechanical problem with elliptic potential, has a direct connection with supersymmetric gauge theory.
It has a relativistic generalization, the corresponding spectral problem is a difference equation with coefficients given by elliptic function \cite{vanDiejen1994, Chalykh0702, Ruijsenaars1404}.
Various relativistic or non-relativistic version of two particles quantum mechanical system with elliptic, trigonometric/hyperbolic and rational potentials are obtained from limits of the relativistic-DTV potential.
In the context of Gauge/Bethe correspondence, relativistic generalization of integrable system corresponds to growing a compact spacetime dimension for gauge theory \cite{NS0908}.
Relativistic generalization of Toda class integrable models are related to 5-dimensional N=1 supersymmetric gauge theory compactificated on a circle \cite{Nekrasov9609},
the partition function of the Omega background deformed gauge theory is computed by K-theoretic localization formula \cite{Nekrasov0206, NekrasovOkounkov0306}.
Some recent discussions of this connection are \cite{BullimoreKimKoroteev1412, HatsudaMarino1511, Sciarappa1706, HatsudaSciarappaZakany1809}.
It would be helpful to study the relativistic-DTV potential from the perspective of gauge theory,
a version of relativistic Heun/van Diejen operator indeed shows up in supersymmetric field theory \cite{NazzalRazamat1801}.
Nevertheless, one should be aware that it might be not straightforward to make their connection precise,
variety of twists could arise as we have seen in the relation between non-relativistic modes with elliptic potentials and 4-dimensional super QCD models ($N_f$=4 and N=$2^*$) studied so far.

\appendix
\renewcommand{\appendixname}{Appendix~\Alph{section}}
\renewcommand{\theequation}{\thesection.\arabic{equation}}
\settocdepth{subsection}

\section{Useful relations of elliptic functions}\label{Usefulrelationsofellipticfunctions}
\setcounter{equation}{0}

The following formulae accompany some other relations listed in appendixes of \cite{wh1412}.

\subsection{Simplifying the Weierstrass elliptic function}\label{SimplifyingWeierstrassEllipticFunction}

The Hamiltonian densities of KdV hierarchy $v_{\ell}(x)$ contain various products of the potential $u(x)$ and its derivatives,
up to constant coefficients, they are polynomials with monomials that take the general form
\be
[u(x)]^m\prod_{d}[\p_x^du(x)]^{n_d},
\label{KdVHamiltonianGeneral}
\ee
with integers $m\geqslant 0, d\geqslant 1, n_d\geqslant 0$.
In turn for the DTV potential they contain various products of the shifted Weierstrass elliptic functions $\widetilde{\wp}(x+\omega_s)$ and their derivatives $\p_x^n\widetilde{\wp}(x+\omega_s)$.
They need to be simplified before evaluating integrals to derive the weak coupling spectrum as in subsection \ref{SpectrumAtWeakCoupling}.

A preliminary simplification comes from the general structure of $v_{2\ell}(x)$ that they can be rewritten as the derivative of a function.
Introducing two sets of polynomial functionals $\mathbf{V}_{2\ell-2}(u,\p_xu,\p_x^2u,\cdots)$ and $\mathbf{U}_{2\ell-2}(u,\p_xu,\p_x^2u,\cdots)$ with $\ell\geqslant 1$,
where the subscripts denote the highest order derivative involved, then we can write $v_{\ell}(x)$ as
\begin{align}
v_{2\ell-1}(x)&=\mathbf{V}_{2\ell-2}(u,\p_xu,\p_x^2u,\cdots,\p_x^{2\ell-2}u), \notag\\
v_{2\ell}(x)&=\p_x\mathbf{U}_{2\ell-2}(u,\p_xu,\p_x^2u,\cdots,\p_x^{2\ell-2}u).
\end{align}
The interesting property is that the expressions of $\mathbf{V}_{2\ell-2}$ and $\mathbf{U}_{2\ell-2}$ have the same structure, only differ by numerical coefficients.
The first few functionals $\mathbf{V}_{2\ell-2}$ and $\mathbf{U}_{2\ell-2}$ are
\begin{align}
\mathbf{V}_0&=\f{1}{2}u,\quad\qquad \mathbf{U}_0=-\f{1}{4}u, \notag\\
\mathbf{V}_2&=-\f{1}{8}u^2+\f{1}{8}\p_x^2u,\quad\qquad \mathbf{U}_2=\f{1}{8}u^2-\f{1}{16}\p_x^2u, \notag \\
\mathbf{V}_4&=\f{1}{16}u^3-\f{5}{32}(\p_xu)^2-\f{3}{16}u\p_x^2u+\f{1}{32}\p_x^4u, \notag \\
\mathbf{U}_4&=-\f{1}{12}u^3+\f{5}{64}(\p_xu)^2+\f{1}{8}u\p_x^2u-\f{1}{64}\p_x^4u.
\label{KdVHamiltonianVandU}
\end{align}
They contain monomials that take the general form (\ref{KdVHamiltonianGeneral}) with an even number of the $\p_x$-operators.

Plugging the DTV potential into $\mathbf{V}_{2\ell-2}$ and $\mathbf{U}_{2\ell-2}$ gives polynomials with monomials in the form
\be
\prod_{s=0}^{3}\Big\{\widetilde{\wp}^{\;m_s}(x+\omega_s)\prod_{d_s}[\p_x^{d_s}\widetilde{\wp}(x+\omega_s)]^{n_{d,s}}\Big\},
\label{KdVHamiltonianDTVGeneral}
\ee
where we have dropped the coefficients which are functions of $b_s$.
The monomial (\ref{KdVHamiltonianDTVGeneral}) is decomposed into two types of factors, the ``{\em P(ower)-type}'' factor is
\be
\prod_{s=0}^{3}\widetilde{\wp}^{\;m_s}(x+\omega_s),
\label{KdVHamiltonianPtype}
\ee
with exponents $m_s\geqslant 0$, and the ``{\em D(erivative)-type}'' factor is
\be
\prod_{s=0}^{3}\prod_{d_s}[\p_x^{d_s}\widetilde{\wp}(x+\omega_s)]^{n_{d,s}},
\label{KdVHamiltonianDtype}
\ee
with orders of derivative $d_s\geqslant 1$ and exponents $n_{d,s}\geqslant 0$. A monomial with only P-type (or D-type) factors is called a P-type (or D-type) term.
For the DTV potential $v_\ell(x)$ has a pole of order $\ell+1$ at $x+\omega_s\to 0$, therefore $m_s, d_s$ and $n_{d,s}$ cannot grow arbitrarily.
For the monomial (\ref{KdVHamiltonianDTVGeneral}) derived from $\mathbf{V}_{2\ell-2}$ or $\mathbf{U}_{2\ell-2}$ the bound is $2m_s+\sum_{d_s}(2+d_s)n_{d,s} \leqslant 2\ell$, for each $s=0,1,2,3$.
The principle of simplifying (\ref{KdVHamiltonianDTVGeneral}) is to break products into summations, and to decrease the orders $d_s$ and exponents $m_s, n_{d,s}$.
In the following two steps simplification, we need relations derived from the basic equation for the shifted Weierstrass elliptic function
\be
[\p_x\widetilde{\wp}(x)]^2=4\widetilde{\wp}^{\;3}(x)-12\zeta_1\widetilde{\wp}^{\;2}(x)+(12\zeta_1^2-g_2)\widetilde{\wp}(x)-(4\zeta_1^3-g_2\zeta_1+g_3).
\label{BasicEq4Weierstrass}
\ee

The first step is to decrease $d_s$ and $n_{d,s}$ of the D-type factor in (\ref{KdVHamiltonianDtype}).
When the discussion is limited to a sub-factor with fixed index $s$, the subscript is dropped. To simplify the factor $[\p^d_x\widetilde{\wp}(x)]^{n_d}$,
notice that $\p^d_x\widetilde{\wp}(x)$ with $d\geqslant 2$ can be transformed as
\begin{align}
&\p_x^{2d}\widetilde{\wp}(x)=\sum_{m=0}^{d+1}c_{m}\widetilde{\wp}^{\;m}(x), \notag\\
&\p_x^{2d+1}\widetilde{\wp}(x)=\left[\sum_{m=0}^{d}c_{m}\widetilde{\wp}^{\;m}(x)\right]\p_x\widetilde{\wp}(x),
\label{Derivatives4Weierstrass}
\end{align}
where $c_{m}$ are constant coefficients whose explicit expressions are not important here. For a factor with even $d$, the derivative disappears;
for a factor with odd $d$, a derivative factor $[\p_x\widetilde{\wp}(x)]^{n_d}$ still remains.
Using the basic equation (\ref{BasicEq4Weierstrass}), we can decrease an even exponent $n_d$ until $n_d^{\prime}=0$,
and decrease an odd exponent $n_d$ until $n_d^{\prime}=1$, so a further simplification follows,
\begin{align}
&[\p_x^{2d+1}\widetilde{\wp}(x)]^{2n}=\sum_{m=0}^{(2d+3)n}c_{m}\widetilde{\wp}^{\;m}(x), \notag\\
&[\p_x^{2d+1}\widetilde{\wp}(x)]^{2n+1}=\left[\sum_{m=0}^{(2d+3)n+d}c_{m}\widetilde{\wp}^{\;m}(x)\right]\p_x\widetilde{\wp}(x).
\end{align}
We plug all factors $[\p^{d_s}_x\widetilde{\wp}(x+\omega_s)]^{n_{d,s}}$ with the same $s$-index into the second product $\prod_{d_s}$ in (\ref{KdVHamiltonianDtype}).
If there are $n_s\geqslant 2$ factors with odd numbers $d_s$ and $n_{d,s}$, then a derivative factor $[\p_x\widetilde{\wp}(x+\omega_s)]^{n_s}$ appears.
Using the basic equation (\ref{BasicEq4Weierstrass}) again, we decrease an even exponent $n_s$ to $n_s^{\prime}=0$, or decrease an odd exponent $n_s$ to $n_s^{\prime}=1$ .
After these simplification we get
\be
\prod_{d_s}[\p_x^{d_s}\widetilde{\wp}(x+\omega_s)]^{n_{d,s}}=\left[\sum_{m_s^{\prime}=0}c_{m_s^{\prime}}\widetilde{\wp}^{\;m_s^{\prime}}(x+\omega_s)\right][\p_x\widetilde{\wp}(x+\omega_s)]^{n_s^{\prime}},
\label{KdVHamiltonianDtypeFirstProd}
\ee
for a definite index $s$.

Applying the first product $\prod_s$ to (\ref{KdVHamiltonianDtypeFirstProd}) leads to a polynomial with monomials in the general form
\be
\prod_{s=0}^{3}\widetilde{\wp}^{\;m_s^{\prime}}(x+\omega_s)[\p_x\widetilde{\wp}(x+\omega_s)]^{n_s^{\prime}},
\ee
with $n_s^{\prime}=0$ or $1$. The orders of pole of $\mathbf{V}_{2\ell-2}$ or $\mathbf{U}_{2\ell-2}$ require the exponents to satisfy the condition $2m_s^{\prime}+3n_s^{\prime} \leqslant 2\ell-2m_s$.
If in a monomial all $n_s^{\prime}=0$, there is no derivative factor anymore. Otherwise, it still contains a derivative factor which can be dissolved in pair by
\begin{align}
&\p_x\widetilde{\wp}(x)\p_x\widetilde{\wp}(x+\omega_i)=-4(e_i-e_j)(e_i-e_k)[\widetilde{\wp}(x)+\widetilde{\wp}(x+\omega_i)+e_i-2\zeta_1],\notag\\
&\p_x\widetilde{\wp}(x+\omega_j)\p_x\widetilde{\wp}(x+\omega_k)=-4(e_j-e_i)(e_k-e_i)[\widetilde{\wp}(x+\omega_j)+\widetilde{\wp}(x+\omega_k)+e_i-2\zeta_1],\notag\\
\end{align}
with indices $i,j,k\in\lbrace1,2,3\rbrace$ and $i\ne j\ne k$. For $\sum_s n_s^{\prime}=2$ or $4$, the derivative factor is totally dissolved; for $\sum_s n_s^{\prime}=1$ or $3$,
a single derivative $\p_x\widetilde{\wp}(x+\omega_s)$ is left. In fact, for the D-type factors from $\mathbf{V}_{2\ell-2}$ and $\mathbf{U}_{2\ell-2}$, derivative factors are indeed totally dissolved.
This is because although $\mathbf{V}_{2\ell-2}$ and $\mathbf{U}_{2\ell-2}$ generally contain derivatives $\p_x^{d_s}\widetilde{\wp}(x+\omega_s)$ with both even and odd order $d_s$,
the total order of the $\p_x$-operator in a monomial is always an even number which ranges from 2 to $2\ell-2$, as in the first few expressions (\ref{KdVHamiltonianVandU}).
Every step of the procedure explained above decreases the total order of the $\p_x$-operator by an even number, so eventually no $\p_x$-operator is left.

In summary, the first step simplification transforms the D-type factor (\ref{KdVHamiltonianDtype}), and hence the monomial (\ref{KdVHamiltonianDTVGeneral}), into a polynomial that contains only P-type terms.
At this point, $v_\ell(x)$ are represented in the form
\begin{align}
v_{2\ell-1}(x)&=\sum_{\{m_s\}}c_{\{m_s\}}\prod_{s=0}^{3}\widetilde{\wp}^{\;m_s}(x+\omega_s), \notag\\
v_{2\ell}(x)&=\p_x\left(\sum_{\{m_s\}}c_{\{m_s\}}\prod_{s=0}^{3}\widetilde{\wp}^{\;m_s}(x+\omega_s)\right),
\label{KdVHamiltonianDtypeSimplified}
\end{align}
with $m_s\leqslant \ell$.

The second step is to break the products and decrease the exponent $m_s$ in the P-type terms in (\ref{KdVHamiltonianDtypeSimplified}), until no products of functions $\widetilde{\wp}(x+\omega_s)$ are left.
When there are products of functions with different period shifts $\widetilde{\wp}^{\;m_s}(x+\omega_s)\widetilde{\wp}^{\;m_t}(x+\omega_t)$,
repeatedly use the following relations to break products and decrease indices $m_s$ and $m_t$ until the smaller one reaches zero,
\begin{align}
&\widetilde{\wp}(x)\widetilde{\wp}(x+\omega_i)=(e_i+\zeta_1)[\widetilde{\wp}(x)+\widetilde{\wp}(x+\omega_i)+e_i-2\zeta_1]+(e_j+\zeta_1)(e_k+\zeta_1),\notag\\
&\widetilde{\wp}(x+\omega_j)\widetilde{\wp}(x+\omega_k)=(e_i+\zeta_1)[\widetilde{\wp}(x+\omega_j)+\widetilde{\wp}(x+\omega_k)+e_i-2\zeta_1]+(e_j+\zeta_1)(e_k+\zeta_1).
\end{align}
It leaves power functions with the same period shift $\widetilde{\wp}^{\;m_s}(x+\omega_s)$ with $m_s\geqslant 0$.
All power functions $\widetilde{\wp}^{\;m_s}(x+\omega_s)$ are then traded by functions with derivative terms,
using the following relation derived from (\ref{Derivatives4Weierstrass}),
\be
\widetilde{\wp}^{\;m}(x)=\mbox{constant}+c_0\widetilde{\wp}(x)+\sum_{d=1}^{m-1}c_{d}\p_x^{2d}\widetilde{\wp}(x).
\label{Powers4Weierstrass}
\ee
The second step simplification transforms the P-type terms into linear combinations of constants,
$\widetilde{\wp}(x+\omega_s)$ and even order derivative terms $\p_x^{2d_s}\widetilde{\wp}(x+\omega_s)$ with $d_s\geqslant 1$.
Then $\p_x(\text{P-type terms})$ are linear combinations of odd order derivative terms $\p_x^{2d_s-1}\widetilde{\wp}(x+\omega_s)$ with $d_s\geqslant 1$.
The orders of derivative rise again, but there is no products anymore, an expression in this form is exactly what we need to evaluate the integral of $v_\ell(x)$.

In the end, $v_\ell(x)$ are simplified to a form ready for computation of the eigenvalue and the eigenfunction, in the shifted Weierstrass zeta function they are
\begin{align}
v_{2\ell-1}(x)&=\text{constant}+\sum_{s=0}^{3}\sum_{d=1}^{\ell}c_{2\ell-1,s,d}\p_x^{2d-1}\widetilde{\zeta}(x+\omega_s), \notag\\
v_{2\ell}(x)&=\sum_{s=0}^{3}\sum_{d=1}^{\ell}c_{2\ell,s,d}\p_x^{2d}\widetilde{\zeta}(x+\omega_s).
\label{KdVHamiltonianDTVFinalForm}
\end{align}
The ``constant'' part in $v_{2\ell-1}(x)$ equals the integral $\varepsilon_\ell$ of eq. (\ref{IntegralDensityOfHamiltonians}).
The coefficients $c_{\ell,s,d}$ are related to, but not identical with the coefficients $c_{l,s,d}$ in the wave function (\ref{EigenfunctionWeakCoupling}) despite they are represented by similar symbols.
The upper bound of $d$ comes from the order of poles for $v_\ell(x)$.

\subsection{Landen transformations}\label{LandenTransformations}

Landen transformations relate elliptic functions with different nome/modulus and arguments, there are three types, the {\em Descending, Ascending} and {\em Irrational} transformations.
See chapter 13.23 of \cite{Erdelyi2}.
The descending transformation is associated with the limits $b_0=b_1=\mathbf{b}, b_2=b_3=0$ and $b_0=b_1=0, b_2=b_3=\mathbf{b}$.
The ascending transformation, also called {\em Gauss' transformation} in \cite{Erdelyi2}, is associated with the limits $b_0=b_2=\mathbf{b}, b_1=b_3=0$ and $b_0=b_2=0, b_1=b_3=\mathbf{b}$.
The irrational transformation, also called {\em generalized transformation} in \cite{NISTDLMF}, is associated with the limits $b_0=b_3=\mathbf{b}, b_1=b_2=0$ and $b_0=b_3=0, b_1=b_2=\mathbf{b}$.

For the shifted Weierstrass elliptic function $\widetilde{\wp}(x;2\omega_1,2\omega_2)$ the three transformations are the following in order\footnote{The irrational Landen transformation is wrongly missed in the discussion of the previous paper \cite{wh1306}.}, with periods explicitly displayed,
\begin{align}
&\widetilde{\wp}(x;\omega_1,2\omega_2)=\widetilde{\wp}(x;2\omega_1,2\omega_2)+\widetilde{\wp}(x+\omega_1;2\omega_1,2\omega_2),\notag\\
&\widetilde{\wp}(x;2\omega_1,\omega_2)=\widetilde{\wp}(x;2\omega_1,2\omega_2)+\widetilde{\wp}(x+\omega_2;2\omega_1,2\omega_2),\notag\\
&\widetilde{\wp}(x;2\omega_1,\omega_3)=\widetilde{\wp}(x;2\omega_1,2\omega_2)+\widetilde{\wp}(x+\omega_3;2\omega_1,2\omega_2).
\label{LandenWeierstrasselliptic}
\end{align}
The constant $\zeta_1$ is given by $\zeta_1\equiv\zeta(\omega_1;2\omega_1,2\omega_2)/\omega_1=\zeta_1(2\omega_1,2\omega_2)$.
To derive the relations (\ref{LandenWeierstrasselliptic}), three relations $\zeta_1(\omega_1,2\omega_2)=2\zeta_1+e_1, \zeta_1(2\omega_1,\omega_2)=2\zeta_1+e_2$ and $\zeta_1(2\omega_1,\omega_3)=2\zeta_1+e_3$ are used.

To take the limit for the weak coupling wave function (\ref{EigenfunctionWeakCoupling}),
we need the transformations expressed in the shifted Weierstrass zeta function,
\begin{align}
&\widetilde{\zeta}(x;\omega_1,2\omega_2)=\widetilde{\zeta}(x;2\omega_1,2\omega_2)+\widetilde{\zeta}(x+\omega_1;2\omega_1,2\omega_2)-\zeta(\omega_1;2\omega_1,2\omega_2),\notag\\
&\widetilde{\zeta}(x;2\omega_1,\omega_2)=\widetilde{\zeta}(x;2\omega_1,2\omega_2)+\widetilde{\zeta}(x+\omega_2;2\omega_1,2\omega_2)-\zeta(\omega_2;2\omega_1,2\omega_2),\notag\\
&\widetilde{\zeta}(x;2\omega_1,\omega_3)=\widetilde{\zeta}(x;2\omega_1,2\omega_2)+\widetilde{\zeta}(x+\omega_3;2\omega_1,2\omega_2)-\zeta(\omega_3;2\omega_1,2\omega_2).
\label{LandenWeierstrasszeta}
\end{align}
Notice that $\widetilde{\zeta}(x+\omega_s;2\omega_1,2\omega_2)=\zeta(x+\omega_s;2\omega_1,2\omega_2)-\zeta_1x$.
To take the limit for the instanton partition function given in section \ref{InstantonPartitionFunctionNf=4SuperQCD}, we need the transformations  expressed in the elliptic theta function,
\begin{align}
&\f{\vartheta_1(z,p)\vartheta_2(z,p)}{\vartheta_1(2z,p^2)}=\f{\vartheta_3(z,p)\vartheta_4(z,p)}{\vartheta_4(2z,p^2)}=\vartheta_4(0,p^2),\notag\\
&\f{\vartheta_1(z,p)\vartheta_4(z,p)}{\vartheta_1(z,\sqrt{p})}=\f{\vartheta_2(z,p)\vartheta_3(z,p)}{\vartheta_2(z,\sqrt{p})}=\f{1}{2}\vartheta_2(0,\sqrt{p}),\notag\\
&\f{\vartheta_1(z,p)\vartheta_3(z,p)}{\vartheta_1(z,-\sqrt{p})}=\f{\vartheta_2(z,p)\vartheta_4(z,p)}{\vartheta_2(z,-\sqrt{p})}=\f{1}{2}\vartheta_2(0,-\sqrt{p}).
\label{LandenJacobitheta}
\end{align}
See section 20.7 of \cite{NISTDLMF}, and \cite{Walker2012}.
Notice that the nome of $\vartheta_r(z,-\sqrt{p})$ in the third transformation of (\ref{LandenJacobitheta}) looks different from the nome used in \cite{NISTDLMF} and \cite{Walker2012},
because we follow the convention of theta function used in \cite{wh1412} so that $\vartheta_r(z,\mathfrak{q}_{\mathrm{here}})=\vartheta_r(z,\mathfrak{q}^2_{\mathrm{nist}})$.
Landen transformation for the Jacobian elliptic function can be found in \cite{Erdelyi2}.

\section{Effective coupling constants $\Omega_m$ in super QCD masses}\label{EffectiveCouplingOmegaMassRelation}
\setcounter{equation}{0}

The effective coupling constants $\Omega_m$ and $\mho_m$ are complicated functions of $b_s$ and $q$, after substituting the relations (\ref{b2murelation}) for $b_s$,
they become rational functions of $\mu_i^{\f{1}{2}},q$ and $\epsilon$.
When $\epsilon$ is treated as a quantization parameter, the classical limit of $b_s$ are $b_0\sim\epsilon^{-2}(\mu_1-\mu_2)^2$ and similar for $b_1, b_2, b_3$.
We give $q$-series of the effective coupling constants $\Omega_m$ up to $m=4$ in the leading order of the $\epsilon$-expansion.

In the following expressions indices $i,j,k\in\lbrace 1,2,3,4\rbrace$.
The vacuum energy at $\eta_{*5}$ is
\begin{align}
\epsilon^2\Lambda_*=&-8(\mu_1\mu_2\mu_3\mu_4)^{\f{1}{2}} q^{\f{1}{2}}+\frac{q}{\mu_1\mu_2\mu_3\mu_4}\Bigg(\sum\limits_{i<j<k}\mu_i^2\mu_j^2\mu_k^2-\mu_1\mu_2\mu_3\mu_4\sum\limits_{i}\mu_i^2\Bigg) \notag\\
&+\frac{q^{\f{3}{2}}}{2^2(\mu_1\mu_2\mu_3\mu_4)^{\f{5}{2}}}\Bigg(\sum\limits_{i<j<k}\mu_i^4\mu_j^4\mu_k^4-2(\mu_1\mu_2\mu_3\mu_4)^2\sum\limits_{i<j}\mu_i^2\mu_j^2+8(\mu_1\mu_2\mu_3\mu_4)^4\Bigg) \notag\\
&+\mathcal{O}(q^2).
\end{align}
The effective coupling constants are expanded as $q$-series whose coefficients are functions of masses. They are given by
\begin{align}
\epsilon^2\Omega_2=&-16(\mu_1\mu_2\mu_3\mu_4)^{\f{1}{2}}q^{\f{1}{2}}+\frac{2^3q}{\mu_1\mu_2\mu_3\mu_4}\sum\limits_{i<j<k}\mu_i^2\mu_j^2\mu_k^2 +\frac{q^{\f{3}{2}}}{2(\mu_1\mu_2\mu_3\mu_4)^{\f{5}{2}}}\Bigg(7\sum\limits_{i<j<k}\mu_i^4\mu_j^4\mu_k^4\notag\\
&-14(\mu_1\mu_2\mu_3\mu_4)^2\sum\limits_{i<j}\mu_i^2\mu_j^2+24(\mu_1\mu_2\mu_3\mu_4)^3\Bigg)+\mathcal{O}(q^2).\\
\epsilon^2\Omega_3=&-\f{2^3i q}{\mu_1\mu_2\mu_3\mu_4}\sum\limits_{i<j<k}(-1)^j\mu_i^2\mu_j^2\mu_k^2 -\f{2^2iq^{\f{3}{2}}}{(\mu_1\mu_2\mu_3\mu_4)^{\f{5}{2}}}\Bigg(\sum\limits_{i<j<k}(-1)^j\mu_i^4\mu_j^4\mu_k^4\notag\\
&-4(\mu_1\mu_2\mu_3\mu_4)^2(\mu_1^2\mu_2^2-\mu_3^2\mu_4^2)\Bigg)+\mathcal{O}(q^2).\\
\epsilon^2\Omega_4=&-\frac{2^3q}{\mu_1\mu_2\mu_3\mu_4}\sum\limits_{i<j<k}\mu_i^2\mu_j^2\mu_k^2 -\frac{2^2q^{\f{3}{2}}}{(\mu_1\mu_2\mu_3\mu_4)^{\f{5}{2}}}\Bigg(\sum\limits_{i<j<k}\mu_i^4\mu_j^4\mu_k^4-2(\mu_1\mu_2\mu_3\mu_4)^2\sum\limits_{i<j}\mu_i^2\mu_j^2\notag\\
&+2(\mu_1\mu_2\mu_3\mu_4)^3-7(\mu_1\mu_2\mu_3\mu_4)^2(\mu_1^2\mu_2^2+\mu_3^2\mu_4^2)\Bigg)+\mathcal{O}(q^2).
\end{align}
The coupling constants $\mho_m$ are obtained from eq. (\ref{mhofromOmega}).
It can be shown that $\mho_m$ and $\Omega_m$ with the same $m$-index have the same order of magnitude.

\section{More about the spectrum at $\eta_{*6}$}\label{MoreAboutDyonicSpectrum}
\setcounter{equation}{0}

The second strong coupling solution of the DTV potential is associated with the critical point $\eta_{*6}$.
The locally valid effective potential $u(\varkappa)$ is used to compute the spectral solution.
The integrand of the unnormalized wave function $\psi(\varkappa)=\exp(\int v(\varkappa)d\varkappa)$ satisfies the equation $v^2(\varkappa)+\p_\varkappa v(\varkappa)=u(\varkappa)+\sigma$.
For the series solution expanded as in (\ref{Vlargemho2expansion}), the general form of $v_{2\ell}(\varkappa)$ is
\be
v_{2\ell}(\varkappa)=\p_\varkappa\left(\tilde{c}_{2\ell}^{\;\prime}\ln\cn\varkappa+\sum_{n=-\infty}^{2\ell}\f{\tilde{c}_{2\ell,2n}^{\;\prime}}{\cn^{2n}\varkappa}\right)
+\p^2_\varkappa\left(\tilde{c}_{2\ell}^{\;\prime\prime}\ln\cn\varkappa+\sum_{n=-\infty}^{2\ell-3}\f{\tilde{c}_{2\ell,2n}^{\;\prime\prime}}{\cn^{2n}\varkappa}\right),\label{vevendyonic}
\ee
where the coefficient of constant terms $\tilde{c}_{2\ell,0}^{\;\prime}$ and $\tilde{c}_{2\ell,0}^{\;\prime\prime}$ are all zero, and $\tilde{c}_{2\ell}^{\;\prime}=0$ for $\ell \geqslant 1$.
The general form of $v_{2\ell+1}(\varkappa)$ is
\be
v_{2\ell+1}(\varkappa)=\sum_{n=-\infty}^{2\ell+1}\f{\tilde{c}_{2\ell+1,2n+1}}{\cn^{2n+1}\varkappa}
+\p_\varkappa\left(\sum_{n=-\infty}^{2\ell-2}\f{\tilde{c}_{2\ell+1,2n+1}^{\;\prime}}{\cn^{2n+1}\varkappa}\right).
\label{vodddyonic}
\ee
The monodromy relation $\mu(\sigma)$ is given by the integral of $v(\varkappa)$ over an interval of the period $2\mathbf{K}+2i\mathbf{K}^{\prime}$,
from which the eigenvalue expansion $\sigma(\mu)$ is computed.
According to the prescription of contour integration explained in \cite{wh1412}, only terms in the first part of $v_{2\ell+1}(\varkappa)$ with $2n+1>0$ contribute non-vanishing integrals, they lead to the relation (\ref{FloquetExponentDefDyonicsimplified}).

The eigenvalue $\sigma$ is obtained from the monodromy relation (\ref{FloquetExponentDefDyonic}), or more directly from application of the transformation (\ref{EigenvalueDyonicFromMagnetic}) to the eigenvalue $\delta$. It can be expressed as
\begin{align}
\sigma=&2\mho_2^{\f{1}{2}}\mu+\f{1}{2^3}(1-2q)\left(\f{4\mu^2}{1-q}+1\right)\notag\\
&-\f{1}{2^5\mho_2^{\f{1}{2}}}\left\{\f{(1-2q)^2}{(1-q)^{\f{1}{2}}}\left(\f{4\mu^3}{(1-q)^{\f{3}{2}}}+\f{3\mu}{(1-q)^{\f{1}{2}}}\right)+4q(1-q)^{\f{1}{2}}\left(\f{4\mu^3}{(1-q)^{\f{3}{2}}}+\f{5\mu}{(1-q)^{\f{1}{2}}}\right)\right\}\notag\\
&+\f{1}{2^{10}\mho_2}\left\{\f{1-2q}{1-q}\left(\f{80\mu^4}{(1-q)^2}+\f{136\mu^2}{1-q}+9\right)-384\mho_4(1-q)\left(\f{4\mu^2}{1-q}+1\right)\right\}\notag\\
&+\mathcal{O}(\mho_2^{-\f{3}{2}}).
\label{EigenvalueStrongCouplingDyonic}
\end{align}
The corresponding unnormalized wave functions are obtained, it is
\begin{align}
\psi_\pm(\varkappa)=&\exp\Bigg(\mp i\mho_2^{\f{1}{2}}q^{-\f{1}{2}}\ln(\dn\varkappa+iq^{\f{1}{2}}\sn\varkappa)\!-\!\f{1}{2}\Big\{\ln\cn\varkappa\mp\f{2\mu}{(1\!-\!q)^{\f{1}{2}}}\ln\f{\dn\varkappa\!+\!(1\!-\!q)^{\f{1}{2}}\sn\varkappa}{\cn\varkappa}\Big\} \notag\\
&\pm\f{i}{2^4\mho_2^{\f{1}{2}}}\Big\{\f{\pm8i(1-q)\mu+i\big[3(1-q)+4\mu^2\big]\sn\varkappa\dn\varkappa}{(1-q)\cn^2\varkappa}+2q^{\f{1}{2}}\ln(\dn\varkappa+iq^{\f{1}{2}}\sn\varkappa) \notag\\
&\quad+\f{8i}{3}\mho_3\cn^3\varkappa-4\mho_4\big[iq^{-1}\sn\varkappa\dn\varkappa-q^{-\f{3}{2}}(1-2q)\ln(\dn\varkappa+iq^{\f{1}{2}}\sn\varkappa)\big]+\mathcal{O}(\mho_5)\Big\} \notag\\
&+\f{1}{2^6\mho_2}\Big\{\f{(1-q)\big[12(1-q)+32\mu^2\big]\pm\big[38(1-q)\mu+8\mu^3\big]\sn\varkappa\dn\varkappa}{(1-q)\cn^4\varkappa} \notag\\
&\quad+\!\f{(1\!-\!2q)\big[3(1\!-\!q)+4\mu^2\big]\big[\pm\mu\sn\varkappa\dn\varkappa-2(1\!-\!q)\big]}{(1\!-\!q)^2\cn^2\varkappa}\!+\!16\mho_3(\pm2\mu\!-\!\sn\varkappa\dn\varkappa)\cn\varkappa \notag\\
&\quad+16\mho_4\big[\cn^2\varkappa\pm2i\mu q^{-\f{1}{2}}\ln(\dn\varkappa+iq^{\f{1}{2}}\sn\varkappa)\big]+\mathcal{O}(\mho_5)\Big\}+\mathcal{O}(\mho_2^{-\f{3}{2}})\Bigg).
\label{EigenfunctionStrongCouplingDyonic}
\end{align}

For solution at the dyonic critical point, the path of integration over an interval of $2\mathbf{K}+2i\mathbf{K}^{\prime}$ in the $\varkappa$-plane is chosen to cross the branch cut of $\ln\{[\dn\varkappa+(1-q)^{\f{1}{2}}\sn\varkappa]/\cn\varkappa\}$, give a monodromy $-i\pi$, but avoid the branch cut of $\ln(\dn\varkappa+iq^{\f{1}{2}}\sn\varkappa)$.
See the discussion in appendix B in \cite{wh1412}.
The remaining terms in the wave function contain $\ln\cn\varkappa, \cn^m\varkappa,\sn^{2m}\varkappa$ and $\sn\varkappa\dn\varkappa$, with $m\in\mathbb{Z}_+$, are invariant under the translation $\varkappa\to\varkappa+2\mathbf{K}+2i\mathbf{K}^{\prime}$. Therefore, the monodromy of wave function comes from a single term $\pm (1-q)^{-\f{1}{2}}\mu\ln\{[\dn\varkappa+(1-q)^{\f{1}{2}}\sn\varkappa]/\cn\varkappa\}$. The canonically conjugate coordinate $\theta$ to the quasimomentum $\mu$ is defined by\footnote{The canonical coordinate can be defined in a slightly different way by\be (1-q)^{-\f{1}{2}}\ln\frac{\dn\varkappa+(1-q)^{\f{1}{2}}\sn\varkappa}{\cn\varkappa}=\theta,\nonumber \ee then the wave functions are written in a more familiar form as \be \psi_\pm(\varkappa)=\exp(\pm i\mu\theta)\mathbf{p}_3^{\pm}(\mu,\theta).\nonumber \ee The corresponding periodic translation of coordinate becomes $\theta\to\theta+i(1-q)^{-\f{1}{2}}\pi$.}
\be
\ln\frac{\dn\varkappa+(1-q)^{\f{1}{2}}\sn\varkappa}{\cn\varkappa}=\theta,\label{CanonCoordDyonicDef}
\ee
so that all terms of elliptic function in the wave functions can be rewritten in $\sinh\theta$ and $\cosh\theta$.
Then the wave functions take the Floquet form
\be
\psi_\pm(\varkappa)=\exp[\pm (1-q)^{-\f{1}{2}}\mu\theta]\mathbf{p}_3^{\pm}(\mu,\theta),
\label{DyonicWavefunctionFloquetForm}
\ee
where $\mathbf{p}_3^{\pm}(\mu,\theta)$ are periodic functions $\mathbf{p}_3^{\pm}(\mu,\theta+i\pi)=\mathbf{p}_3^{\pm}(\mu,\theta)$, they are related by $\mathbf{p}_3^{+}(\mu,\theta)=\mathbf{p}_3^{-}(-\mu,\theta)$.

Now the monodromy relation (\ref{FloquetExponentDefDyonic}) can be explained following the line of reasoning similar to that in subsection \ref{TheAsymptoticEigenfunctionEta5}. As $\mu$ and $\theta$  are canonically conjugate, the correct monodromy relation is
\be
\pm\f{i\mu\pi}{(1-q)^{\f{1}{2}}}=\ln\f{\psi_\pm(\theta+i\pi)}{\psi_\pm(\theta)}.
\ee
Or in the $\varkappa$-coordinate it is
\be
\pm\f{i\mu\pi}{(1-q)^{\f{1}{2}}}=\ln\f{\psi_\pm(\varkappa+2\mathbf{K}+2i\mathbf{K}^{\prime})}{\psi_\pm(\varkappa)}.
\ee

\section{Some details of computing the dual prepotential}\label{Details4DualPrepotentials}
\setcounter{equation}{0}

\subsection{Coefficients at the leading order of the $\epsilon$-expansion}

In section \ref{StrongCouplingExpansionPrepotentialNf4} the integral $a_{(0)}$ is expanded in the magnetic region as given by eq. (\ref{a0ExpansionMagnetic}),
the coefficients $\mbox{CE}_{2\ell+1}$ and $\mbox{CK}_{2\ell+1}$ can be written as series for large $\Omega_2$,
\begin{align}
\mbox{CE}_{2\ell+1}&=\sum_{n=-1}^{\infty}\mbox{ce}_{2\ell+1}^{2n+1}(\Omega_{m\geqslant3},q)\Omega_2^{-\f{2n+1}{2}},\notag\\
\mbox{CK}_{2\ell+1}&=\sum_{n=-1}^{\infty}\mbox{ck}_{2\ell+1}^{2n+1}(\Omega_{m\geqslant3},q)\Omega_2^{-\f{2n+1}{2}}.
\end{align}
Similarly, when $a_{(0)}$ is expanded in the dyonic region as given by eq. (\ref{a0ExpansionDyonic}),
the coefficients $\widetilde{\mbox{CE}}_{\ell}$ and $\widetilde{\mbox{CK}}_{\ell+1}$ can be written as series for large $\mho_2$,
\begin{align}
\widetilde{\mbox{CE}}_{\ell}&=\sum_{n=-1}^{\infty}\widetilde{\mbox{ce}}_{\ell}^{2n+1}(\mho_{m\geqslant3},q)\mho_2^{-\f{2n+1}{2}},\notag\\
\widetilde{\mbox{CK}}_{\ell+1}&=\sum_{n=-1}^{\infty}\widetilde{\mbox{ck}}_{\ell+1}^{2n+1}(\mho_{m\geqslant3},q)\mho_2^{-\f{2n+1}{2}}.
\end{align}
The coefficients $\mbox{ce}_{2\ell+1}^{2n+1}, \mbox{ck}_{2\ell+1}^{2n+1}$ are polynomials of higher order coupling constants $\Omega_{m\geqslant3}$ and $q$,
and $\widetilde{\mbox{ce}}_{\ell}^{2n+1}, \widetilde{\mbox{ck}}_{\ell+1}^{2n+1}$ are polynomials of $\mho_{m\geqslant3}$ and $q$.
For example, the first few $\mbox{ce}_{1}^{2n+1}$ are
\begin{align}
\mbox{ce}_{1}^{-1}&=i\left(1+\f{1}{3}q+\f{1}{5}q^2+\cdots\right),\notag\\
\mbox{ce}_{1}^{1}&=i\left(\f{1}{3}+\f{2}{15}q+\f{3}{35}q^2+\cdots\right)\Omega_4+i\left(\f{4}{15}+\f{4}{35}q+\f{8}{105}q^2+\cdots\right)\Omega_6+\mathcal{O}(\Omega_8),\notag\\
\mbox{ce}_{1}^{3}&=-i\left(\f{1}{60}-\f{1}{210}q-\f{1}{1260}q^2+\cdots\right)\Omega_3^2-i\left(\f{1}{15}+\f{1}{35}q+\f{2}{105}q^2+\cdots\right)\Omega_4^2+\mathcal{O}(\Omega_3\Omega_5),\notag\\
\mbox{ce}_{1}^{5}&=i\left(\f{1}{70}\!-\!\f{1}{210}q\!-\!\f{1}{1155}q^2\!+\!\cdots\right)\Omega_3^2\Omega_4+i\left(\f{1}{35}\!+\!\f{4}{315}q\!+\!\f{2}{231}q^2\!+\!\cdots\right)\Omega_4^3+\mathcal{O}(\Omega_3\Omega_4\Omega_5).
\end{align}
The other three sets of coefficients have the same structure.

In the magnetic dual expansion of prepotential (\ref{MagneticPrepotentialInOmega}), the first few $q$-polynomial coefficients $\mathbf{C}^{n}_{2^{\ell}m_1^{\ell_1}m_2^{\ell_2}\cdots}(q)$ are
\begin{alignat}{3}
\mathbf{C}^1_{2^{-1}}&=i\left(1+\f{1}{3}q+\f{1}{5}q^2+\cdots\right),\quad &\mathbf{C}^1_{2^14^1}&=i\left(\f{1}{3}+\f{2}{15}q+\f{3}{35}q^2+\cdots\right),\notag\\
\mathbf{C}^1_{2^33^2}&=-i\left(\f{1}{60}-\f{1}{210}q-\f{1}{1260}q^2+\cdots\right),\quad &\mathbf{C}^1_{2^34^2}&=-i\left(\f{1}{15}+\f{1}{35}q+\f{2}{105}q^2+\cdots\right),\notag\\
\mathbf{C}^2_{2^0}&=\f{3}{4}+\f{1}{2}q+\f{1}{4}q^2+\cdots,\quad &\mathbf{C}^2_{2^24^1}&=-\f{1}{2}-\f{1}{6}q-\f{1}{10}q^2+\cdots,\notag\\
\mathbf{C}^2_{2^43^2}&=\f{1}{8}-\f{1}{40}q-\f{1}{280}q^2+\dots,&\quad \mathbf{C}^2_{2^44^2}&=\f{1}{4}+\f{1}{10}q+\f{9}{140}q^2+\cdots.
\end{alignat}
In the dyonic dual expansion of prepotential (\ref{DyonicPrepotentialInMho}), the first few $q$-polynomial coefficients $\mathbf{D}^{n}_{2^{\ell}m_1^{\ell_1}m_2^{\ell_2}\cdots}(q)$ are
\begin{alignat}{3}
\mathbf{D}^1_{2^{-1}}&=1+\f{1}{6}q+\f{3}{40}q^2+\cdots,\quad &\mathbf{D}^1_{2^14^1}&=\f{1}{3}+\f{1}{30}q+\f{3}{280}q^2+\cdots,\notag\\
\mathbf{D}^1_{2^33^2}&=-\f{1}{60}+\f{1}{280}q+\f{1}{2016}q^2+\cdots,\quad &\mathbf{D}^1_{2^34^2}&=-\f{1}{15}-\f{1}{210}q-\f{1}{840}q^2+\cdots,\notag\\
\mathbf{D}^2_{2^0}&=-\f{3}{4}+\f{1}{4}q+\f{1}{8}q^2+\cdots,\quad &\mathbf{D}^2_{2^24^1}&=\f{1}{2}-\f{1}{6}q-\f{1}{15}q^2+\cdots,\notag\\
\mathbf{D}^2_{2^43^2}&=-\f{1}{8}+\f{1}{10}q+\f{1}{280}q^2+\dots,&\quad \mathbf{D}^2_{2^44^2}&=-\f{1}{4}+\f{1}{10}q+\f{1}{28}q^2+\cdots.
\end{alignat}
The transformation between the dual expansions of prepotential $\mathcal{F}_D^{(0)}$ and $\mathcal{F}_T^{(0)}$ discussed in the end of subsection \ref{DyonicPrepotential4SuperQCD},
combined with the transformation between coupling constants $\Omega_m$ and $\mho_m$ given by (\ref{mhofromOmega}), leads to relations between the two sets of coefficients $\mathbf{C}^{n}_{2^{\ell}m_1^{\ell_1}m_2^{\ell_2}\cdots}(q)$ and $\mathbf{D}^{n}_{2^{\ell}m_1^{\ell_1}m_2^{\ell_2}\cdots}(q)$.

\subsection{Going to higher orders of the $\epsilon$-expansion}

To extend the computation of subsection \ref{MagneticPrepotential4SuperQCD} to incorporate the $\epsilon$-deformation,
we seek a solution of the equation $v^2(\chi)+\p_\chi v(\chi)=u(\chi)+\delta$ that is expanded as
\be v(\chi)=\sum_{n=-1}^{\infty}v_{(n)}(\chi),\ee
where $v_{(n)}(\chi)$ is of order $\epsilon^n$, the parameter $\epsilon$ is hidden in $\delta\sim\mathcal{O}(\epsilon^{-2})$ and $\Omega_m\sim\mathcal{O}(\epsilon^{-2})$. The first few are
\begin{align}
v_{(-1)}(\chi)&=(\delta+u)^{\f{1}{2}},\notag\\
v_{(0)}(\chi)&=-\f{\p_\chi u}{4(\delta+u)},\notag\\
v_{(1)}(\chi)&=\f{-5(\p_\chi u)^2+4(\delta+u)\p^2_\chi u}{32(\delta+u)^{\f{5}{2}}},\notag\\
v_{(2)}(\chi)&=\f{-15(\p_\chi u)^3+18(\delta+u)\p_\chi u\p^2_\chi u-4(\delta+u)^2\p^3_\chi u}{64(\delta+u)^4}.
\end{align}
The expressions are not ready for evaluating the integral over an interval of $2\mathbf{K}$ because the denominator of $v_{(n)}(\chi)$ is $(\delta+u)^{\f{3}{2}n+1}$, the effective potential $u(\chi)$ contains infinitely many terms.
The computation of the leading order expansion in section \ref{StrongCouplingExpansionPrepotentialNf4} indicates it is crucial to separate $\delta+\Omega_2\sn^2\chi$ from higher order coupling terms,
and use $\delta+\Omega_2\sn^2\chi$ as the expansion quantity. Therefore, $v_{(n)}(\chi)$ can be further expanded as a series for large $(\delta+\Omega_2\sn^2\chi)$.
Then the total integrand $v(\chi)$ is in the form
\be
v(\chi)=\sum_{L=-1}^{\infty}\f{\mathsf{P}^{\alpha}_{L}(\chi)}{(\delta+\Omega_2\sn^2\chi)^{\f{L}{2}}}.
\label{IntegrandExpandEpsilon2K}
\ee
One can also assume the expansion (\ref{IntegrandExpandEpsilon2K}) for $v(\chi)$ from the beginning and solve the equation $v^2(\chi)+\p_\chi v(\chi)=u(\chi)+\delta$ order by order.
Simplifying elliptic functions in the coefficient series $\mathsf{P}^{\alpha}_{L}(\chi)$ using the rules (\ref{substitutionright1}), we get
\begin{align}
\mathsf{P}^{\alpha}_{-1}(\chi)=&1,\quad\quad \mathsf{P}^{\alpha}_{0}(\chi)=0,\quad\quad \mathsf{P}^{\alpha}_{1}(\chi)=\mathsf{P}^{\alpha}_{(0);1}(\chi),\notag\\
\mathsf{P}^{\alpha}_{2}(\chi)=&-\f{1}{2^2}\Big\{2\Omega_2\sn\chi\cn\chi\dn\chi+3\Omega_3\sn^2\chi+4\Omega_4\sn^3\chi\cn\chi\dn\chi+\mathcal{O}(\sn^4\chi)\Big\}, \notag\\
\mathsf{P}^{\alpha}_{3}(\chi)=&\mathsf{P}^{\alpha}_{(0);3}(\chi)\!+\!\f{1}{2^2}\Big\{\Omega_2\!+\!3\Omega_3\sn\chi\cn\chi\dn\chi\!-\!2[(1\!+\!q)\Omega_2\!-\!3\Omega_4]\sn^2\chi\!+\!\mathcal{O}(\sn^3\chi\cn\chi\dn\chi)\Big\},\notag\\
\mathsf{P}^{\alpha}_{4}(\chi)=&\f{1}{2^2}\Big\{2\Omega_2\Omega_3\sn^4\chi+(2\Omega_2\Omega_4+3\Omega_3^2)\sn^5\chi\cn\chi\dn\chi+\mathcal{O}(\sn^6\chi)\Big\}\notag\\
&-\f{1}{2^3}\Big\{3\Omega_3-4[(1+q)\Omega_2-3\Omega_4]\sn\chi\cn\chi\dn\chi+\mathcal{O}(\sn^2\chi)\Big\}, \notag\\
\mathsf{P}^{\alpha}_{5}(\chi)=&\mathsf{P}^{\alpha}_{(0);5}(\chi)-\f{1}{2^5}\Big\{20\Omega_2^2\sn^2\chi+72\Omega_2\Omega_3\sn^3\chi\cn\chi\dn\chi+\mathcal{O}(\sn^4\chi)\Big\}\notag\\
&-\f{1}{2^2}\Big\{[(1+q)\Omega_2-3\Omega_3]+15[(1+q)\Omega_3-\Omega_5]\sn\chi\cn\chi\dn\chi+\mathcal{O}(\sn^2\chi)\Big\}.
\end{align}
$\mathsf{P}^{\alpha}_{(0);2\ell+1}(\chi)$ with $\ell\geqslant 0$ are given by (\ref{p0alphaExpansionMagneticCoeff}).
Notice that $\mathsf{P}^{\alpha}_{L}(\chi)$ are divided up into parts, each in a bracket $\{\cdots\}$, according to their order of magnitude measured by the power of $\epsilon$.
The leading order parts in both $\mathsf{P}^{\alpha}_{2\ell+1}(\chi)$ and $\mathsf{P}^{\alpha}_{2\ell+2}(\chi)$ are of order $\epsilon^{-2\ell-2}$,
terms in subsequent brackets are of higher order $\epsilon^{-2\ell}, \epsilon^{-2\ell+2},\cdots,\epsilon^{-2}$.
For $\mathsf{P}^{\alpha}_{L}(\chi)$ with $L\geqslant 1$, the Taylor series in each bracket contains infinitely many terms because it involves all higher order coupling constants $\Omega_{m\geqslant 3}$.

Now the integrals over an interval of $2\mathbf{K}$ are the same form as those in eqs. (\ref{IntegralS2m+1}) and (\ref{IntegralS2m}) used in section \ref{MagneticPrepotential4SuperQCD},
but with different range of value for the indices.
We denote them by $\mathrm{S}_{L,2m+1}$ and $\mathrm{S}_{L,2m}$, where the index $L$ takes value of all integers $L\geqslant -1$,
and for every $L\geqslant 1$ the index $m$ takes value of some positive integers $m\geqslant 0$.
The integrals needed to compute the dual expansions of prepotential for the $\epsilon$-deformed gauge theory are explained in appendixes \ref{ContourIntegral} and \ref{RecurrenceRelationMellm}.

\section{More prescriptions of contour integrals}\label{ContourIntegral}
\setcounter{equation}{0}

To compute the spectra of elliptic Hill potentials (\ref{EllipticHillOperatorSN}) and (\ref{EllipticHillOperatorCN}) in section \ref{StrongCouplingSpectraAndGaugeTheoryDuality}, we need integrals $\mathrm{I}_{-(2n+1)}$ over an interval of $2i\mathbf{K}^{\prime}$ and $\mathrm{J}_{-(2n+1)}$ over an interval of $2\mathbf{K}+2i\mathbf{K}^{\prime}$.
Integrals over an interval of a period in the $\chi$-coordinate are transformed to integrals along a closed contour in the $\xi$-coordinate through the transformation $\sn^2\chi=\xi$;
similarly integrals over an interval of a period in the $\varkappa$-coordinate are transformed through $\sn^2\varkappa=\xi$. From the first few ones $\mathrm{I}_{-1},\mathrm{I}_{-3}$ and $\mathrm{J}_{-1},\mathrm{J}_{-3}$, other integrals are computed by recurrence relations as already explained in \cite{wh1412}.

To compute the strong coupling expansion of gauge theory prepotential, we encounter some other integrals of the Jacobian elliptic functions over an interval of the period $2\mathbf{K}$.
We explain integrals needed to compute the $\epsilon$-deformed gauge theory by transforming these integrals in the $\chi$-coordinate (or the $\varkappa$-coordinate) to contour integrals in the $\xi$-coordinate.

\subsection{Contour integrals for the magnetic dual expansion}

\begin{figure}[hbt]
\begin{minipage}[t]{0.5\linewidth}
\centering
\includegraphics[width=6cm]{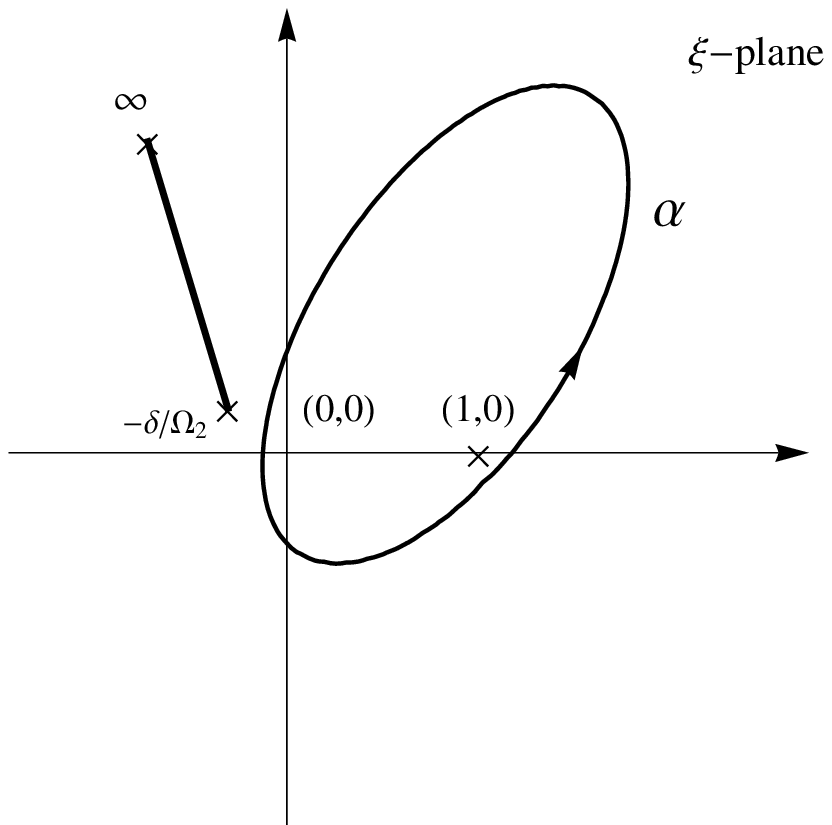}
\caption{Integral contour for (\ref{MagneticAlphaContourIntegral1}).} \label{alphacontourmagnetic1}
\end{minipage}%
\begin{minipage}[t]{0.5\linewidth}
\centering
\includegraphics[width=6cm]{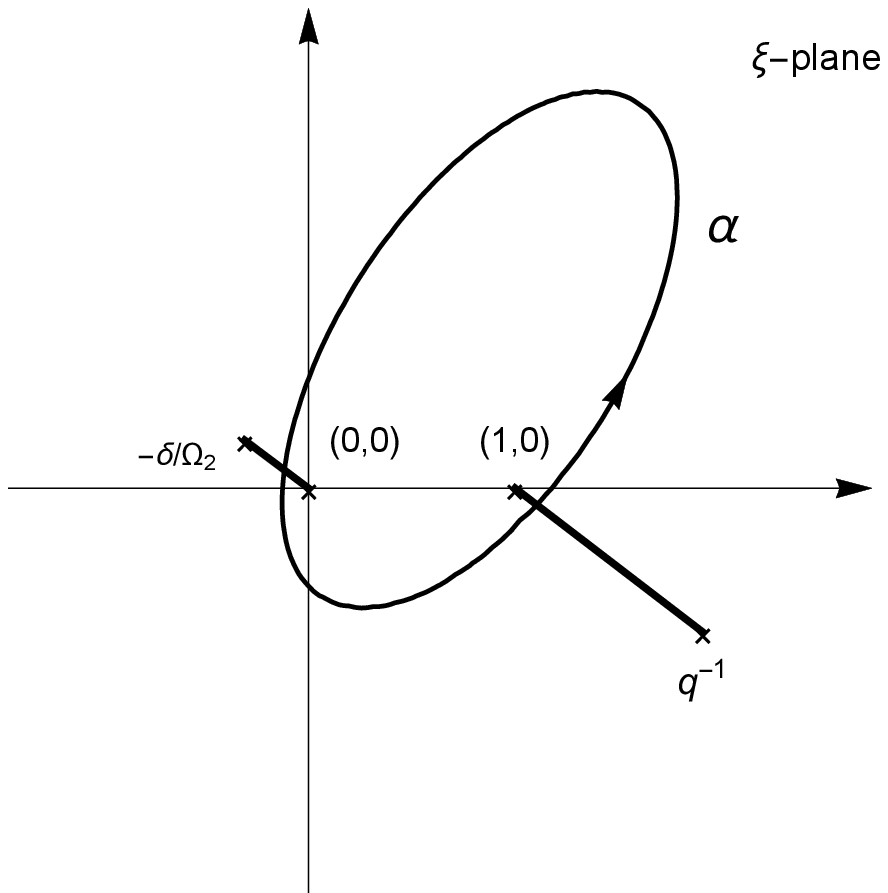}
\caption{Integral contour for (\ref{MagneticAlphaContourIntegral2}).} \label{alphacontourmagnetic2}
\end{minipage}
\end{figure}

By the transformation $\sn^2\chi=\xi$, the first class of integrals in subsection \ref{MagneticPrepotential4SuperQCD} are
\be
\mathrm{S}_{L,2m+1}=\int_{\chi_0}^{\chi_0+2\mathbf{K}}\f{\sn^{2m+1}\chi\cn\chi\dn\chi}{(\delta+\Omega_2\sn^2\chi)^{\f{L}{2}}}d\chi
=\f{1}{2}\oint_\alpha\f{\xi^{m}}{(\delta+\Omega_2\xi)^{\f{L}{2}}}d\xi,
\label{MagneticAlphaContourIntegral1}
\ee
with magnitude $\vert\f{\delta}{\Omega_2}\vert\ll 1$. The $\alpha$-contour is chosen to encircle the points $\xi=0$ and $1$, but avoid the point $\xi=-\f{\delta}{\Omega_2}$.
There are two cases:
\begin{enumerate}
\item For $L=2\ell\geqslant 0$ and $m\geqslant 0$, because the pole at $\xi=-\f{\delta}{\Omega_2}$ is avoided so $\mathrm{S}_{2\ell,2m+1}=0$;
\item For $L=2\ell+1\geqslant -1$ and $m\geqslant 0$, the integrand has a branch cut $[-\f{\delta}{\Omega_2},\infty)$ in the $\xi$-plane.
The contour is arranged to avoid the branch cut, as shown in figure \ref{alphacontourmagnetic1}, so that the integral vanishes, $\mathrm{S}_{2\ell+1,2m+1}=0$.
\end{enumerate}
Therefore, terms with $\sn^{2m+1}\chi\cn\chi\dn\chi$ in the polynomial $\mathsf{P}^{\alpha}_{L}(\chi)$ do not contribute to the integrals over an interval of $2\mathbf{K}$.

The second class of integrals are
\be
\mathrm{S}_{L,2m}=\int_{\chi_0}^{\chi_0+2\mathbf{K}}\f{\sn^{2m}\chi}{(\delta+\Omega_2\sn^2\chi)^{\f{L}{2}}}d\chi
=\f{1}{2}\oint_\alpha\f{\xi^{m-\f{1}{2}}}{(\delta+\Omega_2\xi)^{\f{L}{2}}(1-\xi)^{\f{1}{2}}(1-q\xi)^{\f{1}{2}}}d\xi.
\label{MagneticAlphaContourIntegral2}
\ee
The points $\xi=0,1,\f{1}{q}$ are branch points; the point $\xi=-\f{\delta}{\Omega_2}$ is a pole when $L$ is even, a branch point when $L$ is odd. Then the integrals are:
\begin{enumerate}
\item For $L=2\ell\geqslant 0$, the branch cuts are arranged to be $[0,\infty)$ and $[1,\f{1}{q}]$, the contour crosses both branch cuts so the integral is nonzero.
Rewriting elliptic functions in the amplitude $\varphi$ as in section \ref{MagneticPrepotential4SuperQCD}, we get
\begin{align}
\int_{\chi_0}^{\chi_0+2\mathbf{K}}\f{\sn^{2m}\chi}{(\delta+\Omega_2\sn^2\chi)^{\f{L}{2}}}d\chi=&2\int_0^{\f{\pi}{2}}\f{\sin^{2m}\varphi}{(\delta+\Omega_2\sin^2\varphi)^\ell}\f{d\varphi}{\sqrt{1-q\sin^2\varphi}} \notag\\
=&2\int_0^{\f{\pi}{2}}\f{\sin^{2m}\varphi(1+\f{1}{2}q\sin^2\varphi+\f{3}{8}q^2\sin^4\varphi+\cdots)}{\delta^\ell(1-\mathbf{x}^2\sin^2\varphi)^\ell}d\varphi,
\end{align}
with $\mathbf{x}^2=-\Omega_2/\delta$. To evaluate it the following basis integrals are needed,
\be
\mathrm{M}_{2\ell,2m}=\int_0^{\f{\pi}{2}}\f{\sin^{2m}\varphi}{(1-\mathbf{x}^2\sin^2\varphi)^{\ell}}d\varphi,
\label{IntegralM2L2m}
\ee
with $\vert\mathbf{x}^2\vert\gg 1$ and $m\geqslant 0$;
\item For $L=2\ell+1\geqslant -1$, the branch cuts are arranged to be $[0,-\f{\delta}{\Omega_2}]$ and $[1,\f{1}{q}]$ as shown in figure \ref{alphacontourmagnetic2},
the contour crosses both branch cuts so the integral is also nonzero.
They are given by the basis integrals $\mathrm{M}_{2\ell+1,2m}$ whose recurrence relations are given in appendix \ref{RecurrenceRelationMellm}.
\end{enumerate}

For computation for the leading order of the $\epsilon$-expansion, only integrals with $L=2\ell+1$ are needed. Their integral contours are shown in figure \ref{alphacontourmagnetic1} and figure \ref{alphacontourmagnetic2}.

\subsection{Contour integrals for the dyonic dual expansion}

The dyonic dual expansion of prepotential can be computed following a similar route as the magnetic dual expansion.
Here we explain only integrals used to compute the leading order of the $\epsilon$-expansion.

\begin{figure}[hbt]
\begin{minipage}[t]{0.5\linewidth}
\centering
\includegraphics[width=6cm]{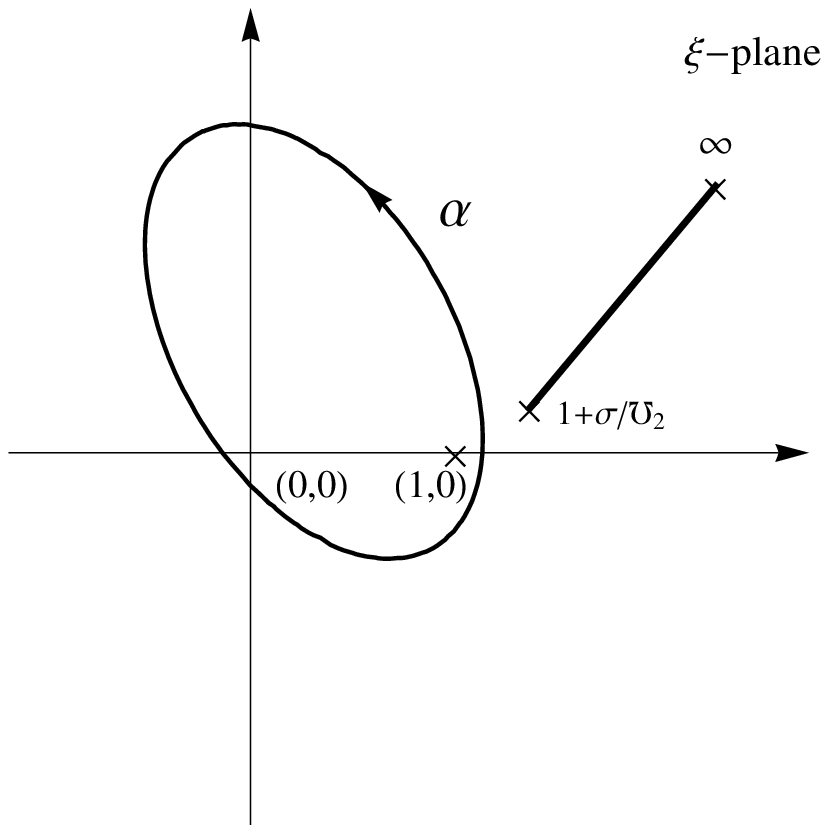}
\caption{Integral contour for (\ref{DyonicAlphaContourIntegral1}).} \label{alphacontourdyonic1}
\end{minipage}%
\begin{minipage}[t]{0.5\linewidth}
\centering
\includegraphics[width=6cm]{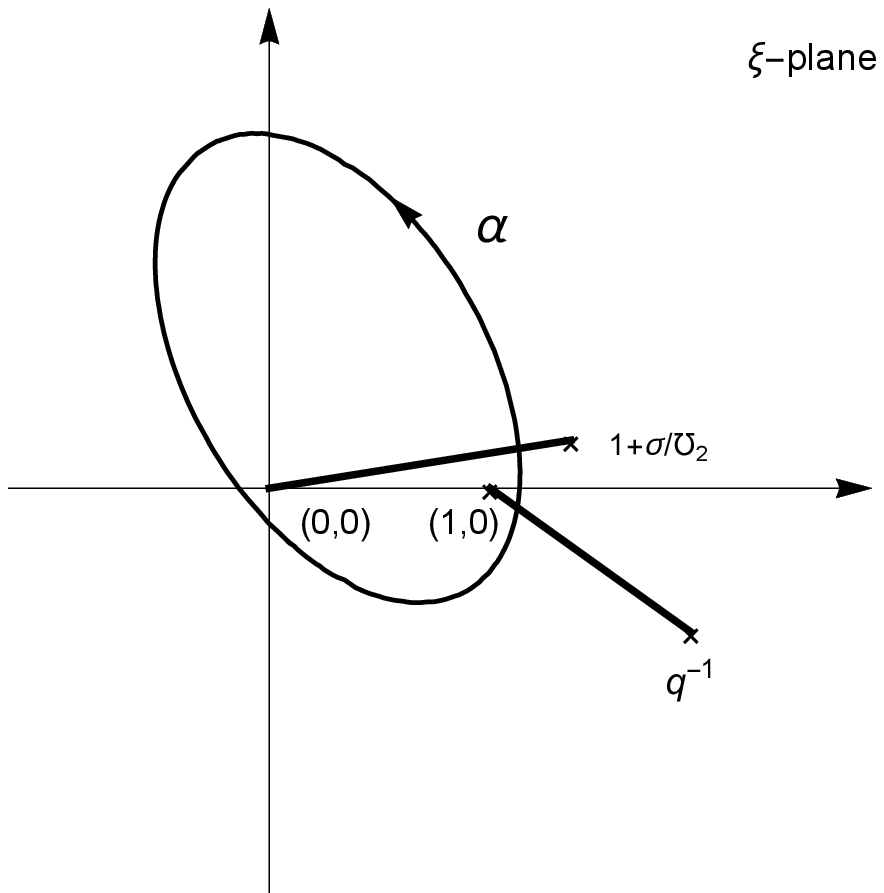}
\caption{Integral contour for (\ref{DyonicAlphaContourIntegral2}).} \label{alphacontourdyonic2}
\end{minipage}
\end{figure}

By transformation $\sn^2\varkappa=\xi$, the first class integrals $\mathrm{T}_{2\ell+1,2m+1}$ in subsection \ref{DyonicPrepotential4SuperQCD} are
\be
\int_{\varkappa_0}^{\varkappa_0+2\mathbf{K}}\f{\cn^{2m+1}\varkappa\sn\varkappa\dn\varkappa}{(\sigma+\mho_2\cn^2\varkappa)^{\f{2\ell+1}{2}}}d\varkappa
=\f{1}{2}\oint_\alpha\f{(1-\xi)^{m}}{(\sigma+\mho_2-\mho_2\xi)^{\f{2\ell+1}{2}}}d\xi,
\label{DyonicAlphaContourIntegral1}
\ee
with magnitude $\vert\f{\sigma}{\mho_2}\vert\ll 1$, the integrand has a branch cut $[1+\f{\sigma}{\mho_2},\infty)$. The second class integrals $\mathrm{T}_{2\ell+1,2m}$ are
\be
\int_{\varkappa_0}^{\varkappa_0+2\mathbf{K}}\f{\cn^{2m}\varkappa}{(\sigma+\mho_2\cn^2\varkappa)^{\f{2\ell+1}{2}}}d\varkappa
=\f{1}{2}\oint_\alpha\f{(1-\xi)^{m-\f{1}{2}}}{(\sigma+\mho_2-\mho_2\xi)^{\f{2\ell+1}{2}}\xi^{\f{1}{2}}(1-q\xi)^{\f{1}{2}}}d\xi,
\label{DyonicAlphaContourIntegral2}
\ee
the integrand has two branch cuts $[0,1+\f{\sigma}{\mho_2})$ and $[1,\f{1}{q}]$.
The integral contour is arranged as shown in figure \ref{alphacontourdyonic1} and figure \ref{alphacontourdyonic2}, to ensure that the integral (\ref{DyonicAlphaContourIntegral1}) vanishes because the contour avoids the branch cut $[1+\f{\sigma}{\mho_2},\infty)$, but the integral (\ref{DyonicAlphaContourIntegral2}) is nonzero because the contour crosses the branch cuts $[0,1+\f{\sigma}{\mho_2}]$  and $[1,\f{1}{q}]$.

The non-vanishing integrals (\ref{DyonicAlphaContourIntegral2}), after rewriting elliptic functions in the amplitude $\varphi$, are computed from the basis integrals $\mathrm{M}_{2\ell+1,2m}$ with modulus $\mathbf{x}^2=\mho_2/(\mho_2+\sigma)\sim 1$.

\section{Recurrence relations for $\mathrm{M}_{L,2m}$}\label{RecurrenceRelationMellm}
\setcounter{equation}{0}

By the transformation $\sin^2\varphi=\xi$, the integrals $\mathrm{M}_{L,2m}$ are given by a universal expression with the Gamma function and the hypergeometric function,
\begin{align}
\mathrm{M}_{L,2m}=&\int_0^{\f{\pi}{2}}\f{\sin^{2m}\varphi}{(1-\mathbf{x}^2\sin^2\varphi)^{\f{L}{2}}}d\varphi
=\f{1}{2}\int_0^1\xi^{m-\f{1}{2}}(1-\xi)^{-\f{1}{2}}(1-\mathbf{x}^2\xi)^{-\f{L}{2}} \notag\\
=&\f{\Gamma(\f{1}{2})\Gamma(m+\f{1}{2})}{2\Gamma(m+1)} {}_2F_1(\f{L}{2},m+\f{1}{2};m+1;\mathbf{x}^2).
\end{align}

\subsection{$L=2\ell+1$}

For the integrals $\mathrm{M}_{2\ell+1,2m}$ of (\ref{IntegralM2L+12m}), by the relations of the contiguous hypergeometric functions ${}_2F_1(a,b;c;\mathbf{x}^2)$ and ${}_2F_1(a\pm1,b;c;\mathbf{x}^2)$, ${}_2F_1(a,b\pm1;c;\mathbf{x}^2)$, ${}_2F_1(a,b;c\pm1;\mathbf{x}^2)$,
see section 15.5 of \cite{NISTDLMF}, the arguments $\ell$ and $m$ can be decreased so that in the end all hypergeometric functions are represented as linear combinations of ${}_2F_1(\pm\f{1}{2},\f{1}{2};1;\mathbf{x}^2)$.
The relations of the hypergeometric functions lead to the recurrence relations for $\mathrm{M}_{2\ell+1,2m}$ explained below, with which the expression (\ref{a0ExpansionMagnetic}) is obtained.

The first two initial integrals of the recurrence relations are the complete elliptic integrals of the first and second kinds,
\be
\mathrm{M}_{-1,0}=\mathbf{E}(\mathbf{x}^2),\qquad
\mathrm{M}_{1,0}=\mathbf{K}(\mathbf{x}^2).
\ee
Three recurrence relations given below generate all other integrals from $\mathrm{M}_{-1,0}$ and $\mathrm{M}_{1,0}$.
For $m=0$, integrals $\mathrm{M}_{2\ell+1,0}$ with $\ell\geqslant 1$ are generated by the first recurrence relation
\be \mathrm{M}_{2\ell+1,0}=\f{2(\ell-1)(2-\mathbf{x}^2)}{(2\ell-1)(1-\mathbf{x}^2)}\mathrm{M}_{2\ell-1,0}-\f{2\ell-3}{(2\ell-1)(1-\mathbf{x}^2)}\mathrm{M}_{2\ell-3,0}.
\ee
For $m=1$, integrals $\mathrm{M}_{2\ell+1,2}$ with $\ell\geqslant 0$ are generated by the second recurrence relation
\be
\mathrm{M}_{2\ell+1,2}=\f{\mathrm{M}_{2\ell+1,0}-\mathrm{M}_{2\ell-1,0}}{\mathbf{x}^2}.
\ee
Then the integrals with fixed $\ell$-index and increasing $m$-index are obtained from $\mathrm{M}_{2\ell+1,0}$ and $\mathrm{M}_{2\ell+1,2}$ by the third recurrence relation
\be
\mathrm{M}_{2\ell+1,2m}=\f{2[m-1+(m-\ell-1)\mathbf{x}^2]\mathrm{M}_{2\ell+1,2m-2}-(2m-3)\mathrm{M}_{2\ell+1,2m-4}}{(2m-2\ell-1)\mathbf{x}^2},
\ee
where $\ell\geqslant -1, m\geqslant 2$.

\subsection{$L=2\ell$}

For the integrals $\mathrm{M}_{2\ell,2m}$ of (\ref{IntegralM2L2m}) with  $\ell\geqslant 0, m\geqslant 0$, they are rational functions of $\mathbf{x}$ and $(1-\mathbf{x}^2)^{\f{1}{2}}$.

\end{document}